\documentclass[fleqn,usenatbib]{mnras}

\usepackage{newtxtext,newtxmath}
\usepackage[T1]{fontenc}
\usepackage{ae,aecompl}

\usepackage{graphicx}	% Including figure files
\usepackage{blindtext}  % Generates random text for testing
\usepackage{natbib}
\usepackage{rotating,times,graphicx,latexsym}
\usepackage{color}
\usepackage{longtable}
\usepackage{lscape}
\usepackage{lipsum} % To generate text
\usepackage{array}
\usepackage{flafter}

\usepackage{soul} % To add highlighting of text
\usepackage{xargs} % Use more than one optional parameter in a new commands
\usepackage[pdftex,dvipsnames]{xcolor}  % Coloured text etc.
\usepackage[colorinlistoftodos,prependcaption]{todonotes}
\newcommandx{\tobedone}[2][1=]{\todo[linecolor=red,backgroundcolor=red!25,bordercolor=red,inline,#1]{#2}}
\newcommandx{\changed}[2][1=]{\todo[linecolor=blue,backgroundcolor=blue!25,bordercolor=blue,inline,#1]{#2}\noindent}
\newcommandx{\thiswillnotshow}[2][1=]{\todo[disable,#1]{#2}}
\newcommandx{\carys}[2][1=]{\todo[linecolor=Orchid,backgroundcolor=Orchid!25,bordercolor=Orchid,inline,#1]{#2}\noindent}
\newcommandx{\dirk}[2][1=]{\todo[linecolor=BurntOrange,backgroundcolor=BurntOrange!25,bordercolor=BurntOrange,inline,#1]{#2}\noindent}
\newcommandx{\bst}[2][1=]{\todo[linecolor=SpringGreen,backgroundcolor=SpringGreen!40,bordercolor=SpringGreen,inline,#1]{#2}\noindent}
\newcommandx{\jochen}[2][1=]{\todo[linecolor=BurntOrange,backgroundcolor=SpringGreen!40,bordercolor=SpringGreen,inline,#1]{#2}\noindent}
\newcommandx{\josch}[2][1=]{\todo[linecolor=BurntOrange,backgroundcolor=BurntOrange!40,bordercolor=BurntOrange,inline,#1]{#2}\noindent}
\newcommand{\ha}{{H$\alpha$}}

\graphicspath{ {images/} }
\newcommand{\jaavso}{{The Journal of the American Association of Variable Star Observers}}

\title[$\sigma$\,Ori Variables]{A survey for variable young stars with small telescopes: V - Analysis of TX\,Ori, V505\,Ori, and V510\,Ori, the HST ULLYSES targets in the $\sigma$\,Ori cluster}

\author[Dirk Froebrich et al.]{Dirk Froebrich$^{1}$\thanks{E-mail: df@kent.ac.uk},
Jochen Eisl\"offel$^{2}$, 
Bringfried Stecklum$^{2}$,
Carys Herbert$^{3}$,
\newauthor
Franz-Josef Hambsch$^{4,5,6}$
\\
$^{1}$School of Physical Sciences, University of Kent, Canterbury CT2 7NH, UK\\
$^{2}$Th\"{u}ringer Landessternwarte, Sternwarte 5, 07778 Tautenburg, Germany\\
$^{3}$Centre for Astrophysics and Planetary Science, School of Physical Sciences, University of Kent, Canterbury CT2 7NH, UK\\
$^{4}$Vereniging Voor Sterrenkunde (VVS), Oostmeers 122 C, 8000 Brugge, Belgium \\
$^{5}$Bundesdeutsche Arbeitsgemeinschaft f\"{u}r Ver\"{a}nderliche Sterne e.V. (BAV), Munsterdamm 90, D-12169 Berlin, Germany\\
$^{6}$American Association of Variable Star Observers (AAVSO), 49 Bay State Road, Cambridge, MA 02138, USA\\
}

\date{Accepted XXX. Received YYY; in original form ZZZ}

\pubyear{2020}

\begin{document}
\label{firstpage}
\pagerange{\pageref{firstpage}--\pageref{lastpage}}
\maketitle

\begin{abstract}

Investigations of the formation of young stellar objects (YSOs) and planets require the detailed analysis of individual sources as well as statistical analysis of a larger number of objects. The {\it Hubble UV Legacy Library of Young Stars as Essential Standards} (ULLYSES) project provides such a unique opportunity by establishing a UV spectroscopic library of young high- and low-mass stars in the local universe. Here we analyse optical photometry of the three ULLYSES targets (TX\,Ori, V505\,Ori, V510\,Ori) and other YSOs in the $\sigma$\,Ori cluster taken at the time of the HST observations to provide a reference for those spectra. We identify three populations of YSOs along the line of sight to $\sigma$\,Ori, separated in parallax and proper motion space. The ULLYSES targets show typical YSO behaviour with pronounced variability and mass accretion rates of the order of 10$^{-8}$\,M$_\odot$/yr. Optical colours do not agree with standard interstellar reddening and suggest a significant contribution of scattered light. They are also amongst the most variable and strongest accretors in the cluster. V505\,Ori shows variability with a seven day period, indicating an inner disk warp at the co-rotation radius. Uncovering the exact nature of the ULLYSES targets will require improved detailed modelling of the HST spectra in the context of the available photometry, including scattered light contributions as well as non-standard reddening.

\end{abstract}

\begin{keywords}
stars: formation, pre-main sequence -- stars: variables: T\,Tauri, Herbig Ae/Be -- stars: rotation
\end{keywords}

%%%%%%%%%%%%%%%%%%%%%%%%%%%%%%%%%%%%%%%%%%%%%%%%%%

%%%%%%%%%%%%%%%%% BODY OF PAPER %%%%%%%%%%%%%%%%%%

\section{Introduction}

% ... lizard eats paper, this @HoysSpace paper disproves Spock, Spock vaporises rock, ... 

The formation of young stellar objects (YSOs) and their potential planetary systems is a highly complex process \citep[e.g.,][]{2007A&A...463.1017B, 2016ARA&A..54..135H}. While it is mostly driven by accretion of matter from a circumstellar disk onto the protostellar surface, a variety of other mechanisms are playing important roles as well \cite[e.g., ][]{2014prpl.conf..451F, 2017RSOS....470114E}. The ejection of matter and thus angular momentum from the system in winds and outflows may regulate the spin-up of the protostar \citep[e.g., ][]{1995ApJ...452..736H, 2006A&A...453..785F, 2016ApJ...818..152B, 2018A&A...609A..87N} and at the same time may have profound implications for the evolution of the disk \citep[e.g.,][]{2016MNRAS.460.3472E, 2013ApJ...772...60R}. Obscuration and shadowing of parts of the system caused by its geometry and motion \citep[e.g.,][]{1999A&A...349..619B,2007A&A...463.1017B, 2018A&A...614A.108S} are rendering this highly variable system even more complex, which makes disentangling the relevant processes a difficult task. Thus, to study how young stars accrete matter and how they grow, how their surrounding disks evolve and possibly form planets, coordinated approaches, observing a multitude of tracers of the various physical processes are indispensable to gain detailed insights.

A once-in-a-lifetime opportunity for such an in-depth study is provided to the community by the {\it Hubble UV Legacy Library of Young Stars as Essential Standards} (ULLYSES)\footnote{https://ullyses.stsci.edu} Director's Discretionary Program  \citep{2020AAS...23523203R}. About 500 orbits of HST time are made available to study about 70 YSOs with low- and medium-resolution spectra, which are covering the wavelength range from $\sim$ 140\,nm to $\sim$\,1\,$\mu$m. As a long-lasting legacy of HST, they are providing unprecedented views of the accretion and outflow tracers in the UV wavelength range (Espaillat et al. 2021 subm.).

This effort is supported by many other ground- and space-based programs to complement and extend this data set to other wavelengths and higher spectral resolution. Foremost PENELLOPE, a $\sim$\,250\,h Large Program at the ESO Very Large Telescope (VLT) is providing quasi-contemporaneous optical high-resolution (R\,$>$\,100.000) and medium-resolution flux-calibrated optical and near-infrared (up to 2.5\,$\mu$m, at R\,$>$\,10.000) spectroscopy \citep{2021A&A...650A.196M}.

One of the star forming regions covered by ULLYSES, the $\sigma$\,Ori region, is a target field of the Hunting Outbursting Young Stars (HOYS) citizen science project \citep{2018MNRAS.478.5091F}. This project is obtaining high-quality long-term photometric data in various filters with small telescopes, mostly contributed by amateur astronomers. All three of the ULLYSES target stars in the $\sigma$\,Ori region are covered by HOYS data. Here we are using these data to study the photometric variability of these three objects in the longer term to describe the state in which they were during the HST and VLT observations and to place their activity into context.

Our paper is organised as follows: We give a brief overview of the literature on the $\sigma$\,Ori star forming region and the three $\sigma$\,Ori ULLYSES targets in Sect.\,\ref{target}. A description of our observational data are presented in Sect.\,\ref{data}, followed by the detailed analysis of the GaiaEDR3 data and selection of $\sigma$\,Ori YSOs in the field (Sect.\,\ref{gaia_ana}). In Sect.\,\ref{hoys_ana} we discuss all the HOYS data for the three ULLYSES targets in detail, and place them in context of the other YSOs in the cluster in Sect.\,\ref{results_context}.

\begin{table*}
\caption{Main data for the $\sigma$\,Ori cluster ULLYSES targets. We list the name adopted by us and the ULLYSES project; the coordinates; the parallax values from GaiaEDR3 \citep{2020yCat.1350....0G}; the spectral type, optical extinction, mass and mass accretion rates from \citet{2014ApJ...794...36H} and \citet{2018ApJ...859....1M} (columns labelled 1) and \citet{2021A&A...650A.196M} (columns labelled 2).}
\label{table:1}
\begin{tabular}{|ll|cc|c|cc|cc|cc|cc|} 
\hline
Object & ULLYSES & \multicolumn{2}{c|}{RA/Dec (J2000)}   & Plx. & \multicolumn{2}{c|}{SpT} & \multicolumn{2}{c|}{A$_V$\,[mag]} & \multicolumn{2}{c|}{M\,[M$_{\sun}$]} & \multicolumn{2}{c|}{log($\dot{\rm M}\,[$M$_{\sun}/yr]$)} \\ \hline
       &   ID    & [h:m:s] & [$^\circ$:\arcmin:\arcsec] & [mas] & 1 & 2 & 1 & 2 & 1 & 2 & 1 & 2 \\ \hline
TX\,Ori   & SO583  & 05:38:33.69 &	-02:44:14.1 & 2.4699\,$\pm$\,0.2230 & K4.5\,$\pm$\,1.5 &	K5 & 0.0  &	0.4&	1.087 &	1.09 &	-8.13 &	-7.21 \\ 
V505\,Ori & SO518  & 05:38:27.26 &	-02:45:09.7 & 2.5064\,$\pm$\,0.0250 & K6.0\,$\pm$\,1.0 &	K7 & 0.0  &	1.0  &	0.754 &	0.81 &	-8.54 &	-8.53 \\ 
V510\,Ori & SO1153 & 05:39:39.83 &	-02:31:21.9 & 2.5212\,$\pm$\,0.0268 & K5.5\,$\pm$\,1.0 &	K7 & 0.15 &	0.1  &	0.875 &	0.76 &	-8.38 &	-8.24 \\ 
\hline
\end{tabular}
\end{table*}

\section{The ULLYSES targets and their host cluster in the literature}\label{target}

In this section we give a brief overview of the literature of the $\sigma$\,Ori cluster and the three individual ULLYSES targets.

\subsection{The \texorpdfstring{$\sigma$\,Ori Cluster}{}}

The $\sigma$\,Ori cluster is a $\sim$\,3\,Myr old, nearby ($\sim$\,388\,pc) region \citep{2019A&A...629A.114C}, first identified by \cite{1967PASP...79..433G} as fifteen B type stars including the $\sigma$ Ori AB system. Low mass cluster members were first identified by \cite{phdthesis} and  \cite{1997MmSAI..68.1081W}. The Mayrit catalogue \citep{2008A&A...478..667C} lists 338 members and candidates, with 241 displaying features of youth. \cite{2007ApJ...662.1067H} classed 336 members using data from the Spitzer Space Telescope.  Spectral types are identified in \cite{2014ApJ...794...36H},
and Herschel data has been used to analyse the disks of 32 T\,Tauri stars in \cite{2016ApJ...829...38M}. 
\subsection{ULLYSES targets}

All three ULLYSES targets, TX\,Ori (SO\,583), V505\,Ori (SO\,518), and V510\,Ori (SO\,1153) are identified as YSOs in \citet{2014ApJ...794...36H}, T\,Tauri type stars in \citet{1999AJ....118.1043H}, and are in the General Catalogue of Variable Stars \citep{2017ARep...61...80S}. There are a number of published individual properties for the stars, as well as photometry points. These are listed below. However, no detailed analysis of time series photometry of the stars has been published to the best of our knowledge.  

\noindent {\bf TX\,Ori:} TX\,Ori has an effective temperature of 4020\,K, mass of 1.087\,M$_\odot$ with an optically thick (Type II) disk \citep{2016ApJ...829...38M}. It is assigned the spectral type K1 in \citet{2019A&A...629A.114C} and K4.5 in \citet{2014ApJ...794...36H}. The GaiaDR2 parallax is 1.7486\,$\pm$\,0.2591\,mas \citep{2018yCat.1345....0G}. A wide range of photometry is available: U\,=\,13.75\,mag, B\,=\,13.16\,mag, V\,=\,12.06\,mag, R\,=\,11.38\,mag, I\,=\,10.73\,mag \citep{2014ApJ...794...36H}; G\,=\,12.4140\,mag \citep{2018yCat.1345....0G};
2MASS J\,=\,10.131\,mag, H\,=\,9.280\,mag, K\,=\,8.66\,mag \citep{2003yCat.2246....0C}; In the mid-infrared ALLWISE W1\,=\,7.567\,mag, W2\,=\,7.117\,mag, W3\,=\,5.101\,mag, W4\,=\,3.083\,mag \citep{2014yCat.2328....0C}. The object is also listed in the following: Large-amplitude variables in GaiaDR2 \citep{2021A&A...648A..44M}, Variability properties of TIC sources with KELT \citep{2018AJ....155...39O}, A first catalog of variable stars measured by ATLAS \citep{2018AJ....156..241H} and the AAVSO International Variable Star Index \citep[VSX, ][]{2006SASS...25...47W, 2007JAVSO..35..414W, 2006JAVSO..35..318W, 2012JAVSO..40..431W}.

\noindent {\bf V505\,Ori:} V505\,Ori has an effective temperature of 4244\,K \citep{2019AJ....157..196K}, mass of 0.754\,M$_\odot$ and a Type II disk \citep{2016ApJ...829...38M}. In \citet{2019A&A...629A.114C} it is classified as M0 type, and in \citet{2014ApJ...794...36H} as K6.0. In GaiaDR2 the parallax of V505\,Ori is measured to be 2.514\,$\pm$\,0.0386\,mas which places it at a distance of 397.4\,$\pm$\,6.1\,pc \citep{2018yCat.1345....0G}. V505\,Ori has a number of photometric measurements available: V\,=\,14.16\,mag, R\,=\,13.54\,mag \citep{2014ApJ...794...36H}; G\,=\,14.5688\,mag \citep{2018yCat.1345....0G}; 2MASS J\,=\,11.955\,mag, H\,=\,10.792\,mag, K\,=\,9.44\,mag \citep{2003yCat.2246....0C}; In ALLWISE W1\,=\,8.940\,mag, W2\,=\,8.381\,mag, W3\,=\,6.564\,mag, W4\,=\,4.539\,mag \citep{2014yCat.2328....0C}. The object is also listed in the following: ASAS-SN catalog of variable stars \citep[e.g.][]{2018MNRAS.477.3145J, 2019MNRAS.486.1907J}, Photometric monitoring in $\sigma$\,Ori cluster \citep{2010ApJS..191..389C}, APOGEE-2 survey of Orion Complex (OSFC). I. \citep{2018ApJS..236...27C}, Catalogue of variable stars in open clusters \citep{2012A&A...548A..97Z}, Large-amplitude variables in GaiaDR2 \citep{2021A&A...648A..44M}, A first catalog of variable stars measured by ATLAS \citep{2018AJ....156..241H} and the AAVSO International Variable Star Index \citep{2006SASS...25...47W, 2007JAVSO..35..414W, 2006JAVSO..35..318W, 2012JAVSO..40..431W}.

\noindent {\bf V510\,Ori:} V510\,Ori has an effective temperature of 4357\,K from \citet{2019AJ....157..196K}. In \citet{2016ApJ...829...38M} the effective temperature is 4140\,K, with a mass of 0.875\,M$_\odot$ and it is a Class\,I candidate. It is classified as K2 in \citet{2019A&A...629A.114C} and K5.5 in \citet{2014ApJ...794...36H}. Collimated outflows have been identified in \citet{2004ApJ...606..353A}. The GaiaDR2 parallax of 2.5328\,$\pm$\,0.0353\,mas gives a distance 394.8\,$\pm$\,5.5\,pc \citep{2018yCat.1345....0G}. The available photometry for V510\,Ori is: U\,=\,14.589\,mag \citep{2013MNRAS.434..806B}; V\,=\,14.30\,mag, R\,=\,13.92\,mag \citep{2014ApJ...794...36H}; G\,=\,13.6272\,mag \citep{2018yCat.1345....0G}; I\,=\,13.073\,mag \citep{2008A&A...478..667C}; J\,=\,11.84\,mag, H\,=\,10.901\,mag, K\,=\,10.218\,mag \citep{2003yCat.2246....0C}; g\,=\,14.422\,mag, r\,=\,13.664\,mag, i\,=\,13.219\,mag \citep{2013MNRAS.434..806B}; 2MASS J\,=\,11.842\,mag, H\,=\,10.901\,mag, K\,=\,10.218\,mag \citep{2003yCat.2246....0C}; In ALLWISE W1\,=\,8.955\,mag, W2\,=\,8.234\,mag, W3\,=\,5.598\,mag, W4\,=\,3.258\,mag \cite{2014yCat.2328....0C}. The object is also listed in the following: ASAS-SN catalog of variable stars \citep[e.g.][]{2018MNRAS.477.3145J, 2019MNRAS.486.1907J}, Photometric monitoring in $\sigma$\,Ori cluster \citep{2010ApJS..191..389C}, Catalogue of variable stars in open clusters \citep{2012A&A...548A..97Z}, Large-amplitude variables in GaiaDR2 \citep{2021A&A...648A..44M}, and the AAVSO International Variable Star Index \citep{2006SASS...25...47W, 2007JAVSO..35..414W, 2006JAVSO..35..318W, 2012JAVSO..40..431W}.

\section{Observational Data}\label{data}

In this section we describe the details of the photometry data analysed in this paper. The vast majority of the data are from the HOYS project, but we also utilise additional photometry from GaiaEDR3 and (NEO)WISE. The latter two, well characterised, data sets are briefly described in the sections they are analysed in (Sect.\,\ref{gaia_ana} and Sect.\,\ref{wise_ana}) and we focus in this section on the HOYS observations. All three ULLYSES targets are also observed by the Transiting Exoplanet Survey Telescope \citep[TESS, ][]{2014SPIE.9143E..20R}. The discussion of the TESS data is beyond the scope of this paper and will be presented by a forthcoming paper from the ULLYSES collaboration (Abraham et al., in prep.).

HOYS is a citizen science project that works with amateur astronomers since 2014 \citep{2018MNRAS.478.5091F}. Additionally, some professional and University observatories deliver data. The participants currently monitor 25 nearby (d\,$<$\,1\,kpc) young (age\,$<$\,1\,Myr) clusters and star forming regions as often as possible in optical broad band filters and in the \ha\ narrow band filter. We require all images to have undergone a basic data reduction, i.e. a bias, dark, and flat-field correction has to be applied. Sets of images for the same target and filter in the same night should be stacked prior to submission to our database, unless they are taken several hours apart. We perform astrometry in all images using the {\tt astrometry.net} software package \citep{2008ASPC..394...27H}. 

Photometry is conducted using the {\tt Source Extractor} software \citep{1996A&AS..117..393B}.  For every HOYS field we have obtained a deep image in each of the filters as a reference image against which the photometry is calibrated. These are all taken with the Beacon Observatory (see Sect.\,\ref{beacon}) with the exception of the U-band, which has been taken with the Alfred-Jensch telescope (see Sect.\,\ref{tls}). The instrumental magnitudes in these reference images are converted into apparent magnitudes using the Cambridge Photometric Calibration Server\footnote{\tt \href{http://gsaweb.ast.cam.ac.uk/followup}{http://gsaweb.ast.cam.ac.uk/followup}}. Thus, the reference photometry system of our photometry is Johnson U, B, and V and Cousins R$_{\rm c}$, and I$_{\rm c}$ (R and I , hereafter). The \ha\ images are calibrated into the R reference frame. To correct potential colour terms from observations obtained trough filters with slightly different transmission curves compared to our reference frames, we developed a correction algorithm in \citet{2020MNRAS.493..184E}. It identifies all non-variable stars in every field and utilises their known magnitudes and colours to determine and correct the colour terms for each image.

At the time of writing, the HOYS database contains a total of 2917 images of the $\sigma$\,Ori target region. Of those, we have 36 in U, 721 in B, 789 in V, 647 in R, 679 in I, and 45 in \ha. Thus, there is excellent coverage, mostly concentrated in the last (2020/21) observing season for all the broad band filters, except U. All the U-band images and most (all but three) of the \ha\ frames have also been taken in the 2020/21 observing season. The vast majority of images for this field have been taken with only three observatories and hence have excellent internal consistency and almost no colour terms. Seventeen further observatories have delivered data, but they only amount to 13.8\,\% of all images. In particular all the U-band data and 98\,\% of the \ha\ images are taken with the same telescope. Below we briefly describe the details of the three main observatories used.

\subsection{Main HOYS observatories}

\subsubsection{ROAD}

The Remote Observatory Atacama Desert \citep[ROAD;][]{2012JAVSO..40.1003H} has delivered 61.5\,\% of all images and dominates the B, V, R, I observations for the 2020/21 observing season with almost daily datasets. Its location in Chile also means that we have pairs of observations in those filters almost coincidental with all the VLT spectra obtained for the ULLYSES targets (see Sect.\,\ref{lc_hst_vlt}). The filters are identical to the ones used in the HOYS reference images. The observations were acquired through Astrodon Photometric filters with an Orion Optics, UK Optimized Dall Kirkham 406/6.8 telescope and a FLI 16803 CCD camera. The field of view of the camera is 0.79\,$\times$\,0.79\,sq.deg with 2.1\arcsec\ pixels. Each data set consists of pairs of exposures with 60\,s (B), 45\,s (V), and 30\,s (R$_c$ and I$_c$). Twilight sky-flat images were used for flat-field corrections. 

\subsubsection{Th\"uringer Landessternwarte}\label{tls}

For HOYS the Alfred-Jensch 2-m telescope of Th\"uringer Landessternwarte (TLS) is used in its Schmidt configuration (clear aperture 1.34\,m, mirror diameter 2.00\,m, focal length 4.00\,m). Its CCD camera TAUKAM \citep{2016SPIE.9908E..4US} is equipped with a Sloan u, g, r, i filter set, as well as a Johnson B and an \ha\ filter, and provides a field of view of 1.32\,$\times$\,1.32\,sq.deg with 0.77\arcsec\ pixels. Usually three consecutive images are obtained in the same filter (exposure time per image: 300\,s in u, 120\,s in B, 60\,s in g, 20\,s in r, 20\,s in i, 120\,s in \ha) and co-added after standard reduction. Dark frames and dome-flats are used for image calibration. This observatory has delivered about 14.5\,\% of all images, entirely during the 2020/21 observing season. It is solely responsible for all of the U-band data and 98\,\% of the \ha\ images, including the U-band reference frame used in HOYS.

\subsubsection{Beacon Observatory}\label{beacon}

The Beacon Observatory has contributed 10\,\% of all images used in the analysis, including all of the B, V, R, and I reference frames. It consists of a 17\arcsec\ {\em Planewave} Corrected Dall-Kirkham (CDK) Astrograph telescope and has a 4k\,$\times$\,4k Peltier-cooled CCD camera and a B, V, R$_c$, I$_c$, and \ha\ filter set. The field of view of the camera is 1.09\,$\times$\,1.09\,sq.deg with a pixel scale of 0.96\arcsec. For HOYS we typically take 8 frames with 120\,s exposure times in B, V, R$_c$, and I$_c$. All individual images are dark and bias subtracted and flat-field corrected using sky-flats. A full detailed description of the observatory can be found in \citet{2020MNRAS.493..184E}.

\section{Selection of \texorpdfstring{$\sigma$\,Ori  YSO\lowercase{s}}{}}\label{gaia_ana}

\subsection{Gaia Data analysis}

\begin{figure*}
\centering
\includegraphics[angle=0,width=\columnwidth]{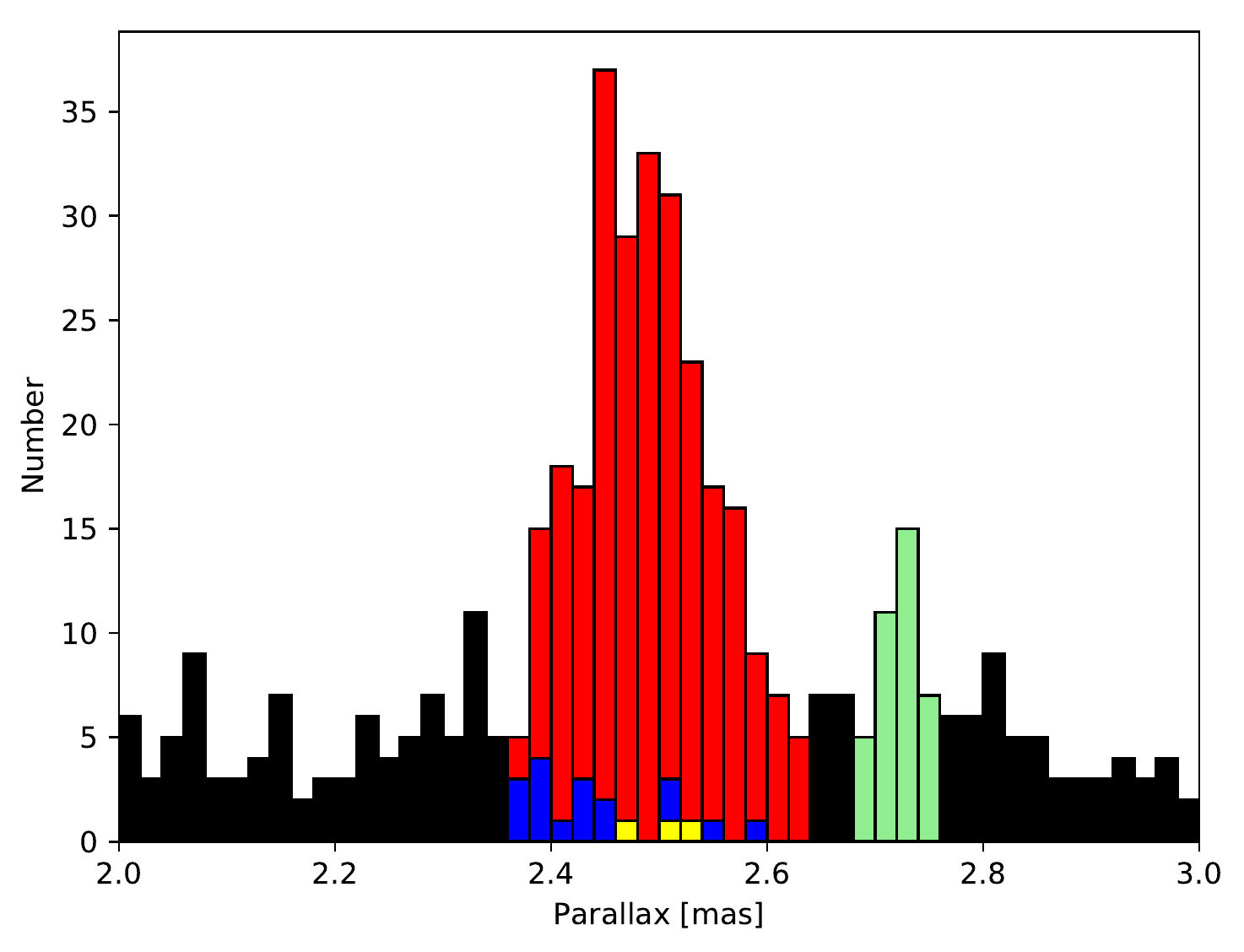} \hfill
\includegraphics[angle=0,width=\columnwidth]{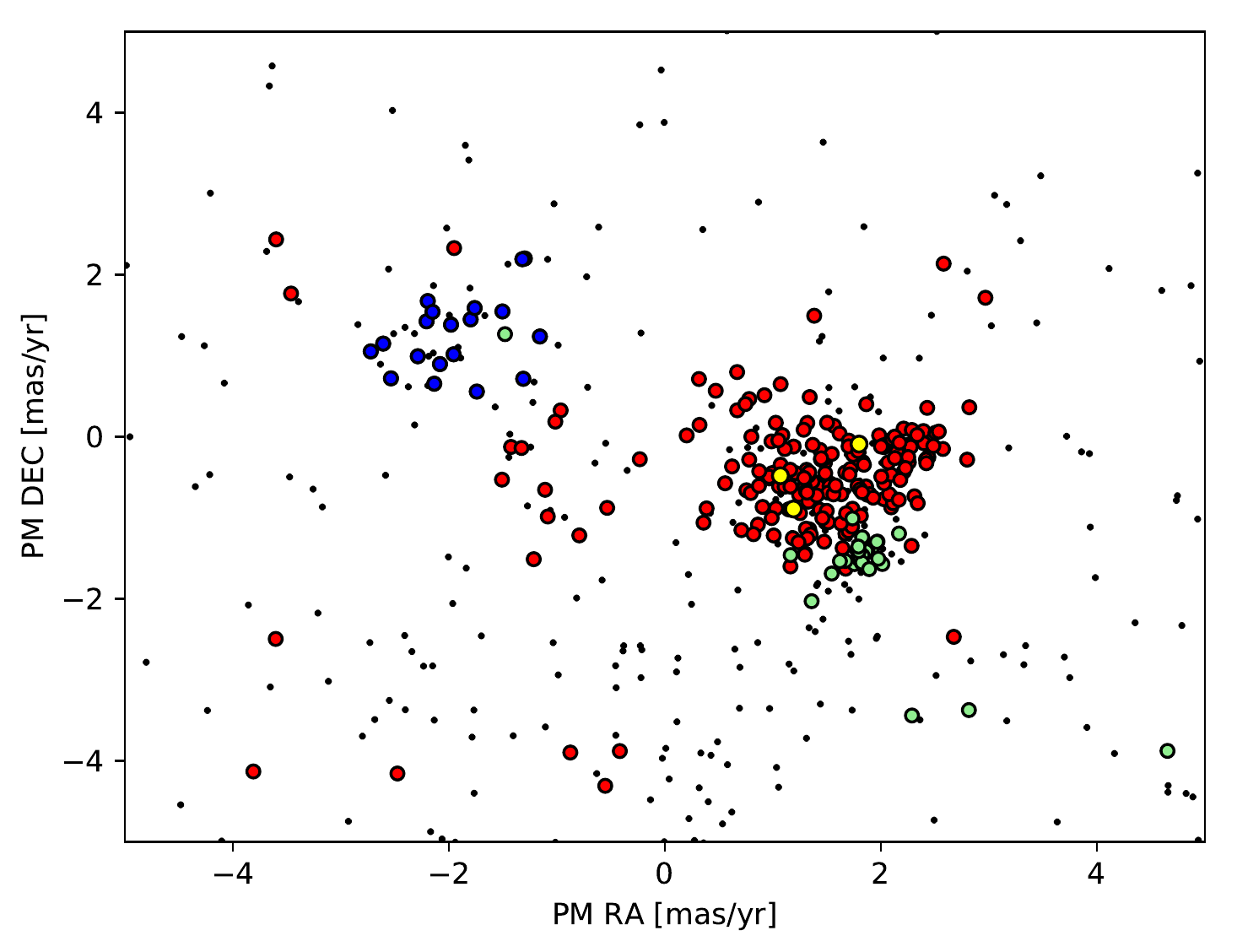} \\
\includegraphics[angle=0,width=\columnwidth]{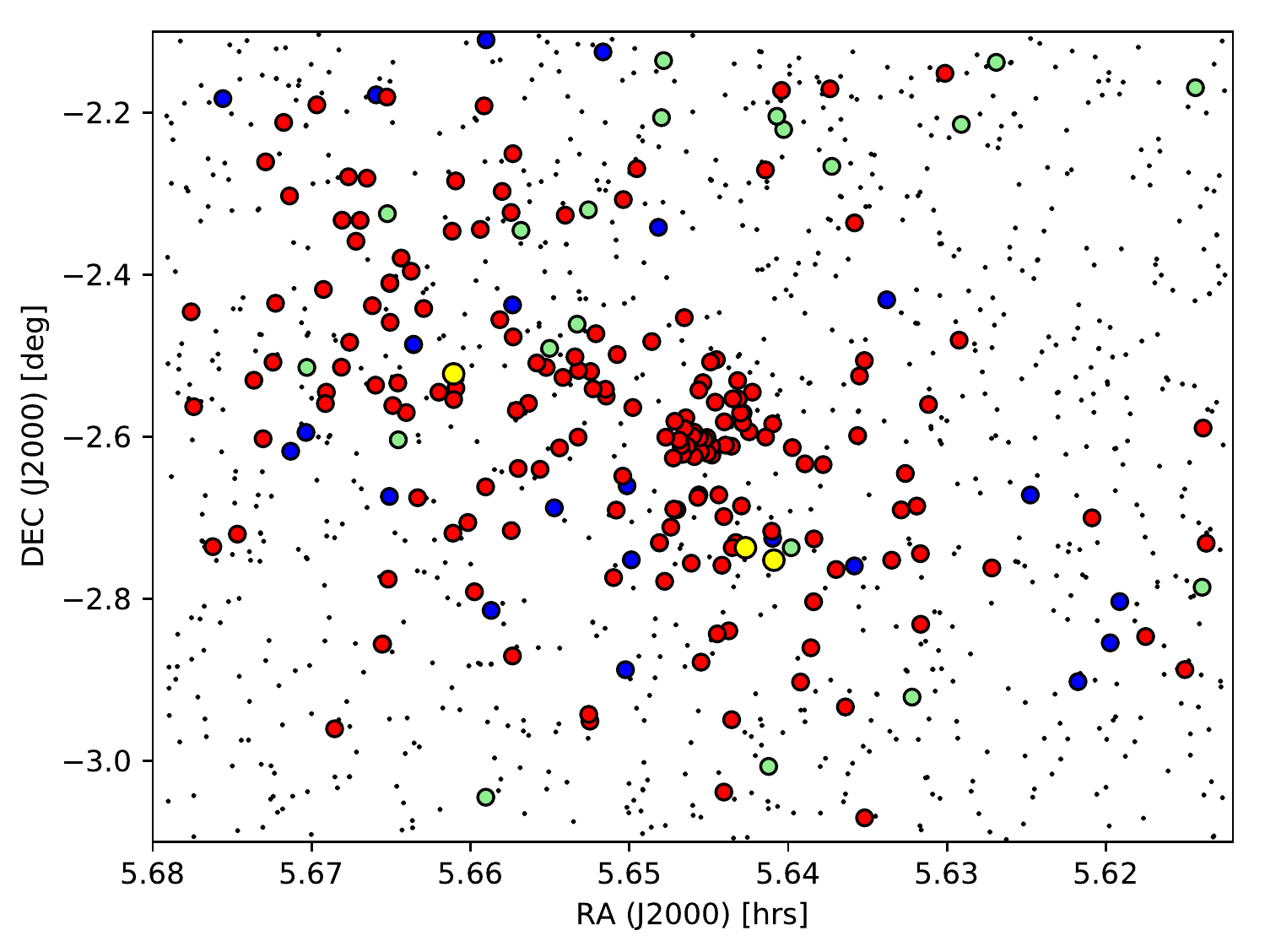} \hfill
\includegraphics[angle=0,width=\columnwidth]{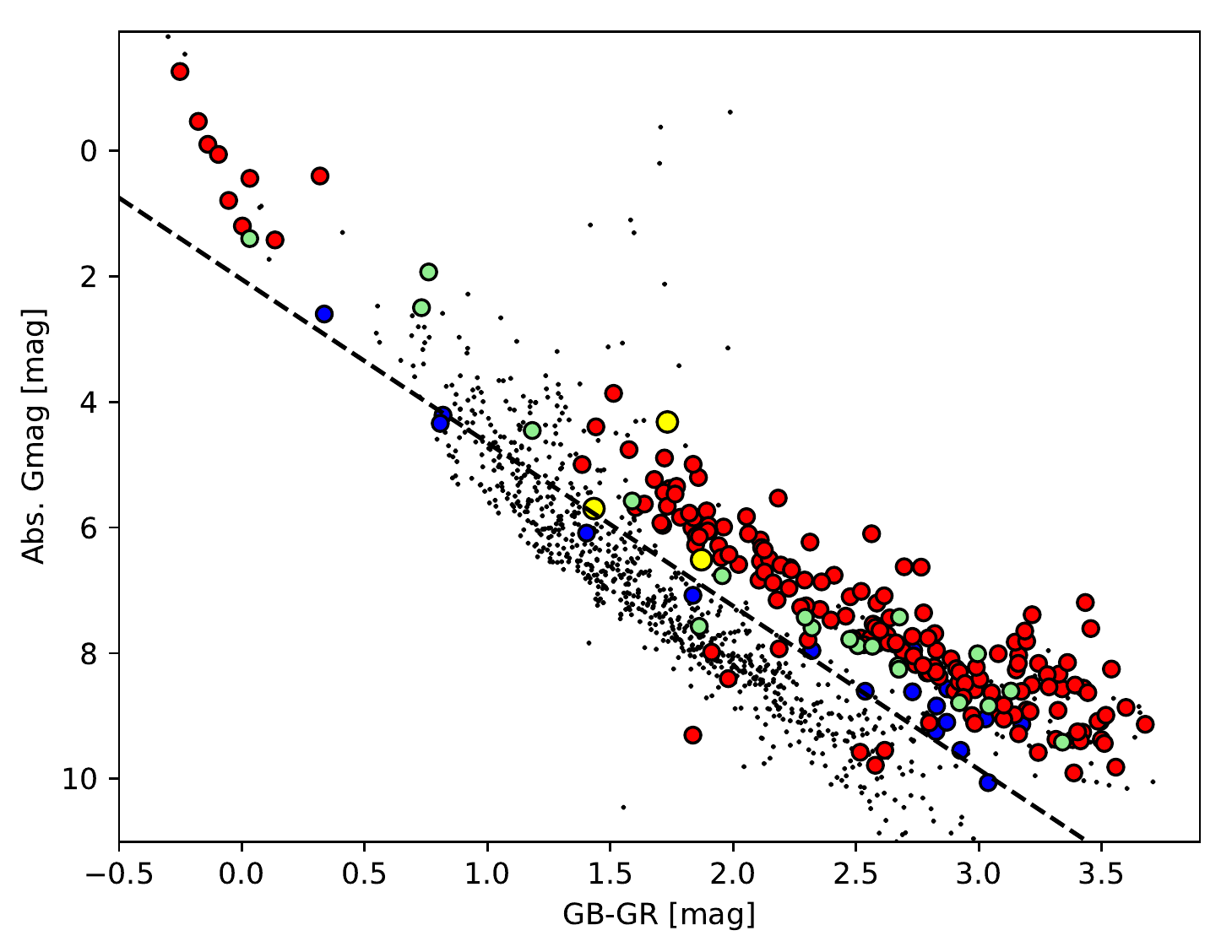} 
\caption{ Analysis of Gaia data of YSOs for the $\sigma$\,Ori field. In all panels the three ULLYSES targets are highlighted in yellow. {\bf Top-Left:} Parallax distribution. The red and light green ranges are the two populations identifiable from parallax alone. {\bf Top-Right:} Proper motions. In black all sources are shown. Red and light green symbols are as in the parallax distribution. The blue symbols indicate the sub-group of the main parallax peak identifiable in proper motion space with PM RA of about $-2$\,mas/yr. {\bf Bottom-Left:} Positions of the stars in the field. Same colour code as in the top right panel with the exception that the red points are selected in parallax and proper motion space (see text for details). {\bf Bottom-Right:} Absolute magnitude vs. Gaia colour diagram. Same colour code as in the bottom left panel. YSO candidates below the dashed line are considered main sequence interlopers.\label{sigori_ysos}}
\end{figure*}

In order to place the three ULLYSES targets observed with the HST and VLT into context of the population of YSOs in the $\sigma$\,Ori field in HOYS, we utilise data from the GaiaEDR3 data release \citep{2021A&A...649A...1G}. We use the central coordinates of our $\sigma$\,Ori field of RA\,=\,05:38:45 and DEC\,=\,$-$02:36:00 (J2000) and extract GaiaEDR3 data of one square degree around it to match the area typically covered in the HOYS data. To ensure a clean sample of YSOs, we only extract sources with a parallax of more than 1\,mas, a signal to noise ratio (SNR) for the parallax of more than three, a Re-normalised Unit Weight Error (RUWE) of less than two and Gmag smaller than 18\,mag.

We first generate a histogram of the parallax values. The part of the histogram with parallax values near the cluster value is shown in the top left panel of Fig.\,\ref{sigori_ysos}. There is a general population, which are most likely field main sequence stars as well as potentially YSOs from a more distributed population in Orion. Two clear peaks in the histogram do indicate two concentrated populations of sources at slightly different distances. The main peak (colour coded in red) ranges from 2.36\,mas to 2.64\,mas with a peak at 2.48\,mas (403\,pc). This main peak also contains the three ULLYSES targets (indicated in yellow). There is a second, less populated peak (indicated in light green) at a parallax of 2.72\,mas (368\,pc). This group of objects is slightly foreground (35\,pc closer) to the main $\sigma$\,Ori cluster.

The proper motions of all GaiaEDR3 sources are shown in the top right panel of Fig.\,\ref{sigori_ysos}. The colour coding is the same as in the left panel of the figure. It is evident that the small foreground cluster occupies a narrow range (less than 0.5\,mas/yr) in proper motion space (1.55\,mas/yr\,$<$\,PM RA\,$<$\,2.10\,mas/yr; $-1.70$\,mas/yr\,$<$\,PM DEC\,$<$\,$-1.20$\,mas/yr) and hence forms indeed a potential cluster of YSOs. The sources belonging to the main peak in the parallax histogram clearly split into two groups in proper motion space. There is a main group (0.00\,mas/yr\,$<$\,PM RA\,$<$\,2.90\,mas/yr; $-1.80$\,mas/yr\,$<$\,PM DEC\,$<$\,0.80\,mas/yr) which also contains the three ULLYSES targets. The second group is much more sparsely populated and at a proper motion range of $-2.90$\,mas/yr\,$<$\,PM RA\,$<$\,$-0.7$\,mas/yr and 0.40\,mas/yr\,$<$\,PM DEC\,$<$\,2.30\,mas/yr. We indicate them and their parallax values in blue in the top left panel of Fig.\,\ref{sigori_ysos}. Both groups overlap in parallax space, but there is a tendency that the second group has slightly smaller parallax values. Thus, these stars are most likely slightly in the background of the main group. Note that there are no further objects in the 2\,mas to 3\,mas parallax range in this (blue) proper motion selection. There is some indication that the main group of objects splits further in proper motion at PM RA\,=\,2.0\,mas/yr, but a detailed investigation of this is beyond the scope of this work. Note that the scatter of proper motions in the three main groups of objects changes from about 0.5\,mas/yr for the foreground group, to 2\,mas/yr for the sparsely populated group and almost 3\,mas/yr for the main group. This indicates a difference in the internal velocity distributions within the groups.

In the bottom row of Fig.\,\ref{sigori_ysos} we investigate the three identified groups of sources. We retain the colour coding from the top row and use blue symbols for the objects belonging to the smaller group in proper motion space that are part of the main peak in the parallax histogram. The bottom left panel in the figure shows the spatial distribution of the objects. The main group clusters in the centre of the field, near $\sigma$\,Ori itself, but the main population is quite distributed. The ULLYSES targets are near the main cluster, but part of the distributed population. The two smaller groupings do not seem to strongly cluster spatially. There is a general paucity of objects in the East and South of the field. 

The bottom right panel of Fig.\,\ref{sigori_ysos} shows the Gaia GB-GR colours against the absolute Gmag, determined from the parallaxes and assuming zero extinction. It is evident in the figure that all three groups of objects lie significantly above the main sequence of objects and hence represent populations of young stars. There are no obvious differences between the positions of the foreground cluster (light green) and the main group (red). The smaller group (blue) seems to be contaminated by some main sequence (binary) interlopers (see Sect.\,\ref{hoys_selection}). A detailed isochrone fitting is needed to determine the ages for the groups, but this is beyond the scope of this work. There is also an apparent lack of sources of roughly 1-2 solar mass objects (absolute Gmag from 2\,mag to 4\,mag). There are a few objects very close to, or on the main sequence. These are most likely interlopers which happen to have the same distance and proper motion. We discuss in the next subsection how we remove them from the sample of $\sigma$\,Ori YSOs that we investigate. Note that there is some scatter in the sequences for the  clusters due to variability. This is already evident for the three ULLYSES targets. We will discuss in Sect.\,\ref{results_context} that the ULLYSES targets are amongst the most variable $\sigma$\,Ori YSOs in our field.

In summary, the GaiaEDR3 data shows that there are three populations of YSOs in our $\sigma$\,Ori field. There is a smaller distributed foreground population at d\,=\,368\,pc. The other two populations are both at a distance of 403\,pc but do have different proper motions by about 4\,mas/yr. At this distance, this corresponds to a transverse velocity difference of about 7.5\,km\,s$^{-1}$. In the next section we discuss our selection criteria for the GaiaEDR3 data and HOYS light curves to obtain a sample of $\sigma$\,Ori YSOs to analyse in order to put the data of the three ULLYSES targets in context.

\subsection{HOYS light curve selection}\label{hoys_selection}

We created a sample of $\sigma$\,Ori YSOs from the GaiaEDR3 sample by selecting all sources that fulfil the criteria outlined above. For all of those we extracted the HOYS light curves. Objects with less than 50 data points in the light curve have been removed from the analysis. This leaves 176 potential YSOs.

\begin{figure}
\centering
\includegraphics[angle=0,width=\columnwidth]{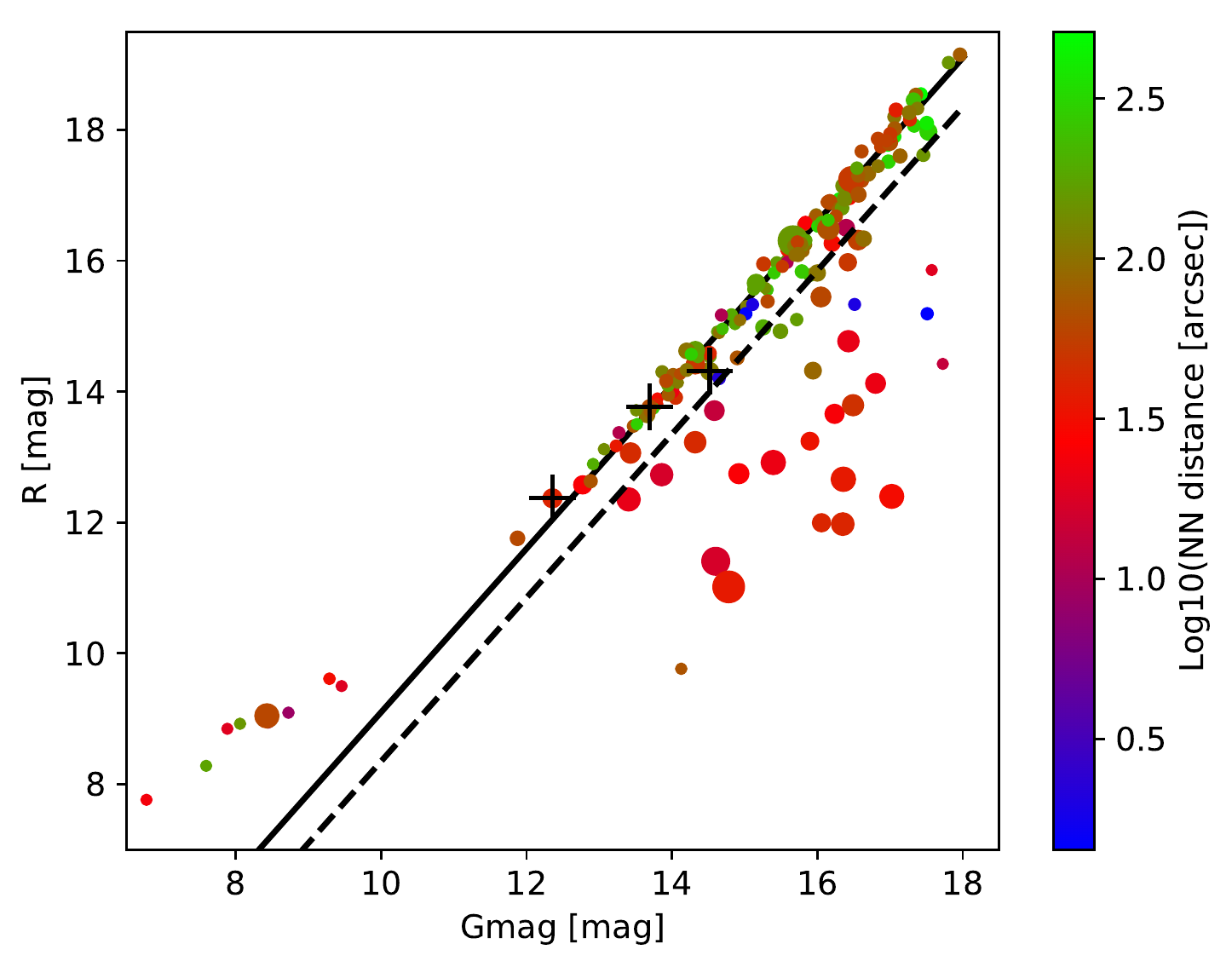} 
\caption{Plot showing the Gaia Gmag vs the median HOYS R magnitudes for all YSOs in the $\sigma$\,Ori cluster, selected based on the Gaia parallax and proper motions as described in the text. The symbol size represents the R-band Stetson index and the colour code the distance to the nearest other YSO in the sample. The crosses indicate the three ULLYSES targets. We indicate the trend of the majority of the fainter targets by the solid black line and the selection cut-off for HOYS-GaiaEDR3 miss-matches by the dashed line. \label{gaia_v_hoys}}
\end{figure}

The typical seeing in the HOYS images is of the order of 3\arcsec\,--\,5\arcsec. Thus, there is a possibility that the Gaia objects merge with neighbouring stars. Hence, those light curves cannot be attributed to a single source. We thus plot the Gaia Gmag values of all selected YSOs against their median R magnitude in the HOYS light curve. This is shown in Fig.\,\ref{gaia_v_hoys}. The symbol size represents the R-band Stetson index \citep{1996PASP..108..851S} of the HOYS light curve. The colour code indicates the distance of the Gaia coordinates to the nearest other Gaia selected YSO in the sample, in order to identify close YSO pairs in the sample. The crosses indicate the three ULLYSES targets. 

In Fig.\,\ref{gaia_v_hoys} we see that generally the R magnitudes correlate well with the Gmag values. The trend is linear for the objects fainter than about 12\,mag. At the bright end the objects deviate from the linear trend due to a change in colour and saturation effects in the HOYS data. There is no systematic indication that variable sources (high Stetson index) do not follow the trend. This is emphasised by the three ULLYSES targets, which are all quite variable but follow the general trend line closely. 

There are three spatially very close pairs of Gaia selected YSOs, visible as blue symbols. In all three cases the HOYS light curve is the same for both Gaia sources due to our worse resolution. The pairs hence have the same R magnitude but different Gmag values. In one pair both Gaia sources are almost identical in brightness and are thus indistinguishable in the plot. We have removed both of these sources from the analysis. In the other two pairs one object dominates the brightness by two to three magnitudes. We have removed the fainter of the Gaia sources in those pairs from the analysis, as the HOYS data will correspond to the brighter of the Gaia sources.

There is a large number of sources significantly below the trend line. Almost all the sources have larger Stetson indices, but are very far away from the nearest Gaia YSOs. These are objects that are very close to other, brighter field stars. Their apparent variability is caused by our mix of good and bad seeing images. Thus, the light curves do at times represent just the Gaia YSO, but at most times the flux of both sources merges together. These light curves clearly do not represent the YSOs. We have hence removed all objects below the dashed line in Fig.\,\ref{gaia_v_hoys} from our analysis. This selection line is shifted by 0.75\,mag from the general trend, hence all sources where the YSO contributes less than 50\,\% of the HOYS flux are removed from the analysis. The selection line follows the equation: $R = -4.15 + 1.25 \times Gmag$. This ensures that most of the blends of Gaia YSOs with other sources, where the Gaia YSO is the fainter source, are removed from our analysis, allowing for some scatter from the trend due to variability. Note this selection also removes the fainter of the two YSO pairs discussed above. The equal brightness pair is just above the selection and has been manually removed from the analysis.

We apply one further selection to the YSO sample. This is indicated in the bottom right panel of Fig.\,\ref{sigori_ysos}. All objects situated below the line $Abs.Gmag = 2.05 + 2.6 \times (GB-GR)$ are removed from the potential YSO list. This ensures the vast majority of sources that are potentially main sequence interlopers with the same parallax and proper motion are removed from the YSO list and not included in the analysis to place the ULLYSES targets into the context of the $\sigma$\,Ori YSOs. In total we are left with 140 YSOs to analyse, plus the three ULLYSES targets. We will discuss the properties of the three objects in the context of the other YSOs in the cluster in Sect.\,\ref{results_context}.

\section{Photometry Analysis of the ULLYSES targets}\label{hoys_ana}

\begin{figure*}
\centering
\includegraphics[angle=0,width=5.85cm]{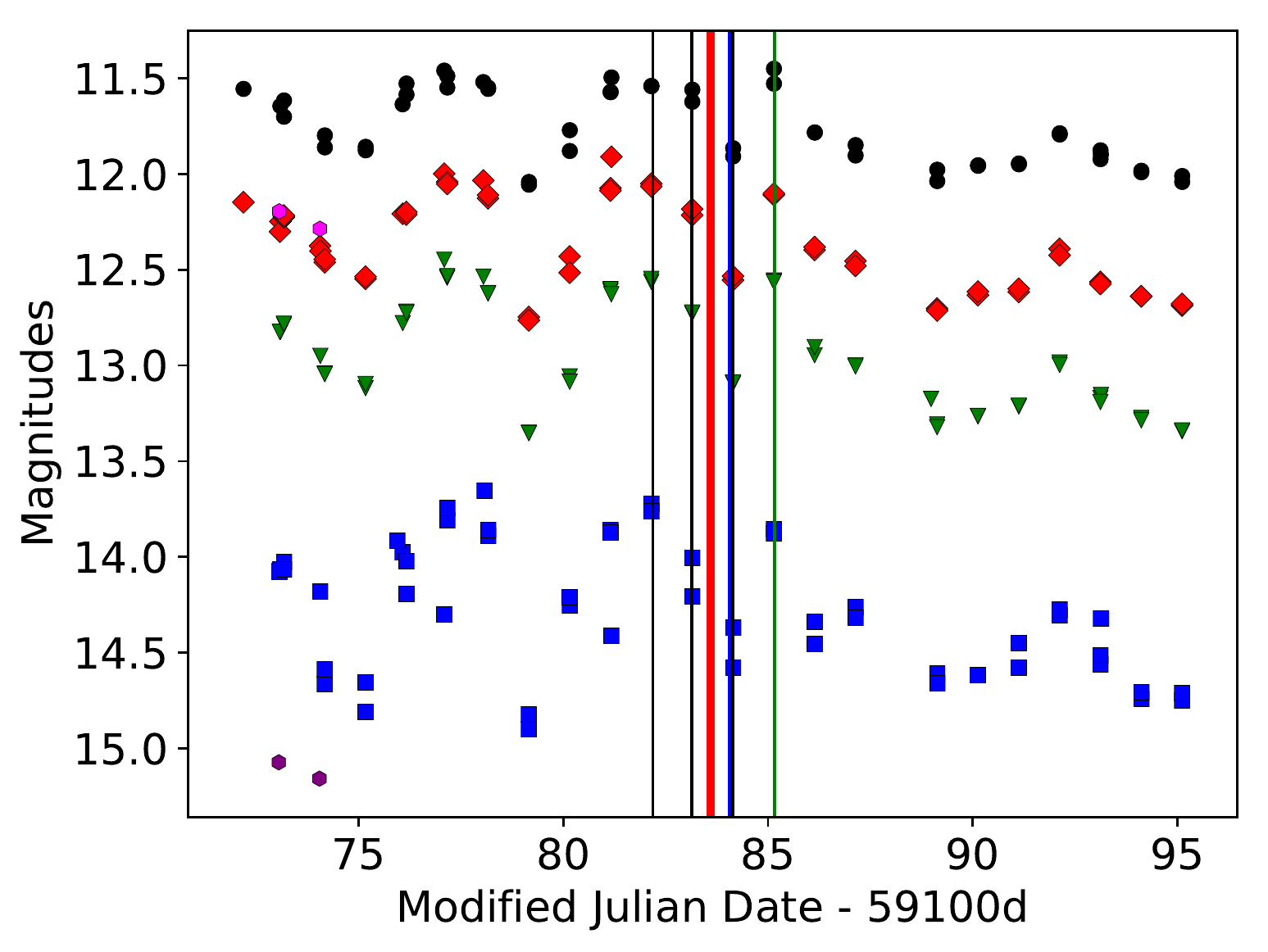} \hfill
\includegraphics[angle=0,width=5.7cm]{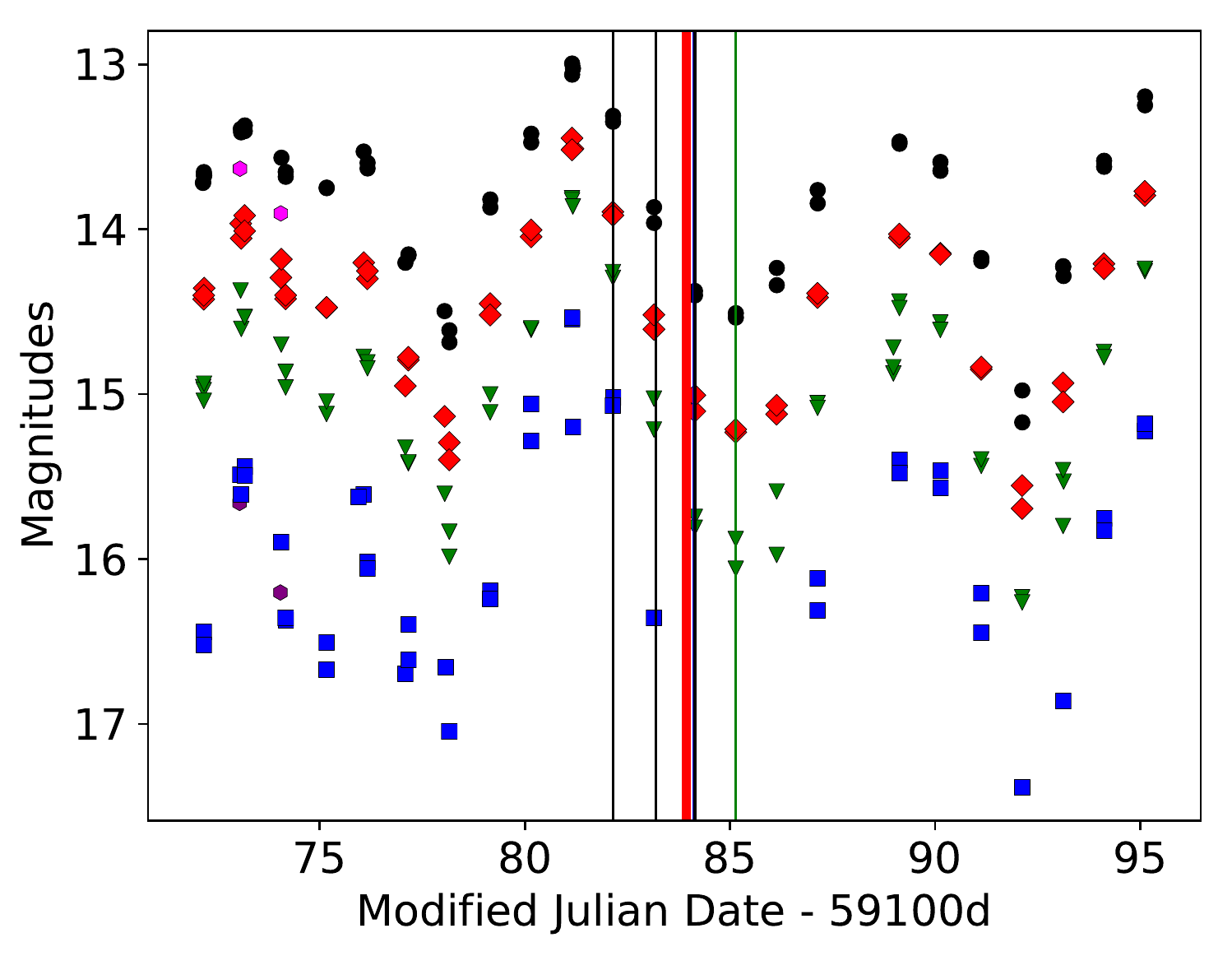} \hfill
\includegraphics[angle=0,width=5.85cm]{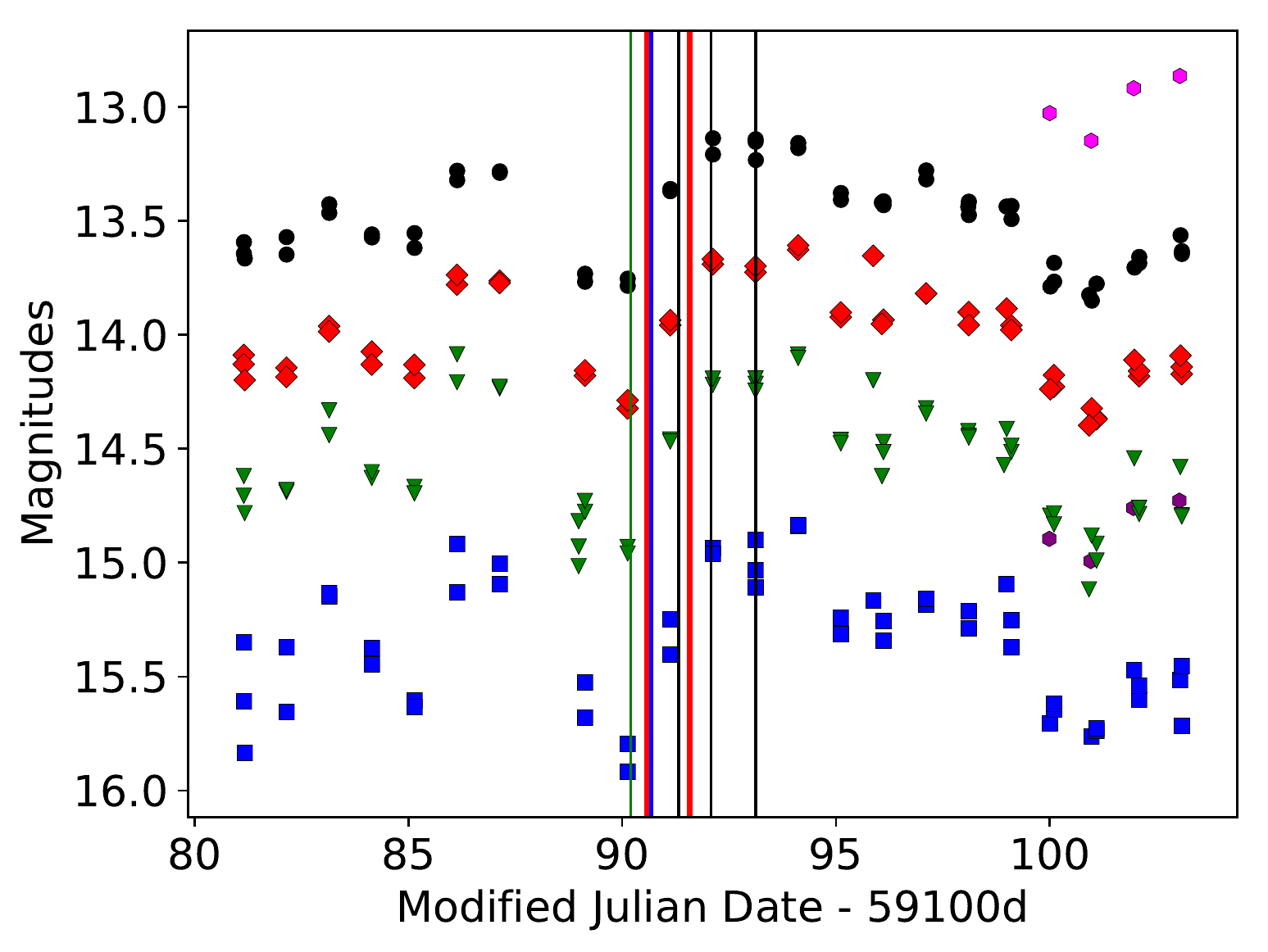} 
\caption{HOYS light curves for TX\,Ori ({\bf left}), V505\,Ori ({\bf middle}), and V510\,Ori ({\bf right}) in all filters (U - purple hexagons, B - blue squares, V - green triangles, R - red diamonds, \ha\ - pink hexagons, I - black circles) during the $\pm 10$\,d of the HST and VLT spectroscopic observations. The vertical coloured lines indicate the times the HST and VLT spectra are taken. The colour code is as follows: blue for STIS, red for COS, green for X-shooter and black for UVES (TX\,Ori, V505\,Ori) or ESPRESSO (V510\,Ori). \label{lc_hst}}
\end{figure*}

\subsection{HOYS Light curves during HST and VLT observations}\label{lc_hst_vlt}

In order to show the state of the ULLYSES targets during the HST observations with COS and STIS, as well as the VLT observations with UVES, X-Shooter, and ESPRESSO, we show the HOYS light curves for the targets for the ten days prior and past these  observations in Fig.\,\ref{lc_hst}. All HOYS data points in U, B, V, R, \ha, and I are shown.  Unfortunately there are very few U and \ha\ points available in this time period. But we have near daily observations in the other broad band filters. The times of the HST and VLT observations are indicated by vertical lines in the plots. Below is a brief discussion of the behaviour of the objects during the HST data taking period. We list all available magnitudes at the time of the HST and VLT observations in Table\,\ref{tab_mags}. We determined the values in all cases by using a linear interpolation of the HOYS data taken at most one day prior or after the HST observations. The typical uncertainties are of the order of 0.1\,mag.

\begin{table}
\setlength{\tabcolsep}{5pt}
\caption{\label{tab_mags} Interpolated magnitudes of the three ULLYSES targets at the times of the HST (COS, STIS) and VLT (UVES, X-Shooter, ESPRESSO) spectroscopic observations. Missing values indicate times where no data for interpolation is available. Uncertainties in the values are of the order of 0.1\,mag. The adopted times are the mid points of the individual spectra. For some targets multiple HST and VLT observations are performed more than an hour apart. These are listed separately.}
\centering
\begin{tabular}{|c|c|c|c|c|c|c|}
\hline
Target & STIS/ & MJD & B & V & R & I \\ 
 & COS & [days] & [mag] & [mag] & [mag] & [mag] \\ \hline
TX\,Ori   & STIS & 59184.56 & 14.44 & 13.06 & 12.52 & 11.86 \\
TX\,Ori   & COS  & 59184.03 & 14.24 & 12.86 & 12.33 & 11.70 \\
TX\,Ori   & COS  & 59184.09 & 14.27 & 12.89 & 12.36 & 11.72 \\
TX\,Ori   & COS  & 59184.16 & 14.29 & 12.91 & 12.38 & 11.74 \\
TX\,Ori   & UVES & 59182.68 & 13.75 & 12.56 & 12.06 & 11.54 \\
TX\,Ori   & UVES & 59183.63 & 14.09 & 12.72 & 12.20 & 11.59 \\
TX\,Ori   & UVES & 59184.63 & 14.47 & 13.09 & 12.54 & 11.88 \\
TX\,Ori   & XSHO & 59185.66 & 13.87 & 12.57 & 12.11 & 11.50 \\  \hline
V505\,Ori & STIS & 59184.62 &  --   & 15.76 & 15.07 & 14.38 \\
V505\,Ori & COS  & 59184.36 &  --   & 15.60 & 14.93 & 14.25 \\
V505\,Ori & COS  & 59184.42 &  --   & 15.63 & 14.96 & 14.28 \\
V505\,Ori & COS  & 59184.49 &  --   & 15.68 & 15.00 & 14.32 \\
V505\,Ori & UVES & 59182.65 & 15.04 & 14.28 & 13.91 & 13.33 \\
V505\,Ori & UVES & 59183.69 & 16.35 & 15.19 & 14.60 & 13.93 \\
V505\,Ori & UVES & 59184.65 &  --   & 15.77 & 15.07 & 14.39 \\
V505\,Ori & XSHO & 59185.64 &  --   & 15.99 & 15.22 & 14.52 \\ \hline
V510\,Ori & STIS & 59191.18 & 15.55 & 14.68 & 14.10 & 13.54 \\
V510\,Ori & COS  & 59191.05 & 15.62 & 14.74 & 14.15 & 13.60 \\ 
V510\,Ori & COS  & 59191.11 & 15.59 & 14.72 & 14.13 & 13.58 \\ 
V510\,Ori & COS  & 59192.04 & 15.17 & 14.36 & 13.84 & 13.28 \\ 
V510\,Ori & COS  & 59192.11 & 15.15 & 14.34 & 13.82 & 13.27 \\
V510\,Ori & ESPR & 59191.82 & 15.25 & 14.42 & 13.89 & 13.32 \\ 
V510\,Ori & ESPR & 59192.58 & 14.97 & 14.22 & 13.69 & 13.18 \\ 
V510\,Ori & ESPR & 59193.62 & 15.01 & 14.22 & 13.71 & 13.16 \\ 
V510\,Ori & XSHO & 59190.70 & 15.80 & 14.91 & 14.27 & 13.74 \\  \hline
\end{tabular}
\end{table}

\noindent {\bf TX\,Ori:} The star shows variations on one day timescales from 0.5\,mag (I) to 1.0\,mag (B) in the days prior to the HST and VLT observations. The observations themselves have been taken during a period of slightly decreased brightness. The STIS and VLT data have been taken almost coincidentally with a B, V, R, I data set in HOYS, so there is a good broad band reference available. The COS data have been taken about half way between two HOYS datasets just prior to the STIS data. The object has decreased its brightness between the COS and STIS observations.

\noindent {\bf V505\,Ori:} The seven day periodic variations of the star (discussed in detail in Sect.\,\ref{v505-var}) have very large amplitudes near the time of the HST and VLT observations and the spectra cover almost the entire decline from maximum to minimum brightness. The amplitudes of the star range from 1.5\,mag (I) to almost 2.5\,mag (B). Similar to TX\,Ori, the HST data has been taken near the faint state of the star, and the VLT spectra during a decline in brightness. As discussed in Sect.\,\ref{colmags}, the optical light of the star in the faint state, and thus during the HST and later VLT observations, is dominated by scattered light. The STIS and VLT data are taken almost coincidental with a set of V, R, I HOYS data, while the COS observations have been taken very slightly prior to a V, R, I data set. Unfortunately, the source is too faint for any B data being available in HOYS near the time of the HST and VLT observations, with the exception of the first two UVES spectra. Based on the faint state at similar magnitudes in the other filters, we estimate that the star should have been roughly B\,=\,17\,mag at the dates where no brightness is listed in Table.\,\ref{tab_mags}. 

\noindent {\bf V510\,Ori:} The star varies by up to 0.5\,mag on a day to day basis in the days near the HST and VLT observations. Unfortunately the HST and earlier VLT data have been taken during a steep increase in brightness. The STIS data have been taken half way between two sets of B, V, R, and I HOYS data sets and can thus be interpolated. There are two sets of COS data. The first almost coincides with the STIS observations. The second has been taken one day later, in between the next set of HOYS data points. All the VLT data have been taken almost coincidental with a HOYS data set. 

\subsection{Colour magnitude diagrams}\label{colmags}

\begin{figure*}
\centering
\includegraphics[angle=0,width=5.8cm]{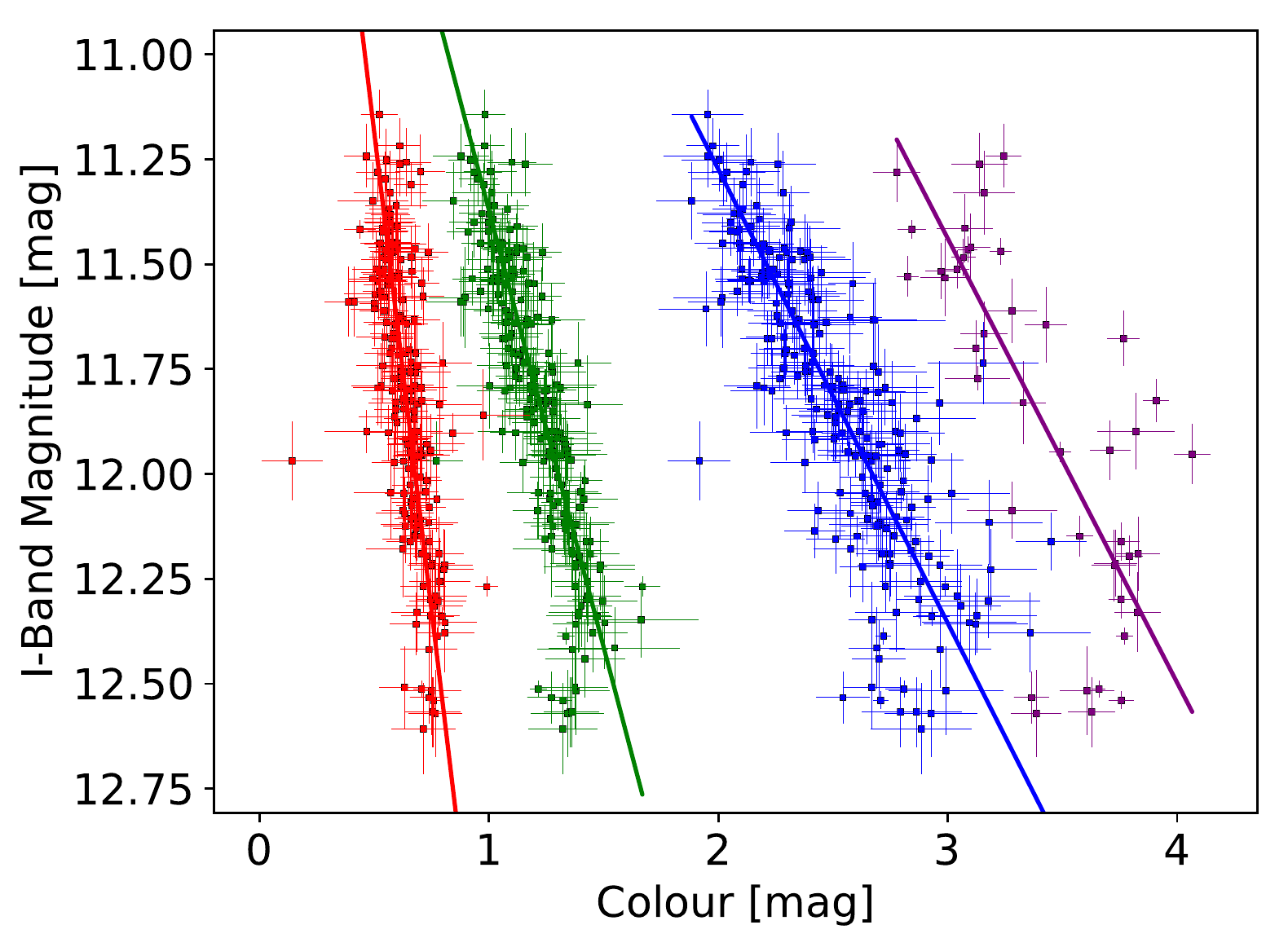} \hfill
\includegraphics[angle=0,width=5.7cm]{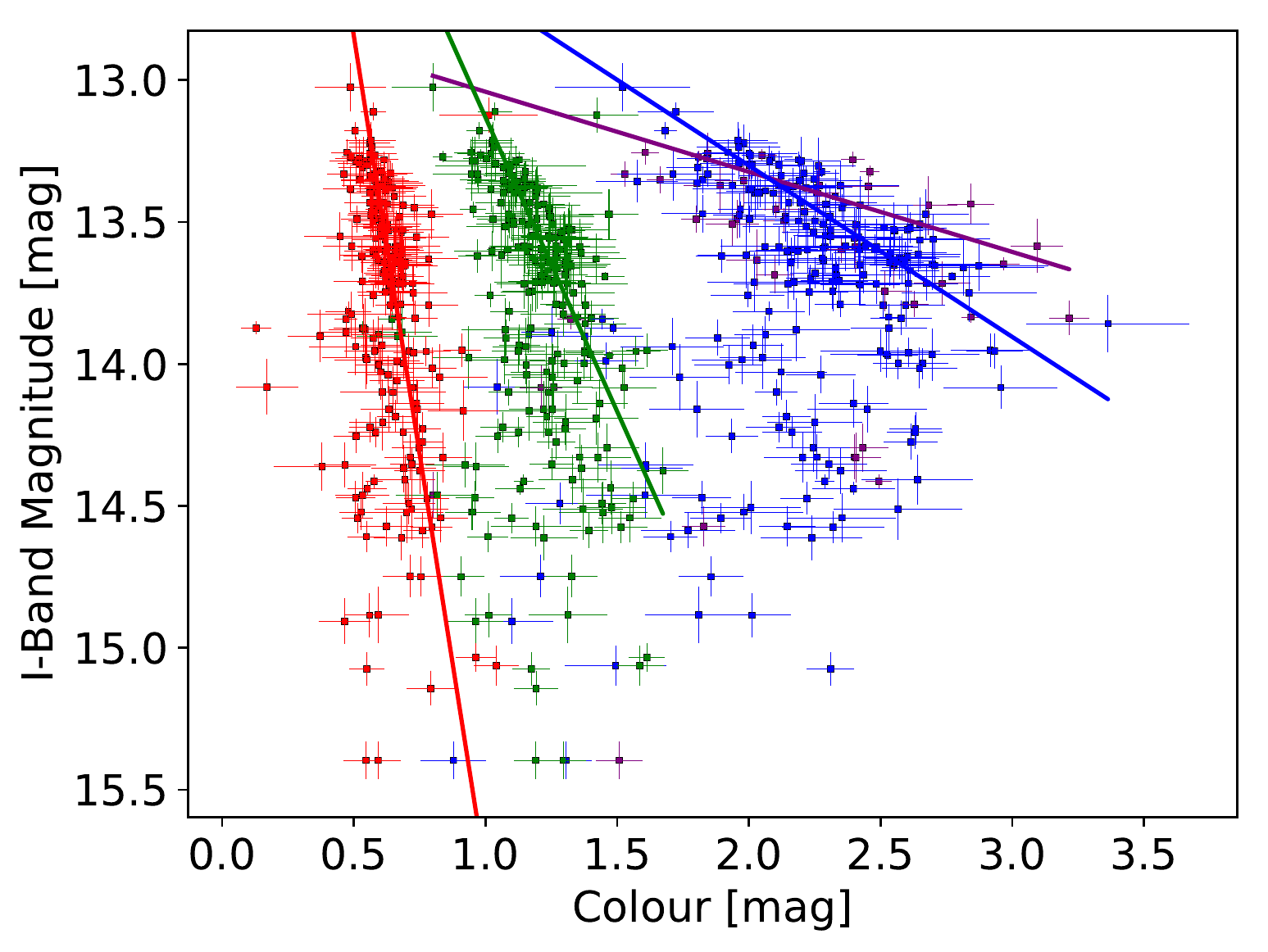} \hfill
\includegraphics[angle=0,width=5.8cm]{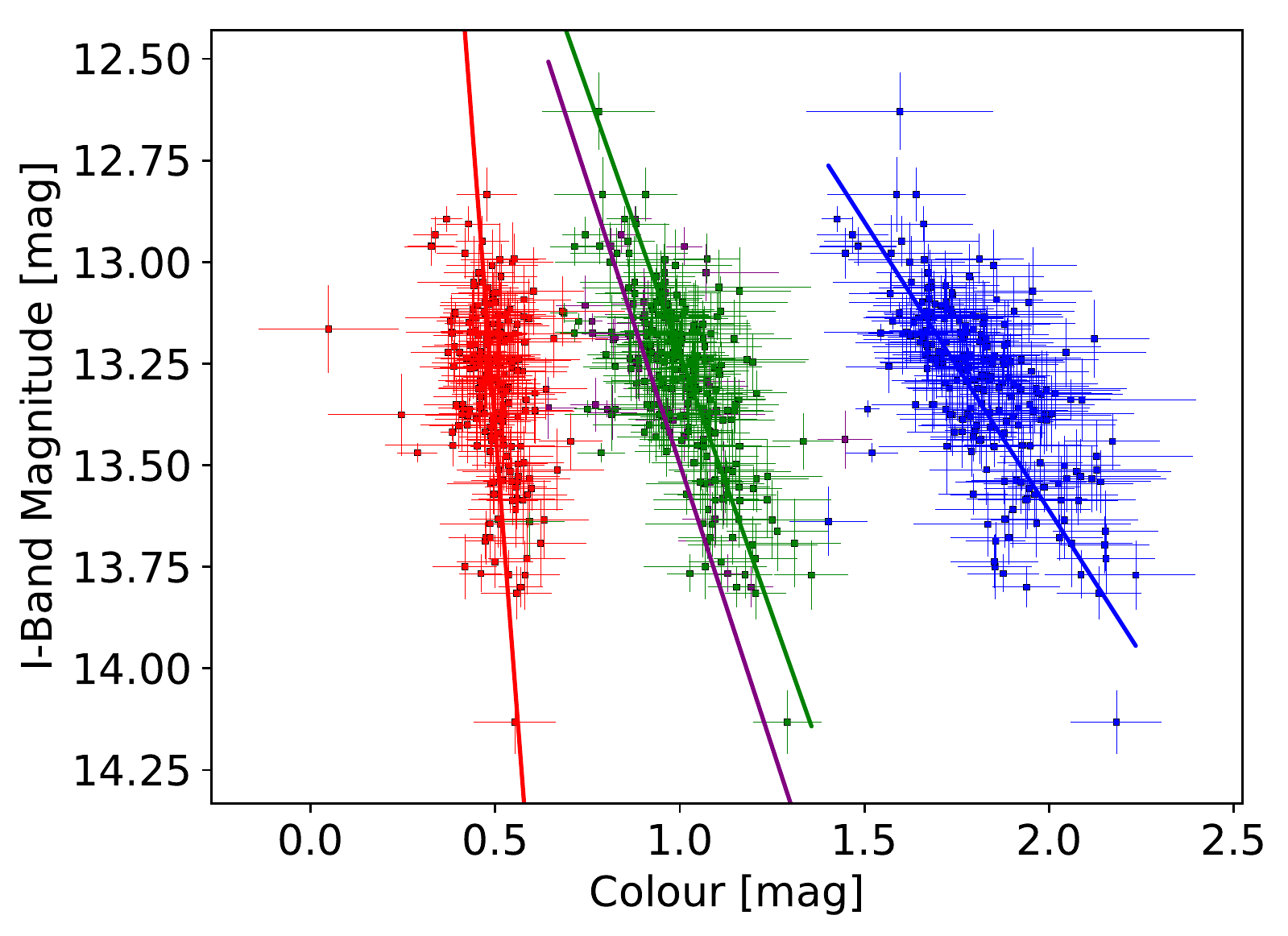} \\
\includegraphics[angle=0,width=5.8cm]{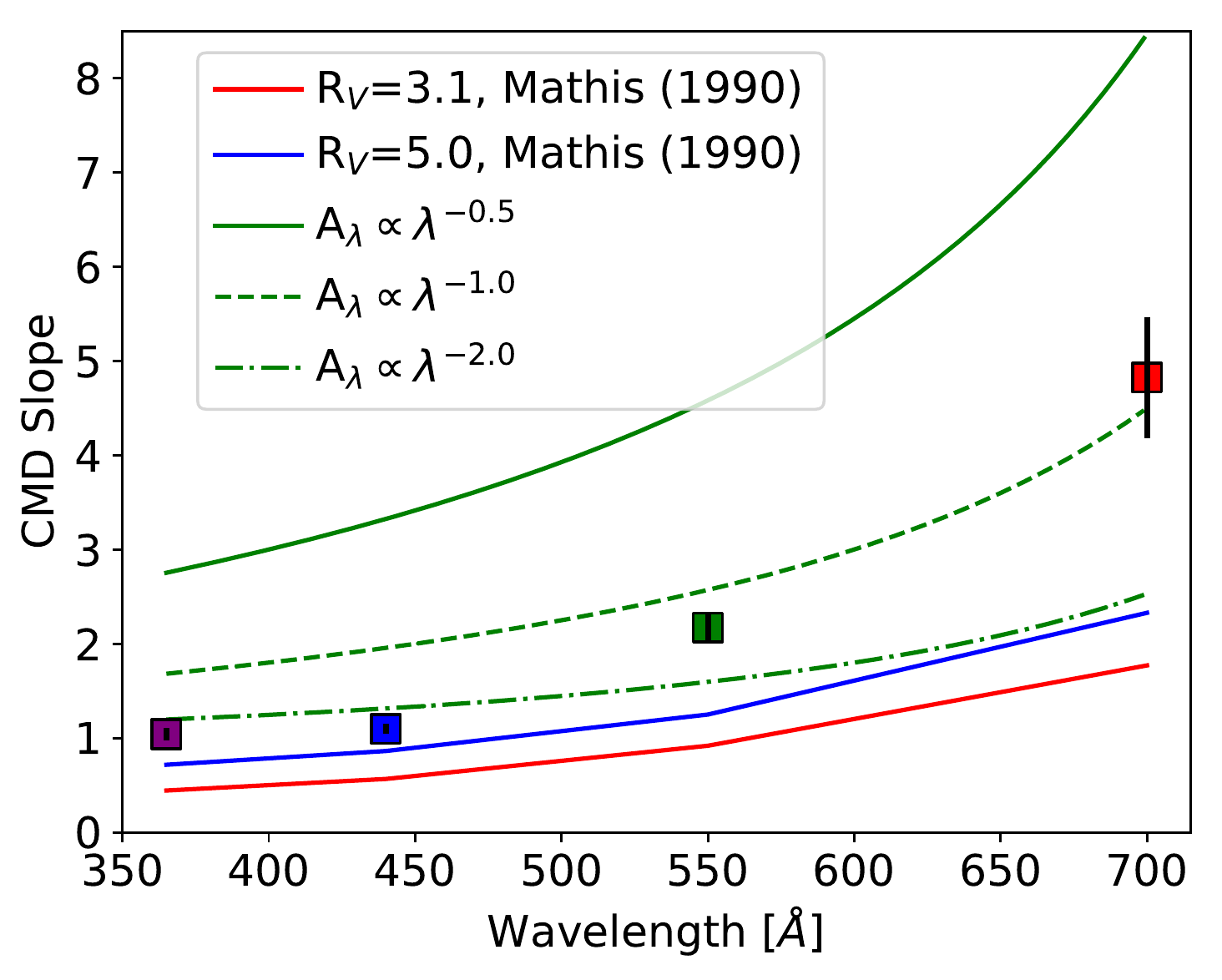} \hfill
\includegraphics[angle=0,width=5.8cm]{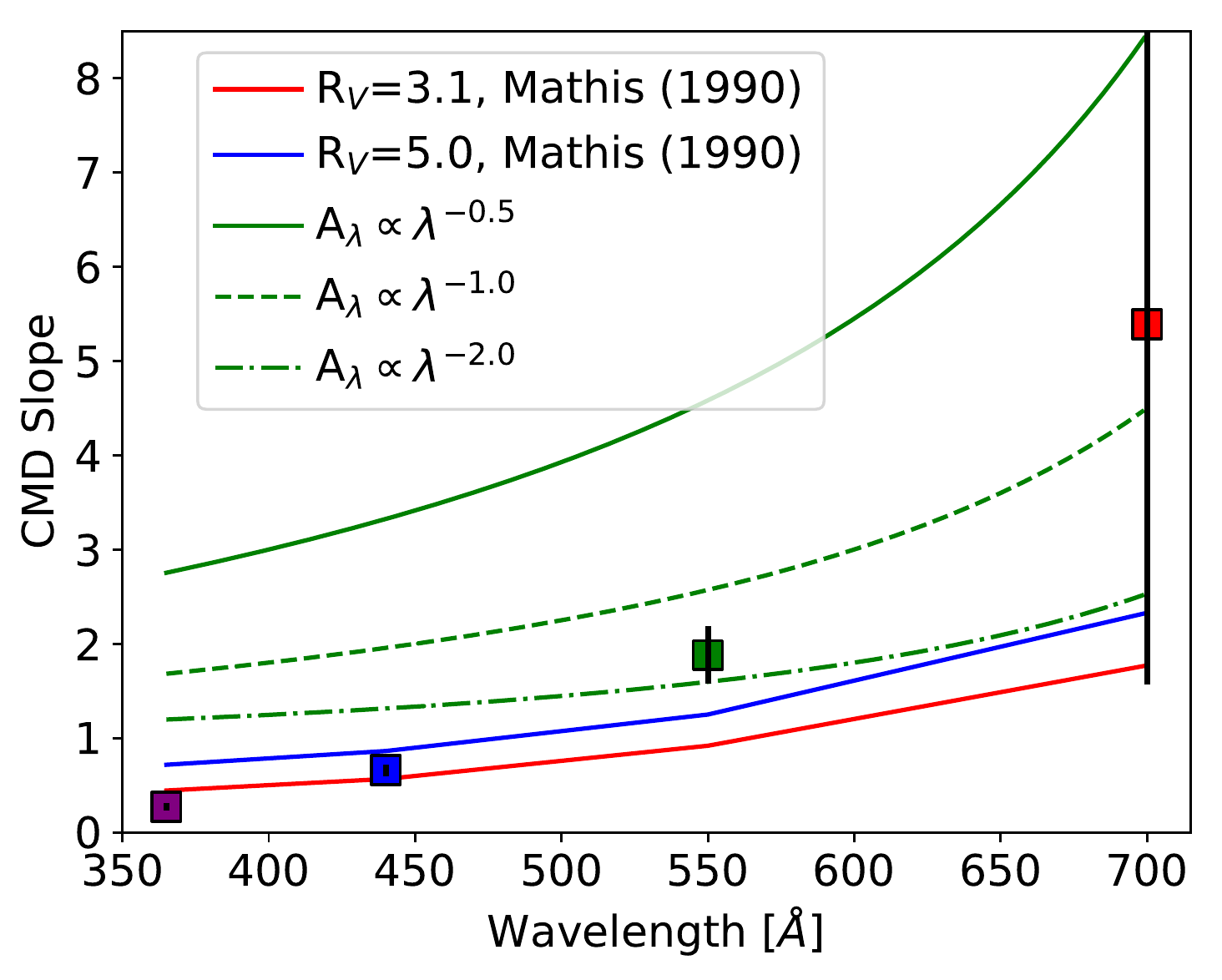} \hfill
\includegraphics[angle=0,width=5.8cm]{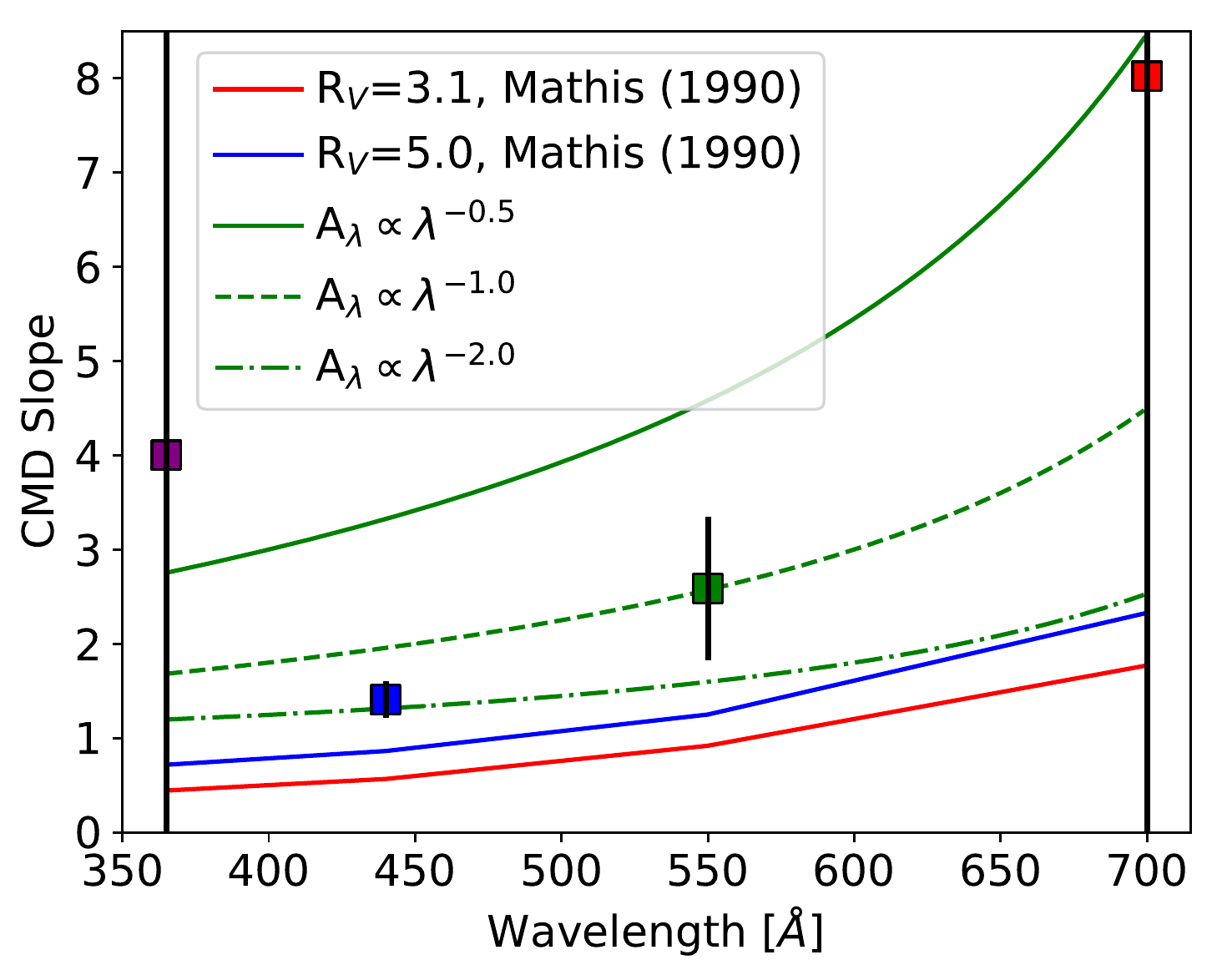} 
\caption{{\bf Top Row:} Colour magnitude diagrams for HST ULLYSES targets. We show the R-I (red), V-I (green), B-I (blue), and U-I (purple) colours against the I magnitude. All HOYS data from the 2020/21 observing season are shown. The over plotted lines indicate the best fit from a perpendicular distance regression to the bright part of the data. {\bf Bottom Row:} Slopes of the colour magnitude diagrams against the shorter wavelength used in the colours. In some cases the uncertainties of the slopes are smaller than the symbol. The red and blue lines represent the extinction model from \citet{1990ARA&A..28...37M}, while the green lines are some basic power law extinction models for different slopes. Each column represents one of the objects. {\bf Left:} TX\,Ori {\bf Middle:} V505\,Ori {\bf Right:} V510\,Ori. \label{cmd_hst}}
\end{figure*}

To investigate the potential causes for the photometric variability of the ULLYSES targets, we have generated colour magnitude diagrams (CMDs) for all objects. Given the partly very rapid variability of all objects, colours are determined for all data points that have been taken less than five hours apart. Typically, at least the B, V, R, and I data sets have been taken with much shorter time gaps. For ease of viewing we show the I-band magnitude against all possible colours containing the I filter, i.e. R-I, V-I, B-I, and U-I. The plots for the three ULLYSES targets are shown in the top panel of Fig.\,\ref{cmd_hst}. We only utilised HOYS data from the 2020/21 observing season, but including the earlier data makes no significant difference to the results.

In order to compare the behaviour of the data with extinction models, the slope in the colour magnitude diagrams was determined for each colour. Since in many cases the data are spread almost vertically in the parameter space, we used an orthogonal minimisation regression. The photometric errors for the magnitudes and colours are considered in the fit. We determine the slope and its uncertainty with a Monte Carlo simulation in the following way. For each magnitude measurement in the CMD we create a set of 2000 Gaussian distributed random numbers with a mean of the magnitude and a sigma of the magnitude uncertainty. The regression to determine the slope is then repeated for each of those randomly generated magnitudes and colours. The median value of all slopes obtained this way is taken as the slope and the root mean square ($RMS$) from the median as the uncertainty. Note that in the case of almost vertical slopes (as for R-I) or sparse data (U-I), the uncertainties can become very large. In some cases, there is a clear deviation from a linear behaviour for fainter magnitudes (see details below). These data have been excluded from the fit. These fits for all colour magnitude plots are shown in Fig.\,\ref{cmd_hst} and the values for the slopes are detailed in Table\,\ref{tab_slopes}.

\begin{table}
\caption{\label{tab_slopes} CMD slopes and their one sigma uncertainties for the ULLYSES targets. For TX\,Ori and V505\,Ori the slopes are determined excluding the faint magnitudes where the colours turn blue (see text for details).}
\centering
\begin{tabular}{|c|c|c|c|c|}
\hline
Target & U-I & B-I & V-I & R-I \\ \hline
TX\,Ori   & 1.04\,$\pm$\,0.06 & 1.10\,$\pm$\,0.05 & 2.18\,$\pm$\,0.14 & 4.79\,$\pm$\,0.61 \\
V505\,Ori & 0.27\,$\pm$\,0.04 & 0.66\,$\pm$\,0.06 & 1.87\,$\pm$\,0.30 & 5.35\,$\pm$\,3.58 \\
V510\,Ori & 4.02\,$\pm$\,13.4 & 1.41\,$\pm$\,0.19 & 2.57\,$\pm$\,0.96 & 8.02\,$\pm$\,22.0 \\ \hline
\end{tabular}
\end{table}

We compare the slopes in the colour magnitude diagrams to the extinction model from \citet{1990ARA&A..28...37M} for $R_V$\,=\,3.1 and 5.0 in the bottom panel of Fig.\,\ref{cmd_hst}. It can be seen that the variability of neither of the sources can be explained by the extinction models. We note that the scatter in the U-I colours is much larger than in the other colours. This colour is potentially influenced by accretion rate variability, which we will investigate in Sect.\,\ref{macc}. The slopes do in all cases lie above the R$_V$\,=\,5 models. They also do not follow a simple power law model. For two sources a deviation from the linear trend towards bluer colours is found for fainter magnitudes. This is usually attributed to the brightness becoming dominated by scattered light \citep[e.g.][]{1991Ap&SS.186..283G}. Below we provide a more in-depth discussion of the CMDs of the individual sources.

\noindent {\bf TX\,Ori:} The object varies by about 1.4\,mag in the I-band. Aside from a few outlying points, the data align well with a straight line in all but the U-I CMD. Thus, the scatter from the straight line fit seems to be entirely caused by the photometry uncertainties. For magnitudes of about I\,$>$\,12.3\,mag, all CMDs show a clear systematic deviation from the straight line behaviour seen for the brighter magnitudes. For these faint states the colours become independent of magnitude and even slightly turn bluer towards fainter magnitudes. This is particularly evident in the V-I colour, but can be seen in all other colours as well. This turning in colour behaviour is usually interpreted as a sign that increased extinction makes the direct line of sight towards the central star optically thick, and the light from the object starts to be dominated by scattered light from the disk \citep[e.g.][]{1991Ap&SS.186..283G}. If one assumes that at this turnover point the scattered light accounts for at least half the radiation received, and the variation in brightness is indeed caused by some form of extinction, then even at the bright state almost 20\,\% of the light is scattered. This might explain why the colours do not fit any extinction model. However, it can also clearly be seen that the scatter of the U-I colours from the linear behaviour in the CMD is far greater than what would be expected from the photometric uncertainties. This could be a sign that the star is undergoing accretion rate variations, as the U-band excess traces $\dot{M}$ \citep{2003ApJ...592..266M}. We discuss this further in Sect.\,\ref{macc}.

\noindent {\bf V505\,Ori:} This object varies by up to 2.3\,mag in I. Similar to TX\,Ori the brighter part of the data follows a straight line in most of the CMDs. However, the range of magnitudes where the behaviour deviates is almost 1.5\,mag and occurs for I\,$>$\,13.7\,mag. The scatter of the data at these fainter magnitudes is much larger than the nominal photometric uncertainties, but the object is much fainter than the other two. The turning towards bluer colours is especially evident in the B-I colour, where the object turns almost one magnitude bluer at the very faintest points, compared to its reddest state. Thus, this source is even more dominated by scattered light during the faint state. Like for TX\,Ori, the U-I colours are extremely variable and seem almost uncorrelated with the I-band magnitude. This might indicate a higher accretion rate variability in this object. The slopes in the CMDs are not in agreement with a reddening law. 

\noindent {\bf V510\,Ori:} This object is the least variable in I, with a range of only 0.8\,mag, ignoring a few single outliers. All but the U-I data do follow a straight line, consistent with the photometric uncertainties. Contrary to the other two sources, there is no indication of a change in slope of the behaviour for fainter magnitudes. Despite this, the slopes are inconsistent with an extinction model, especially in the R-I data. The U-I data scatters to the point that there is no real correlation with the I magnitude. This might again indicate accretion rate variations of the object.

\subsection{Periodic variability of \texorpdfstring{V505\,}{} Ori}\label{v505-var}

\begin{figure*}
\centering
\includegraphics[angle=0,width=0.95\columnwidth]{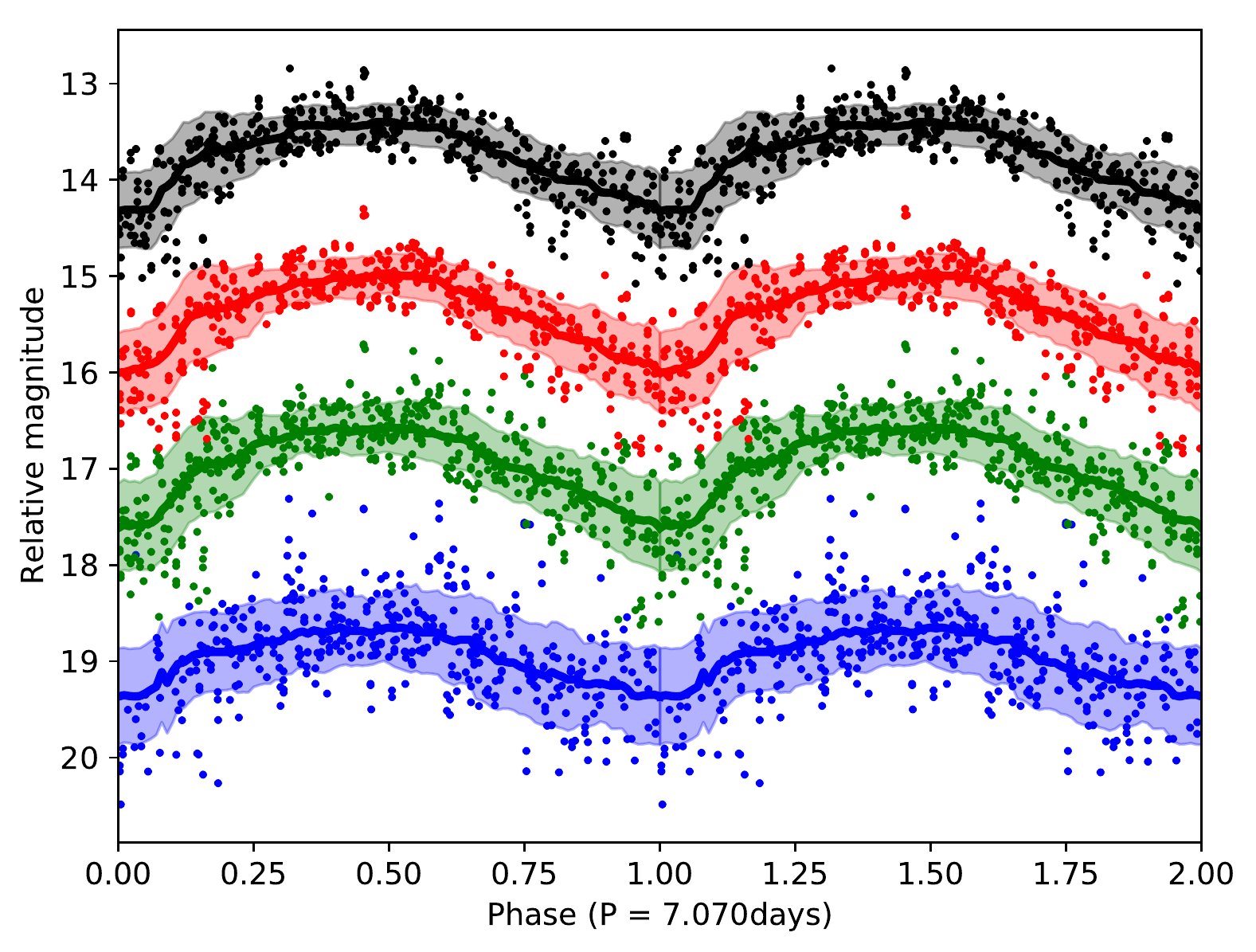} \hfill
\includegraphics[angle=0,width=1.05\columnwidth]{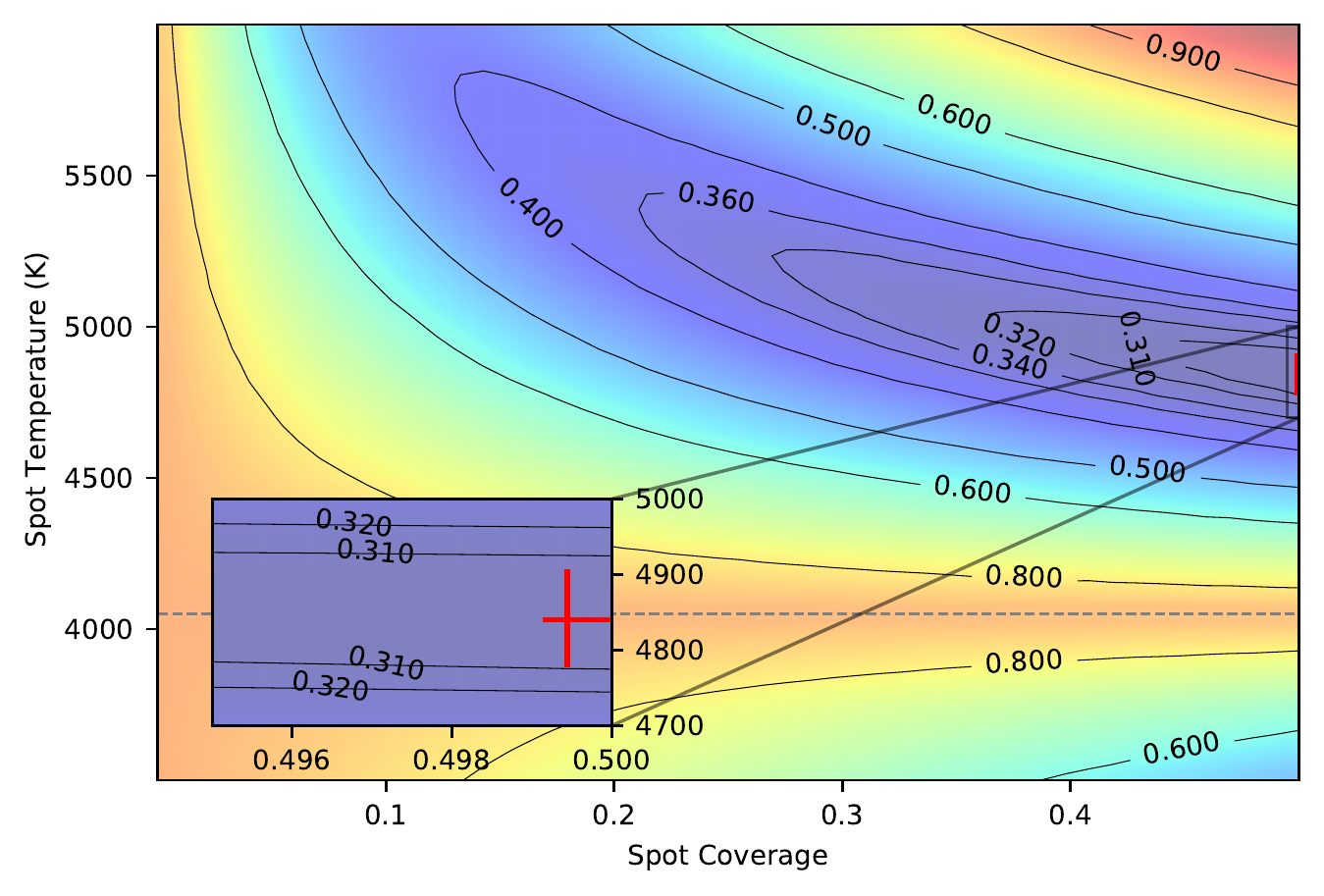}
\caption{{\bf Left:} Phase folded B (blue), V (green), R (red), and I (black) photometry data for V510\,Ori. We over plot a running median and one sigma scatter of the data from the median. The R data are shown at the correct magnitude, the other data are shifted for best visibility. The actual apparent magnitudes and colours of the source can be seen in Figs.\,\ref{lc_hst} and \ref{cmd_hst}. {\bf Right:} Model results to fit the photometry of V505\,Ori with a spot model. The background image shows the $RMS$ of the fit from the model as a function of the spot coverage and temperature. The red cross indicates the best fitting model and its uncertainties. The large $RMS$ values for the best model and its placement at a spot coverage of 0.5 indicate that a spot cannot explain the data. \label{phase_v505ori}}
\end{figure*}

A visual inspection of the V505\,Ori light curve shows a clear apparent periodicity in all filters, which is also visible in Fig.\,\ref{lc_hst}. We hence investigate this in more detail in this subsection. In order to establish the period of the source, we determined a simple Lomb-Scargle periodogram for the B, V, R, and I data separately. Like for most of the other analysis, we only include the data from the 2020/21 observing season, but note that the results do not change when including all data. Each of the four periodograms shows a clear significant peak at a period of seven days. We determine the average and $RMS$ scatter of the four periods as P\,=\,7.070\,$\pm$\,0.007\,d (7d:01h:41m\,$\pm$\,10m).

We show the phase folded light curves for this period in B (blue), V (green), R (red), and I (black) in the left hand panel of Fig.\,\ref{phase_v505ori}. The R data are shown at the correct magnitudes, the data in the other filters are shifted for best visibility. We also over plot a running median as solid lines and the one sigma scatter of the data from the running median as shaded regions. The phase folded plot clearly shows the periodic nature of the light curve. The average brightness minima have an almost triangular shape, while the average maxima are rounded. As evident in Fig.\,\ref{lc_hst}, the individual maxima and minima can be sharper, especially on timescales shorter than our typical cadence of one day. These short term variations are more apparent in the TESS data which will be discussed in Abraham et al. (in prep.). Furthermore, the dips are not symmetric. The decline in brightness occurs slower than the subsequent brightening. 

One potential interpretation of the periodicity is that the dips are caused by an occulting structure in the accretion disk. The most common type of such systems are AA\,Tau objects \citep{2015A&A...577A..11M}. There, an inner disk warp is created by a misalignment of the stellar magnetosphere and the inner disk \citep{2007A&A...463.1017B}. The warps, and thus the occulting material are situated near or at the co-rotation radius \citep{1999A&A...349..619B}. The asymmetry in the phase folded data would then indicate that the occulting structure is asymmetric, i.e. has a sharper leading edge. While the HOYS data from the other observing seasons fit into the phase folded plot, there is an insufficient number of data points to investigate potential changes in the structure of the phase folded light curve. 

Figures\,\ref{phase_v505ori} and \ref{lc_hst} also show that the light curve is not strictly periodic. The height of the maxima varies by about 0.2\,mag, the scatter in the minima is up to one magnitude. This is not entirely reflected by the shaded areas in the phase plot, but is evident by the many data points situated below the nominal scatter around the minima. This is not caused by the photometric uncertainties, which are much smaller than these variations. Thus, if the variations are caused by an inner disk warp, then the column density along the sight line to the central source varies by up to a factor of about two. A detailed inspection of the light curve shows that these variations can occur at the full range from one dip to the next. Thus, material is moved in and out the inner disk structure on timescales as small or smaller than the orbital period. If the object is indeed an AA\,Tau type star, the orbital period would be the same as the rotation period of the star. Note that such a scenario would also suggest a variable mass accretion rate (see Sect.\,\ref{macc}).

However, the CMD plots (Fig.\,\ref{cmd_hst}) for this source do indicate that the variations in colour are not in agreement with an extinction model. This could be caused by the contribution of scattered light but also suggest that the variations could alternatively be caused by a spot on the surface of the rotating star. If the spot properties (coverage, temperature) change on short time scales (less than the rotation period) this could explain the data. We hence attempted to investigate if the amplitudes in the different filters can be explained by a spot model. 

We determined the observed peak-to-peak amplitudes ($\hat{A}^o_\lambda$) in all filters ($\lambda \in$ B, V, R, I) from the phase folded data in Fig.\,\ref{phase_v505ori}. They are shown with their uncertainties in Table\,\ref{v505ori_amp}. For our simple spot model we assume that the amplitude of the variation is caused by a single spot with temperature $T_s$, covering a fraction $f$ of the visible stellar surface. The spot free stellar surface has a temperature $T_*$, producing a flux $F_0^\lambda(T_*)$. The spot reduces the flux of the star by the coverage $f$ and adds its own flux $F_s^\lambda(T_s)$ . The model peak-to-peak amplitude variations between the spot free and the spot covered situation can hence be determined  according to Eq.\,\ref{Fspotstar_c}.

\begin{equation}
\hat{A}^m_{\lambda}  = - 2.5 \times \log \left( \frac{F_0^\lambda(T_*)}{\left( 1 - f \right) \times F_0^\lambda(T_*)   + f \times F_s^\lambda(T_s) } \right) 
\label{Fspotstar_c}
\end{equation}

To model the stellar and spot fluxes $F_0^\lambda$ and $F_s^\lambda$ in all filters, we use the ATLAS9 \citep{2003IAUS..210P.A20C} and PHOENIX \citep{2013A&A...553A...6H} stellar atmosphere models. They are calculated in the BVRI Johnson-Cousin filters \citep{1990PASP..102.1181B} accessed through the {\tt speclite.filters}\footnote{\tt \url{https://github.com/desihub/speclite/blob/master/speclite/filters.py}} in Python. To model V505\,Ori we use the models for a stellar temperature of 4050\,K \citep{2016ApJ...829...38M}, $log(g) = 4.0$, and $[M/H] = 0.0$. These are reasonable values for the star. Their exact choice does not influence our results. All details and uncertainties of our spot fitting method are discuss in Herbert et al. (2022, in prep.).

In order to determine the best fitting spot model for V505\,Ori we generated $5 \times 10^5$ sets of model amplitudes $\hat{A}^m_{\lambda}$. They were generated using randomly selected, homogeneously distributed values for the spot temperature (3500\,K\,$\leq T_s \leq$\,12000\,K) and spot coverage (0.0\,$\leq f \leq$\,0.5). We then identify the best fitting model as the one with the minimum $RMS$ defined in Eq.\,\ref{rms_def}. Similar to the determination of the slopes in the CMD (Sect.\,\ref{colmags}), we repeat the calculations 200 times by randomly varying $\hat{A}^o_{\lambda}$ within their uncertainties. We then take the median best spot properties and their standard deviation from the median as the best value and its uncertainty. 

\begin{equation}
RMS = \frac{1}{2} \sqrt{ \sum_{\lambda} \left( \hat{A}^m_{\lambda}  - \hat{A}^o_{\lambda} \right)^2}
\label{rms_def}
\end{equation}

In the right panel of Fig.\,\ref{phase_v505ori} we show the $RMS$ distribution (in colour and contours) for the measured peak-to-peak amplitudes using the ATLAS9 models as an example. The plot for the PHOENIX models looks very similar. The over plotted red cross (also in the zoomed in panel) shows the position and uncertainties of the best fitting model. It is clear from the position of the best coverage ($f \approx 0.5$) and the large $RMS$ values ($\approx 0.3$\,mag, i.e. about ten times the amplitude uncertainties), that none of the spot models can fit the data. It would require an unrealistically large spot, and still generate very different amplitudes. The same results are obtained for the PHOENIX models. 

Thus, we conclude that the periodic variability of V505\,Ori cannot be explained by a spot on the surface of the star. It is hence probably an AA\,Tau like object with the material in the disk warp being moved in and out on time scales below the rotation period. The scattering properties of the occulting disk material are unusual and/or the brightness of the system is dominated by scattered light, in particular during the dim state.

\begin{table}
\caption{\label{v505ori_amp} Average peak-to-peak amplitudes ($\hat{A}^o_\lambda$) of the variability in V505\,Ori in the different filters and their one sigma uncertainties. There is insufficient data to determine the amplitude in the U-band.}
\centering
\begin{tabular}{|c|c|c|c|c|}
\hline
Filter & B & V & R & I \\  \hline
Amplitude [mag] & 0.698 & 1.048 & 0.931 & 0.872 \\
Uncertainty [mag] & 0.051 & 0.029 & 0.027 & 0.029 \\
\hline
\end{tabular}
\end{table}

\begin{figure*}
\centering
\includegraphics[angle=0,width=2\columnwidth]{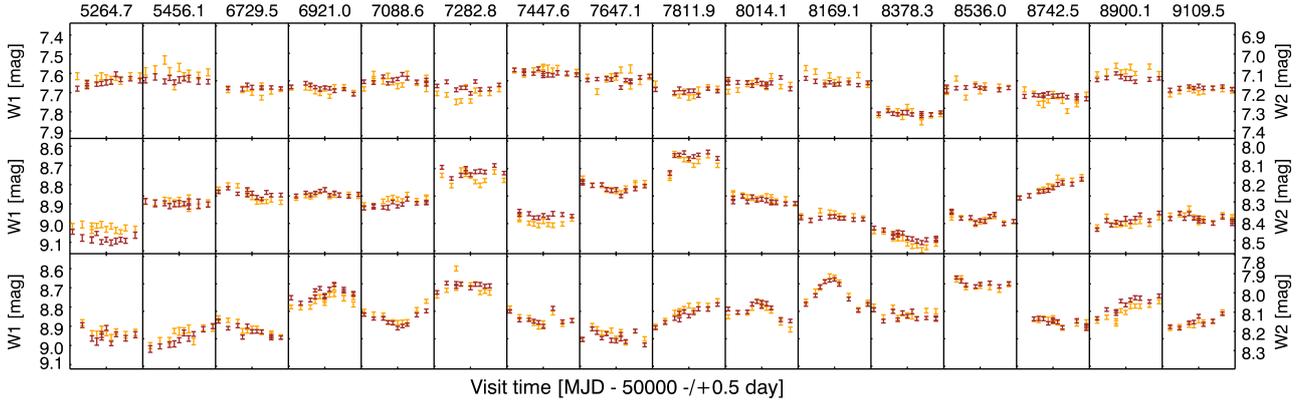} 
\caption{(NEO)WISE W1 (orange) and W2 (brown) light curves for each visit of the ULLYSES targets {\bf Top:} TX\,Ori {\bf Middle:} V505\,Ori {\bf Bottom:} V510\,Ori. The magnitude ranges are mutually shifted for best overlap. Each column shows one day of data with the mid-point modified Julian Date, subtracted by 50,000, indicated for each visit on the top. \label{lc_wise}}
\end{figure*}

\subsection{Long term (NEO)WISE light curves}\label{wise_ana}

To supplement the optical observations and to provide a longer temporal baseline, infrared (IR) photometry for the $\sigma$\,Ori YSOs from the Wide-field Infrared Survey Explorer (WISE) space telescope \citep{2010AJ....140.1868W} and the subsequent (NEO)WISE \citep{2014ApJ...792...30M} mission was retrieved from the NASA/IPAC Infrared Science Archive (IRSA). It covers observations from 2010 until end of 2020, with bi-annual visits of the $\sigma$\,Ori region in spring and autumn, lasting for about one day. Thus, both long-term and intra-day variability is covered by the photometry.  After the termination of the cryogenic WISE mission, only the two shortest bands, W1 (3.4\,$\mu$m) and W2 (4.6\,$\mu$m), are available during the ongoing (NEO)WISE survey. By default, a saturation correction has been applied\footnote{See\,\url{http://wise2.ipac.caltech.edu/docs/release/neowise/expsup/sec2\_1civa.html}} to account for a photometric bias due to warm-up of the detector, of $+0.02$\,mag (W1) and $+0.33$\,mag (W2), respectively. The usefulness of the (NEO)WISE data for the characterisation of YSO variability has been demonstrated by \cite{2021arXiv210710751P}.

The (NEO)WISE light curves for the ULLYSES targets are shown in Fig.\,\ref{lc_wise}. Here, the magnitude ranges were adjusted to match the W1 and W2 bands. In general, the intra-day variability in both bands is well correlated. While, according to the Stetson index (see Sect.\,\ref{context_var}), V505\,Ori shows the largest overall variability in the WISE bands, the intra-day variations are most pronounced for V510\,Ori, as evidenced by the mean scatter of the photometry. This hints at different time scales for the underlying mechanisms causing the variations. We discuss the general variability of the sources in the (NEO)WISE and optical bands in comparison to the other $\sigma$\,Ori cluster members in Sect.\,\ref{HST_context_var}.

\begin{figure*}
\centering
\includegraphics[angle=0,width=0.5\columnwidth]{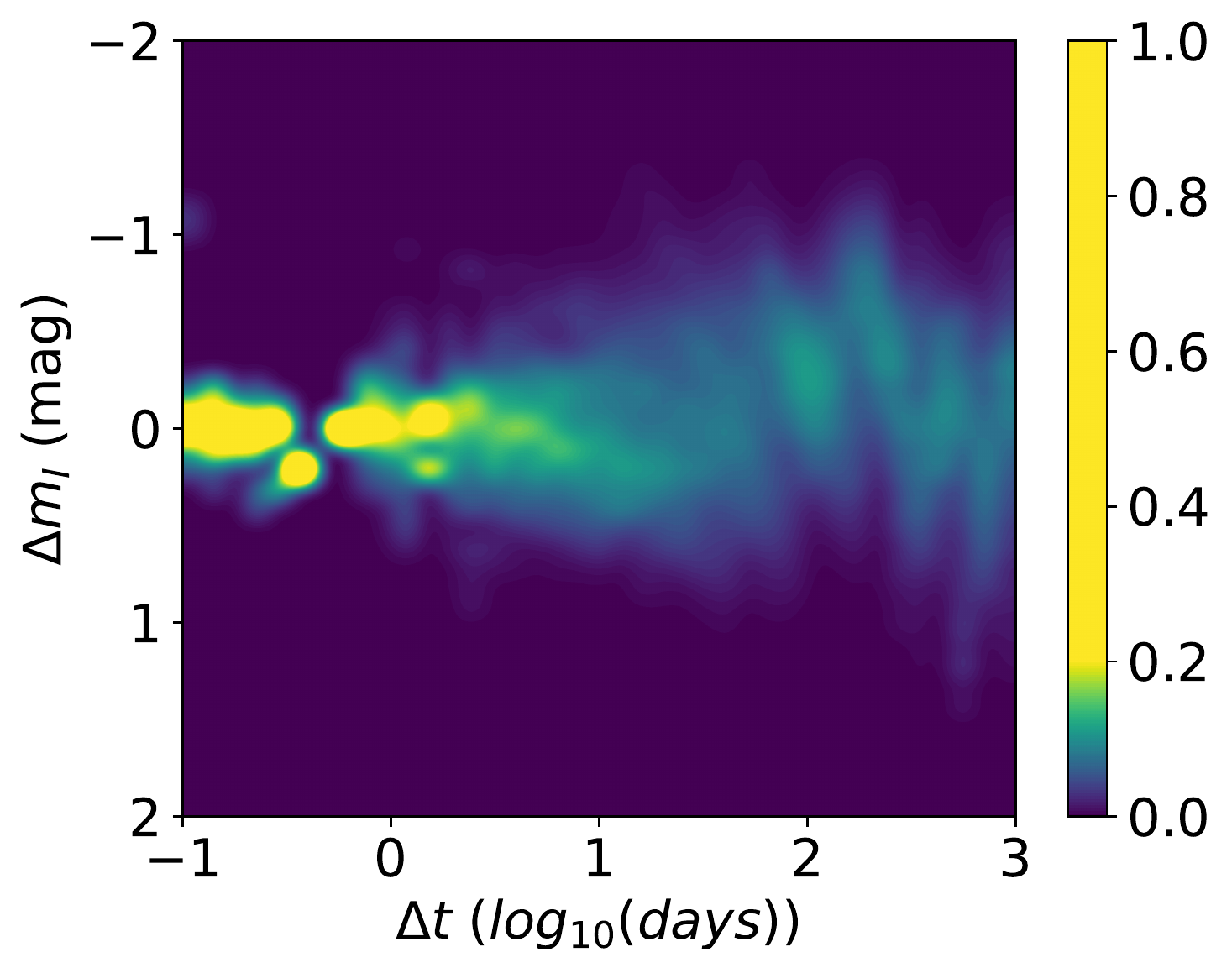} \hfill
\includegraphics[angle=0,width=0.5\columnwidth]{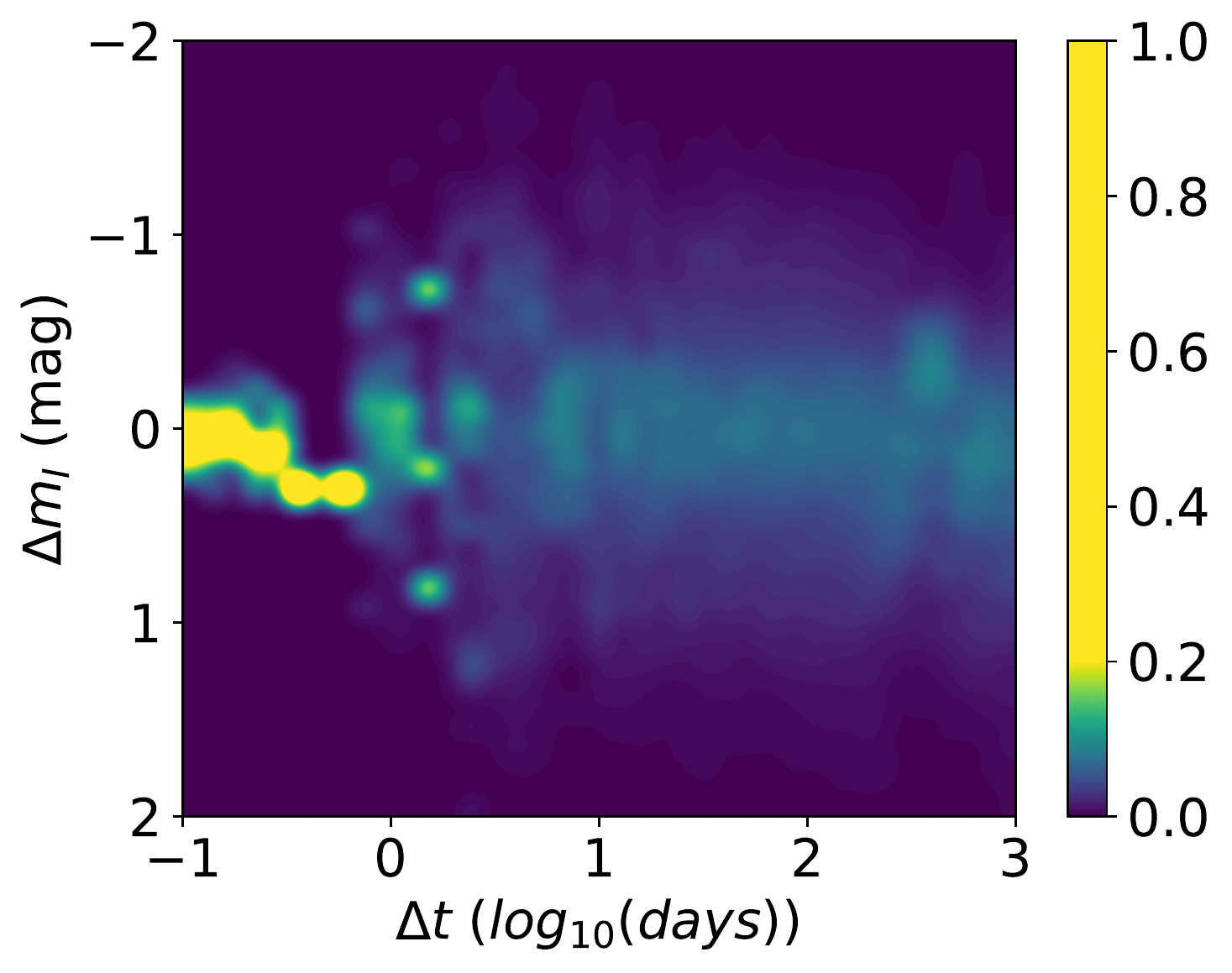} \hfill
\includegraphics[angle=0,width=0.5\columnwidth]{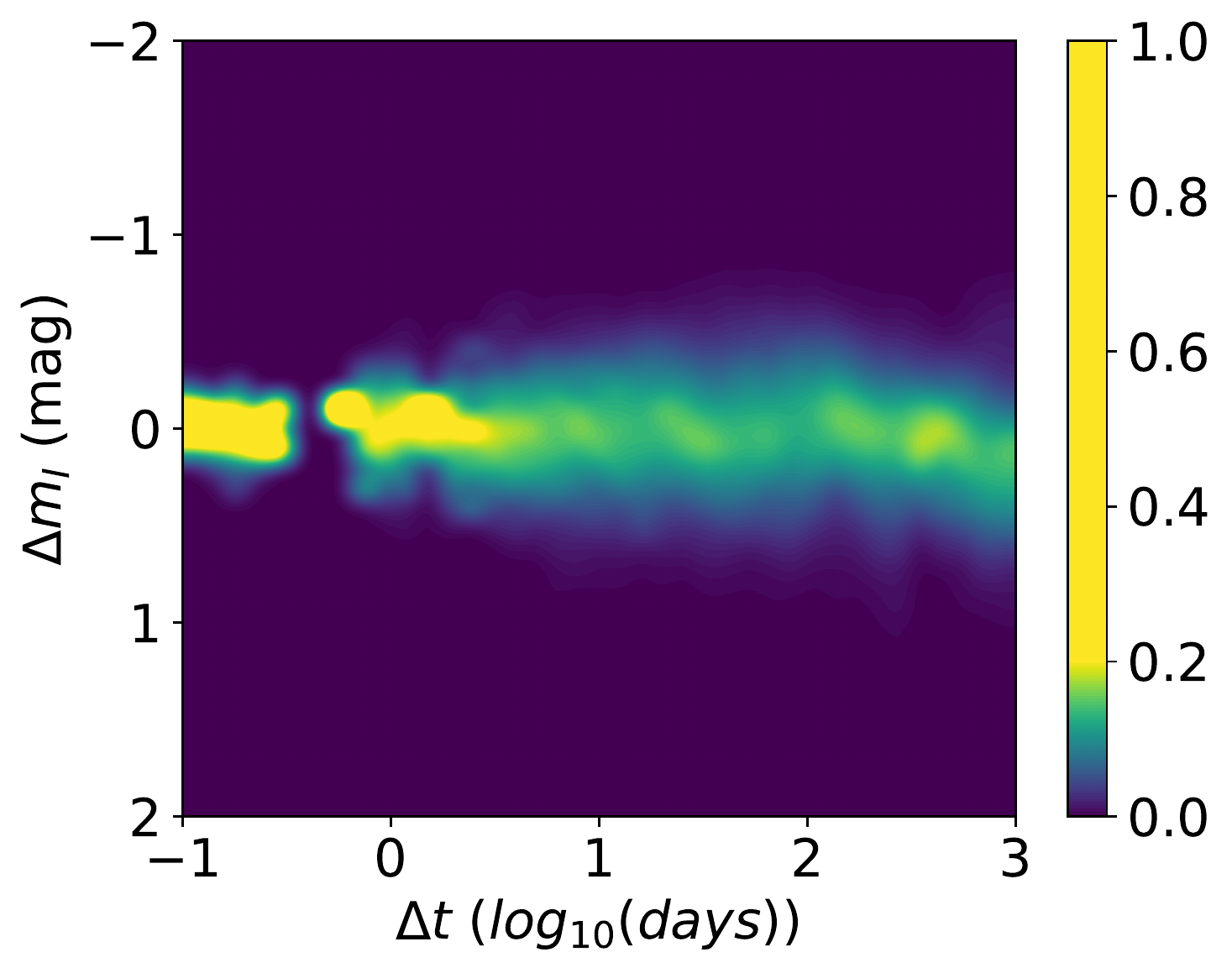} \hfill
\includegraphics[angle=0,width=0.5\columnwidth]{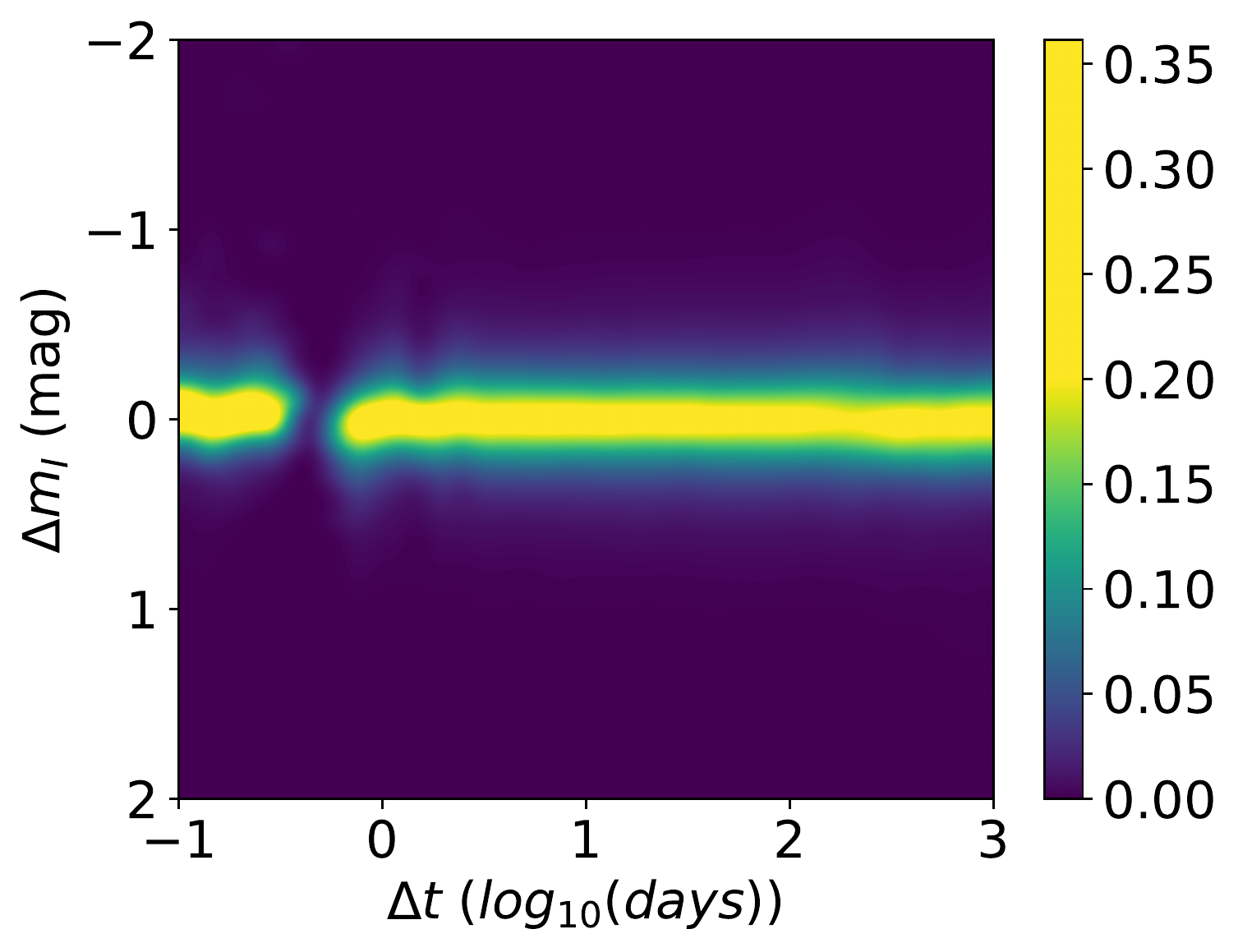} \\
\includegraphics[angle=0,width=0.5\columnwidth]{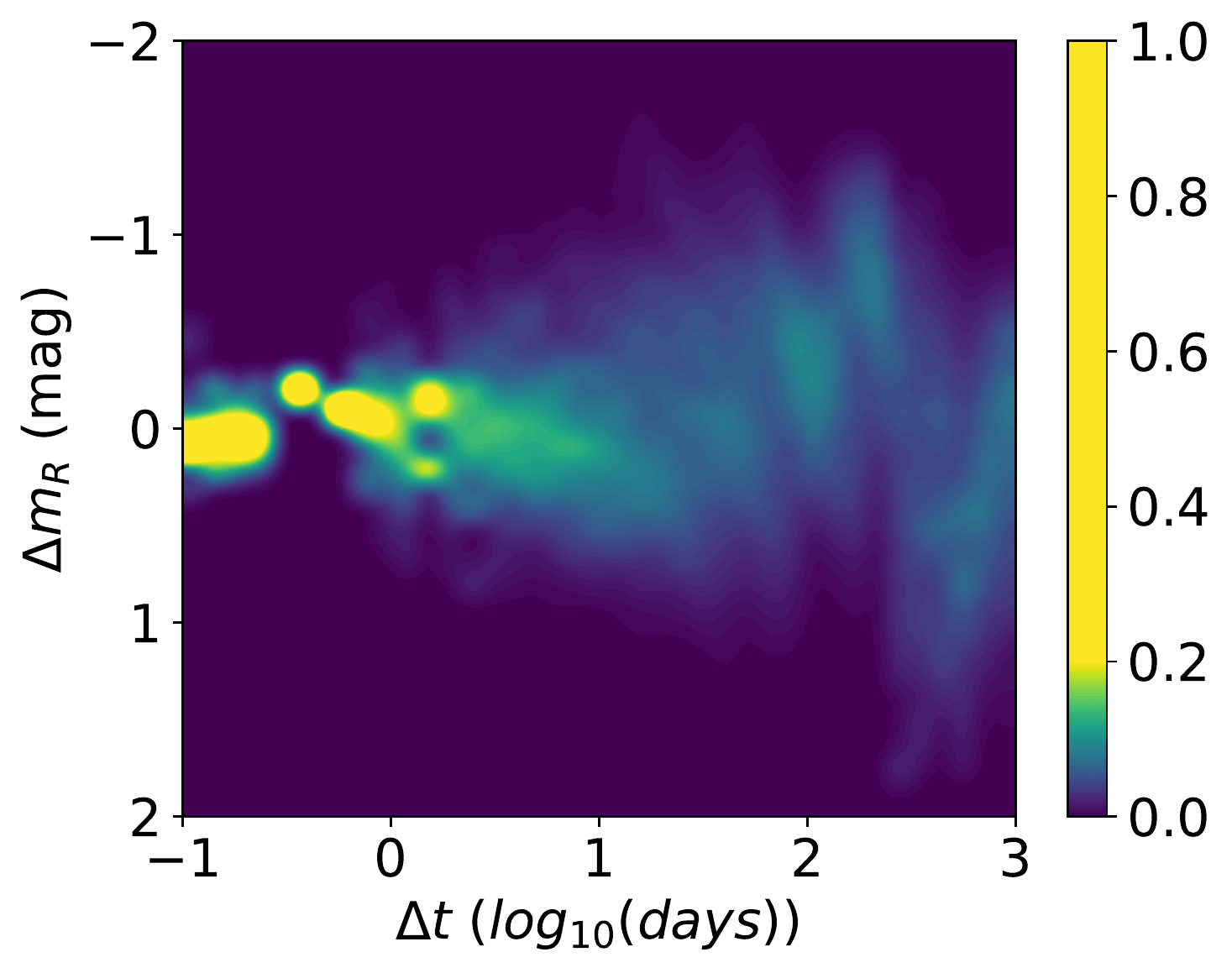} \hfill
\includegraphics[angle=0,width=0.5\columnwidth]{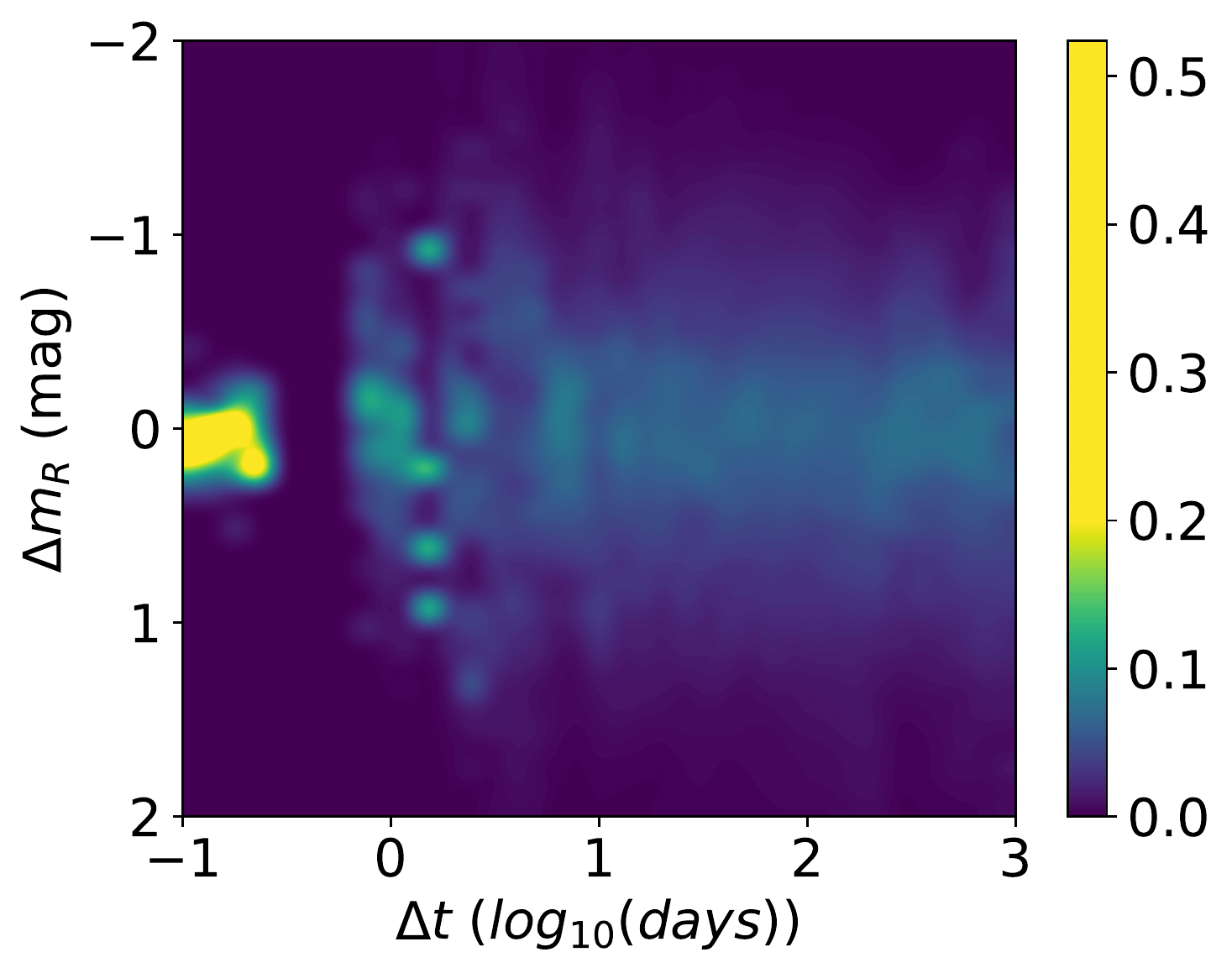} \hfill
\includegraphics[angle=0,width=0.5\columnwidth]{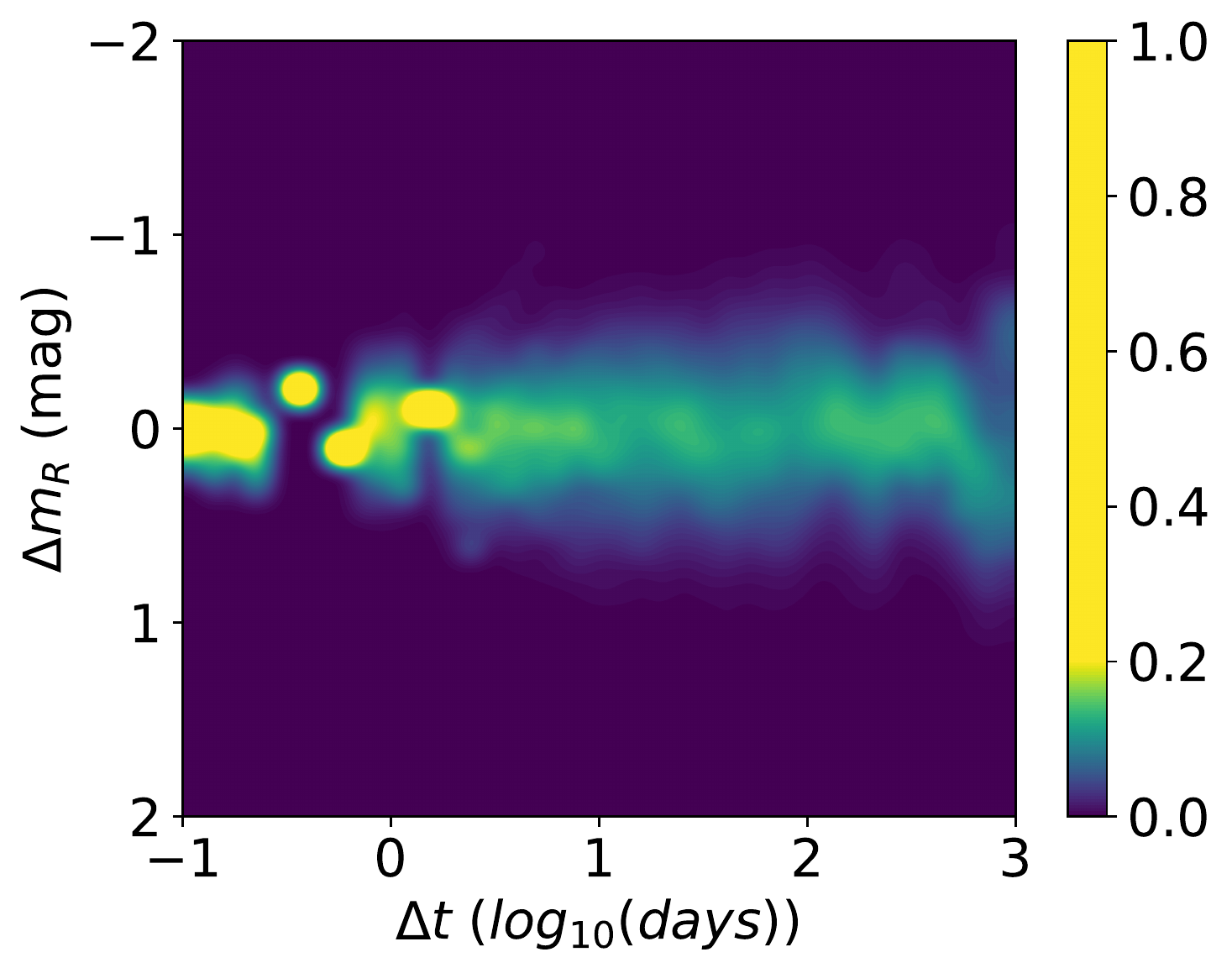} \hfill
\includegraphics[angle=0,width=0.5\columnwidth]{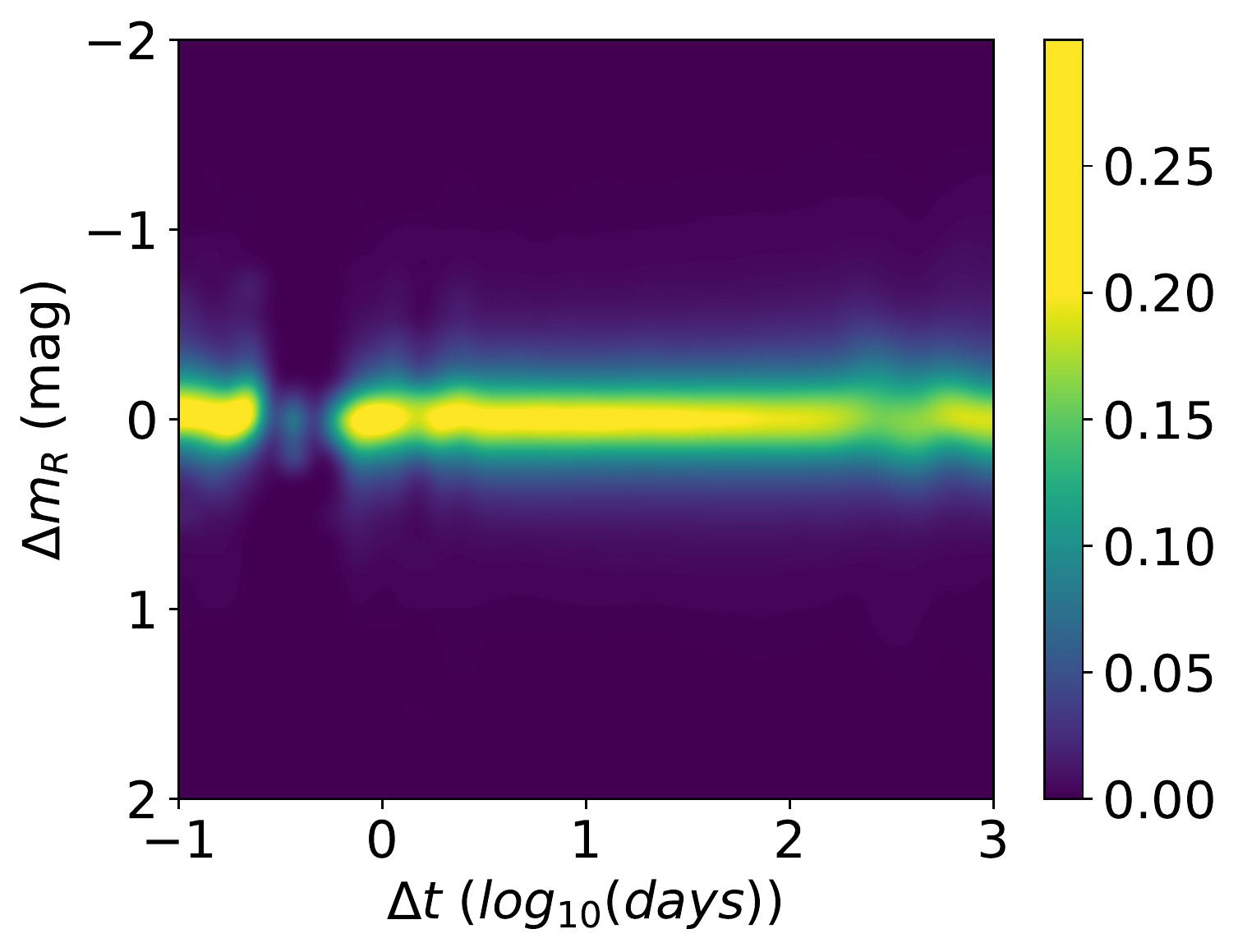} \\
\includegraphics[angle=0,width=0.5\columnwidth]{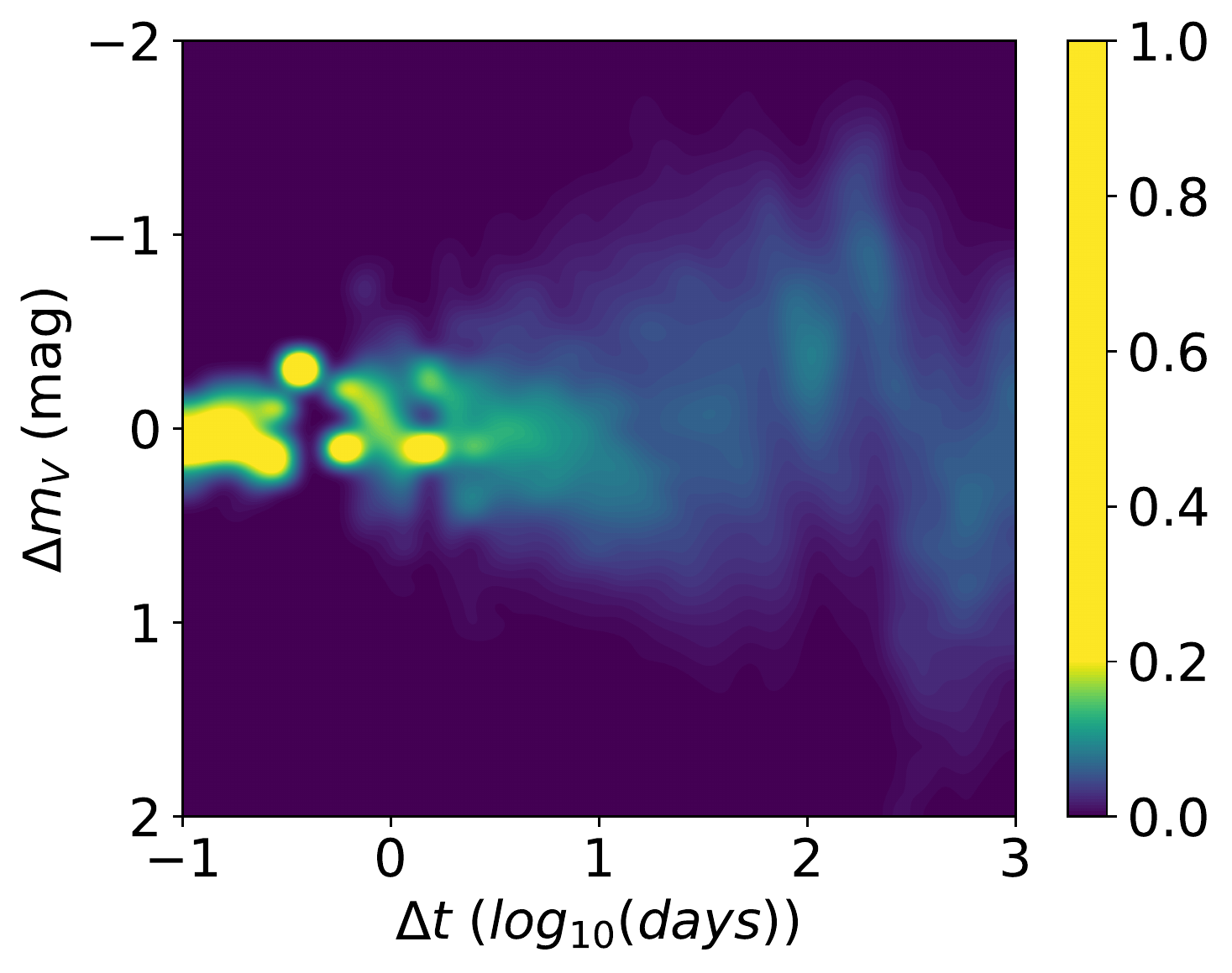} \hfill
\includegraphics[angle=0,width=0.5\columnwidth]{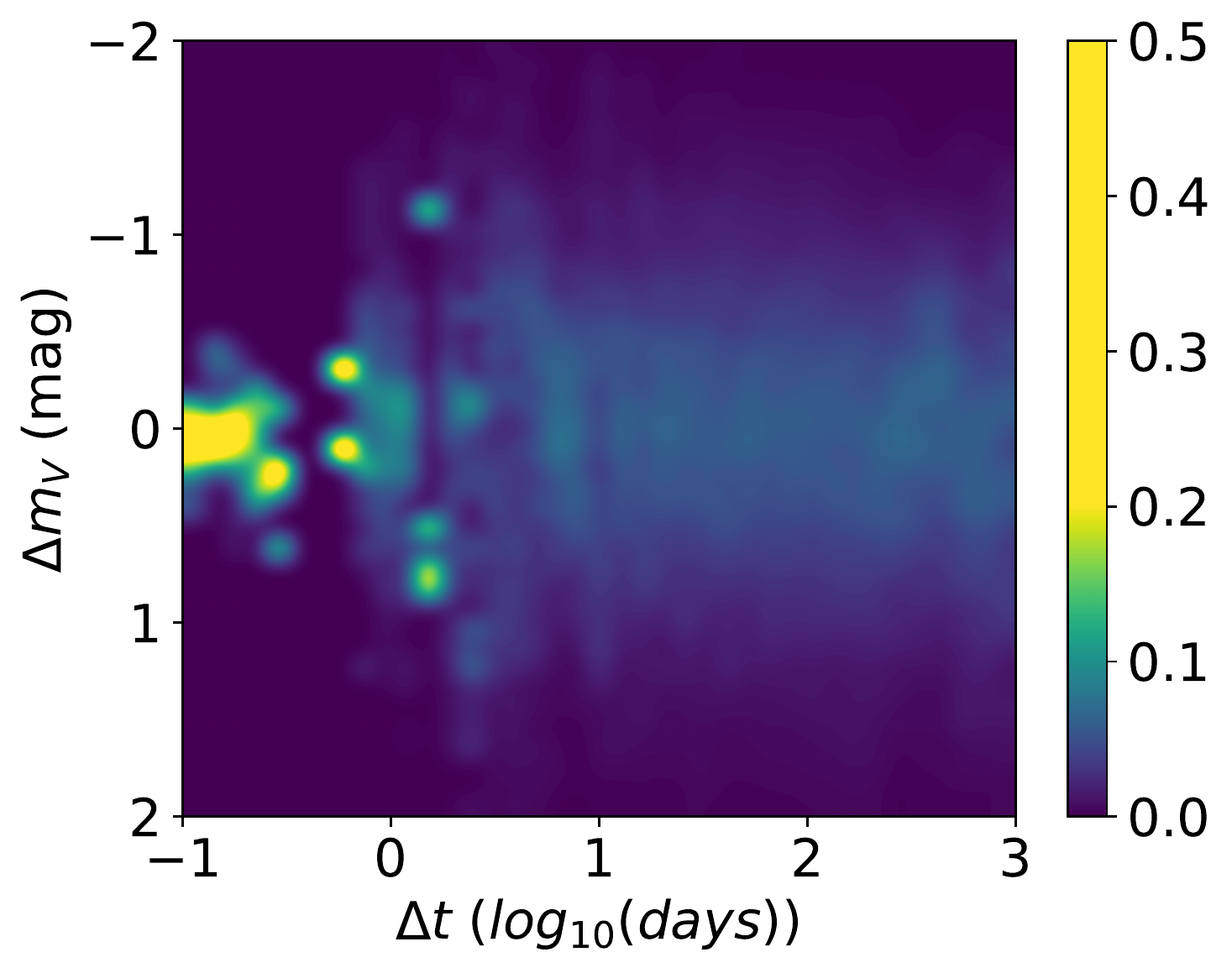} \hfill
\includegraphics[angle=0,width=0.5\columnwidth]{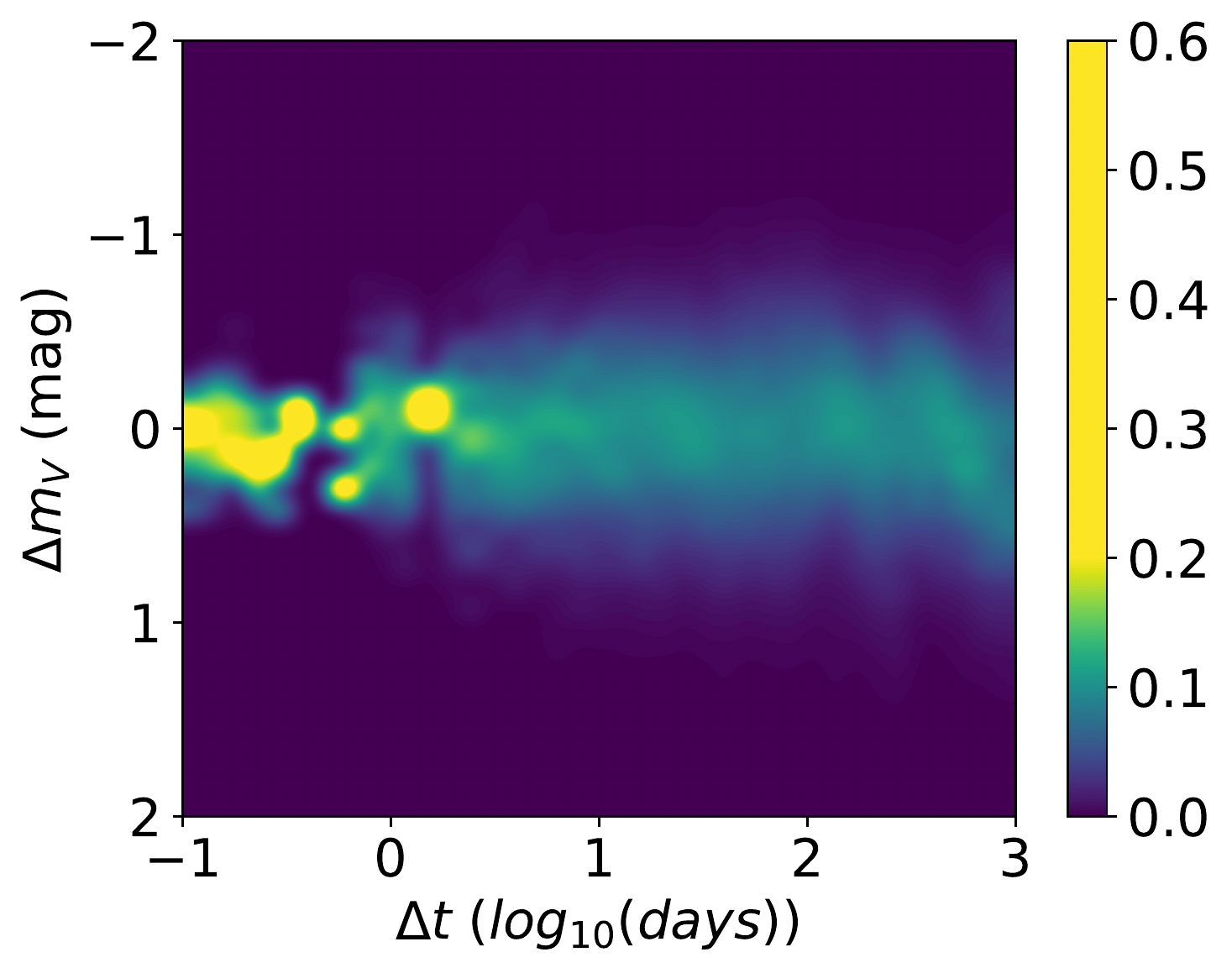} \hfill
\includegraphics[angle=0,width=0.5\columnwidth]{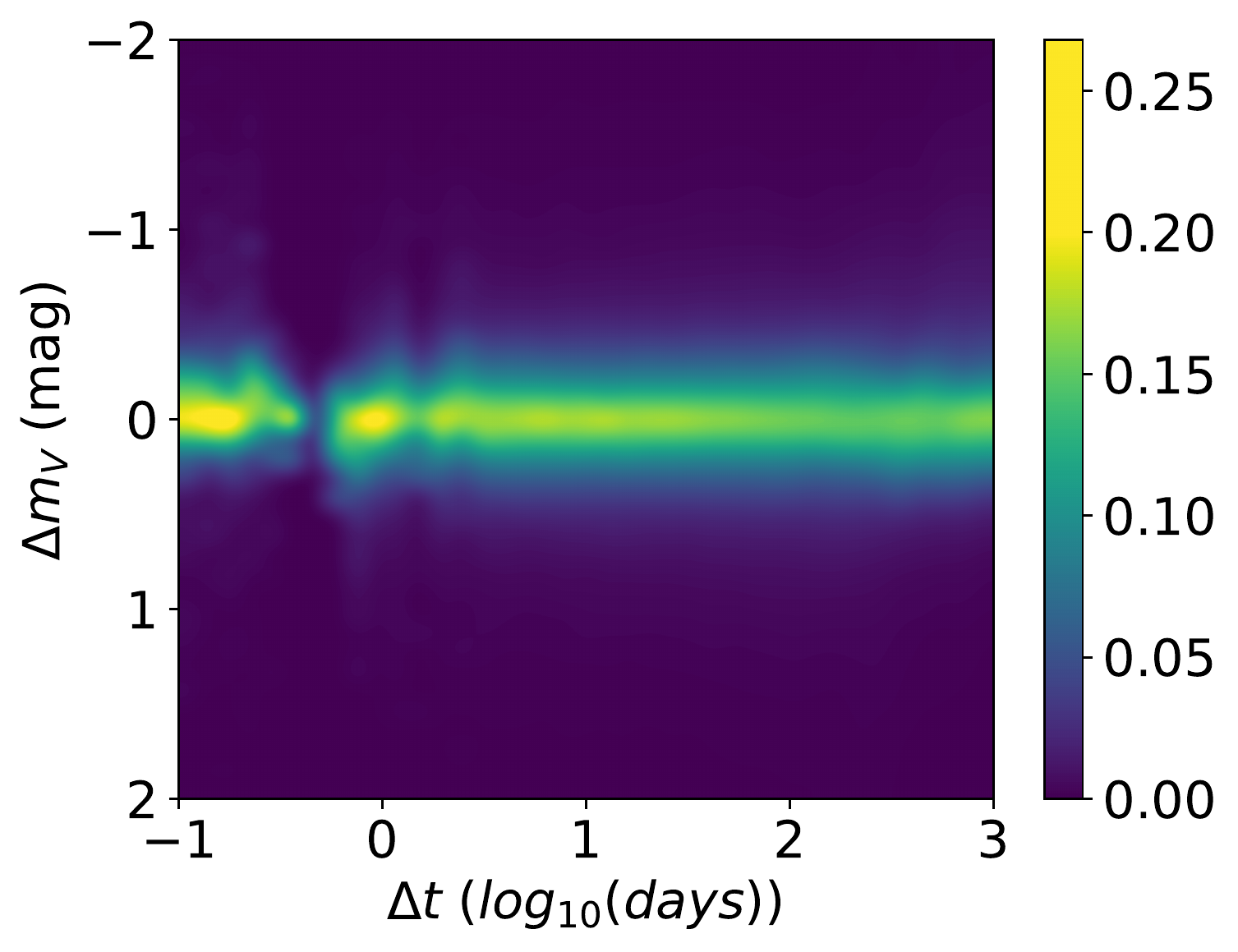} \\
\includegraphics[angle=0,width=0.5\columnwidth]{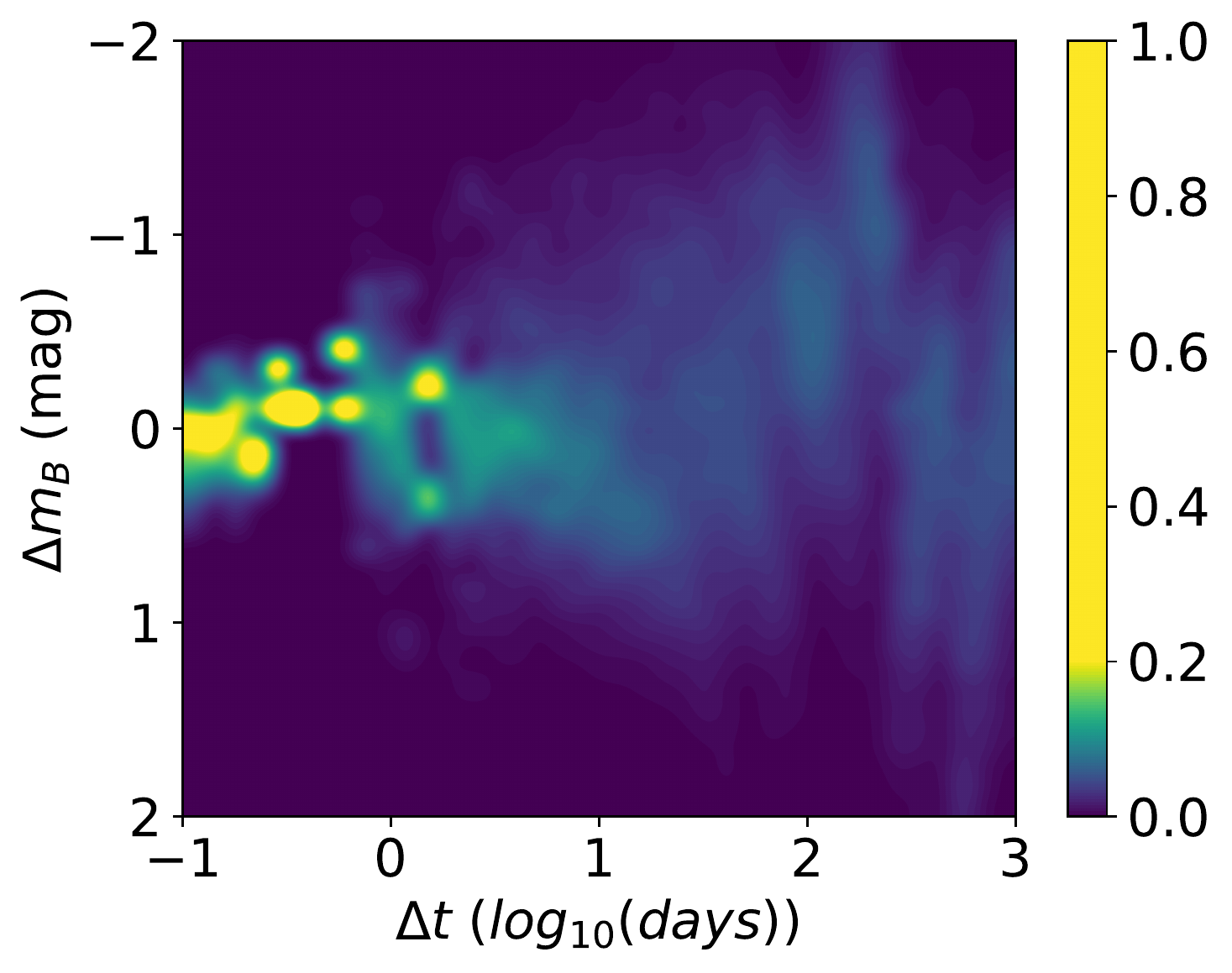} \hfill
\includegraphics[angle=0,width=0.5\columnwidth]{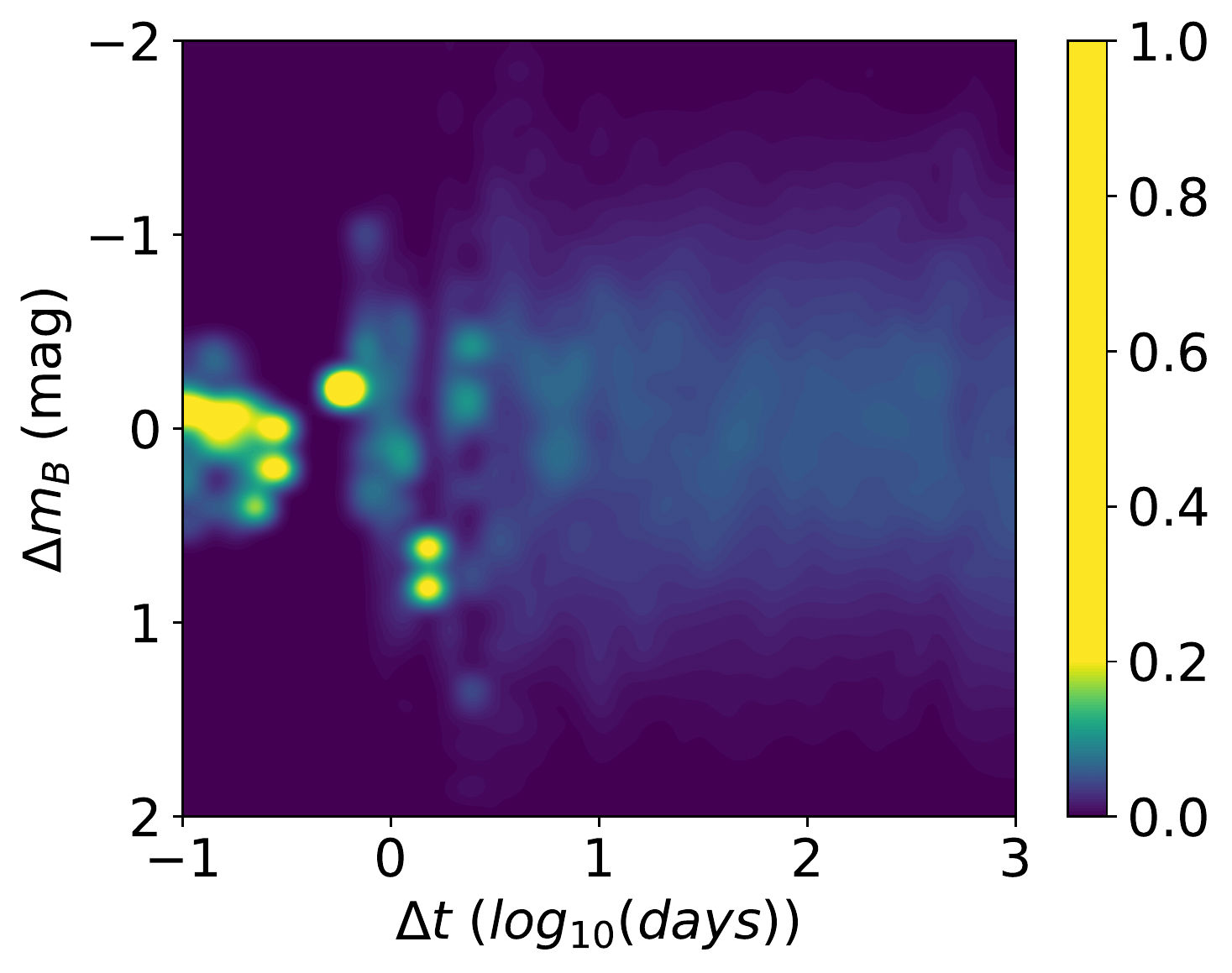} \hfill
\includegraphics[angle=0,width=0.5\columnwidth]{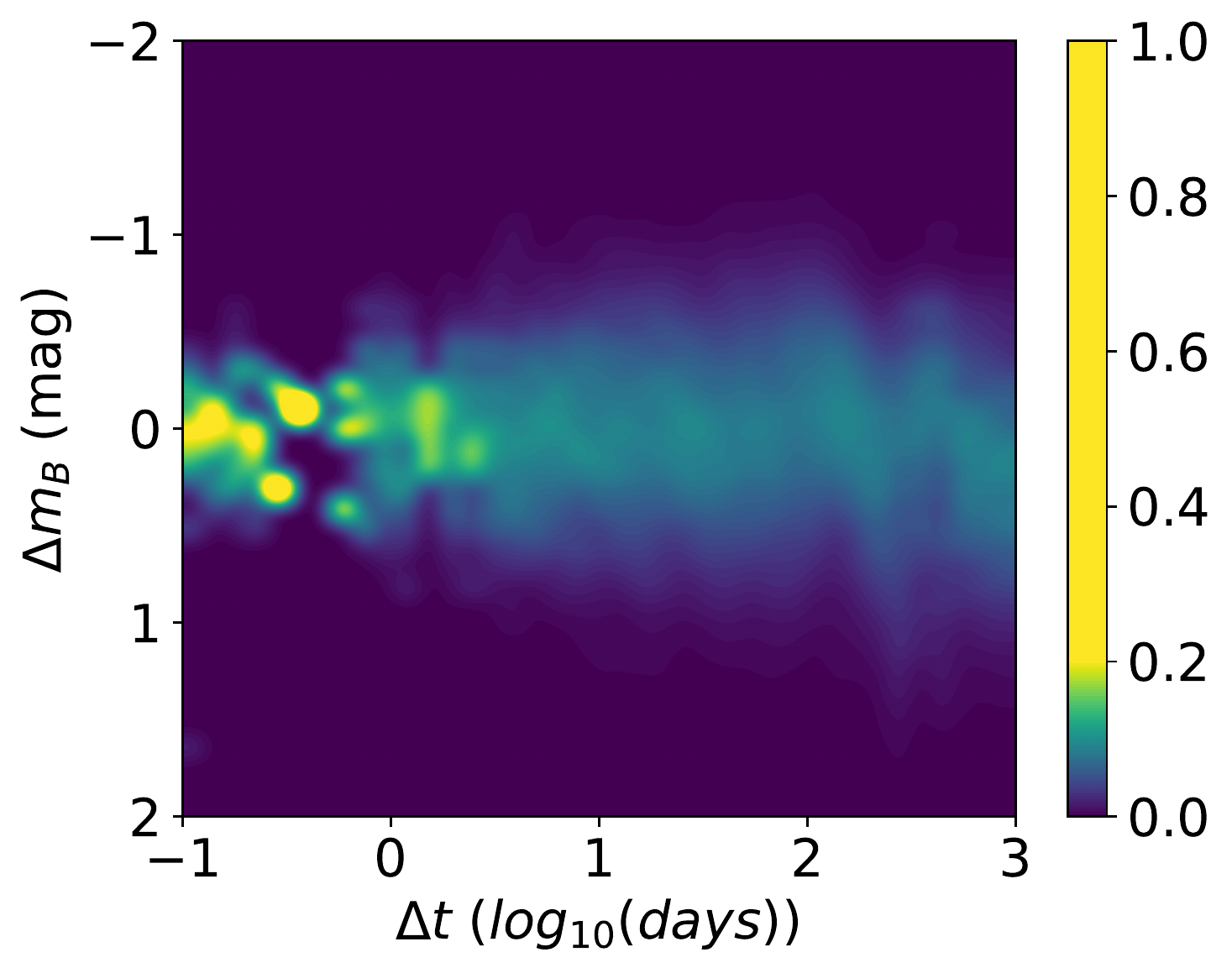} \hfill
\includegraphics[angle=0,width=0.5\columnwidth]{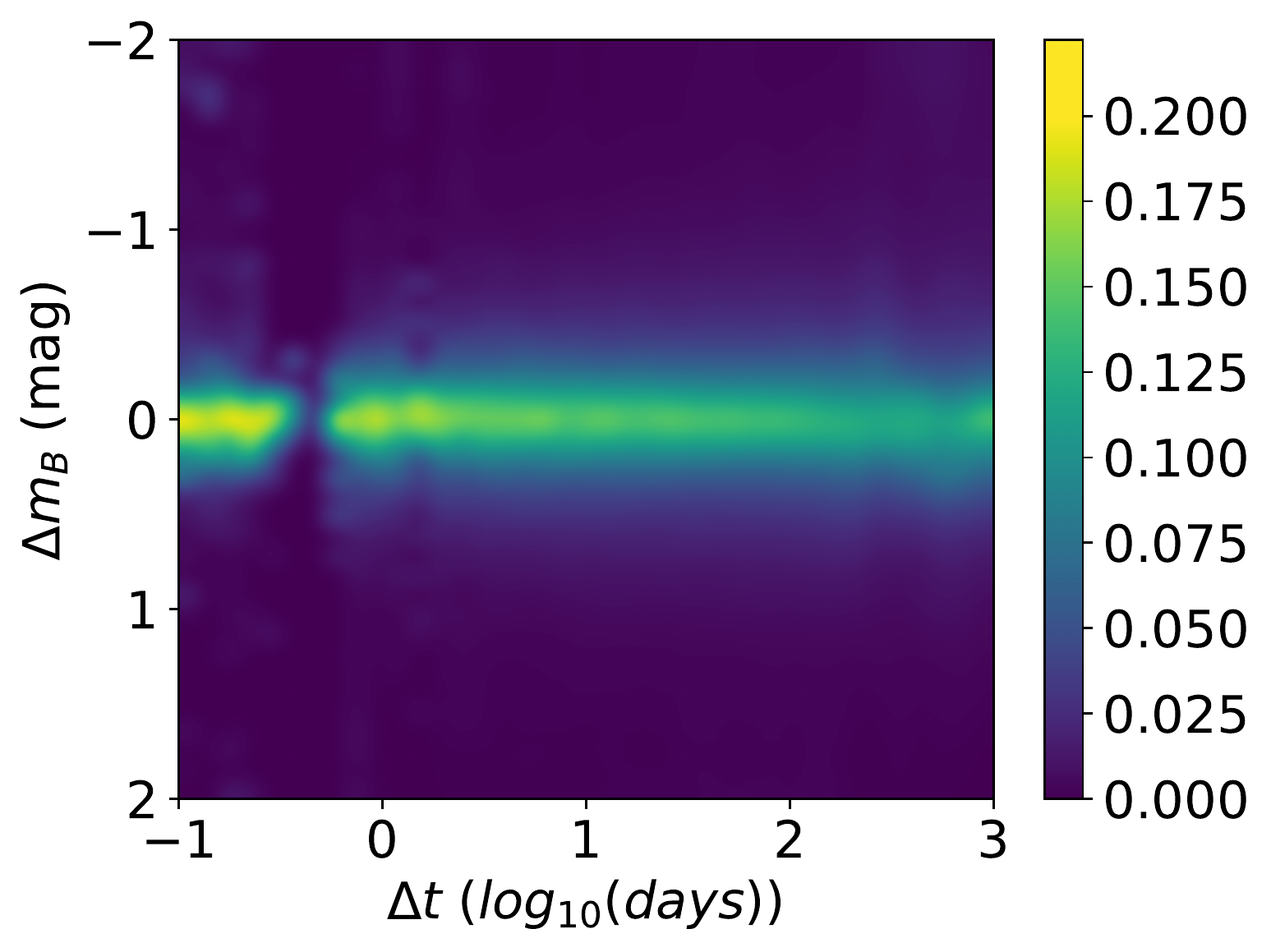} \\
\includegraphics[angle=0,width=0.5\columnwidth]{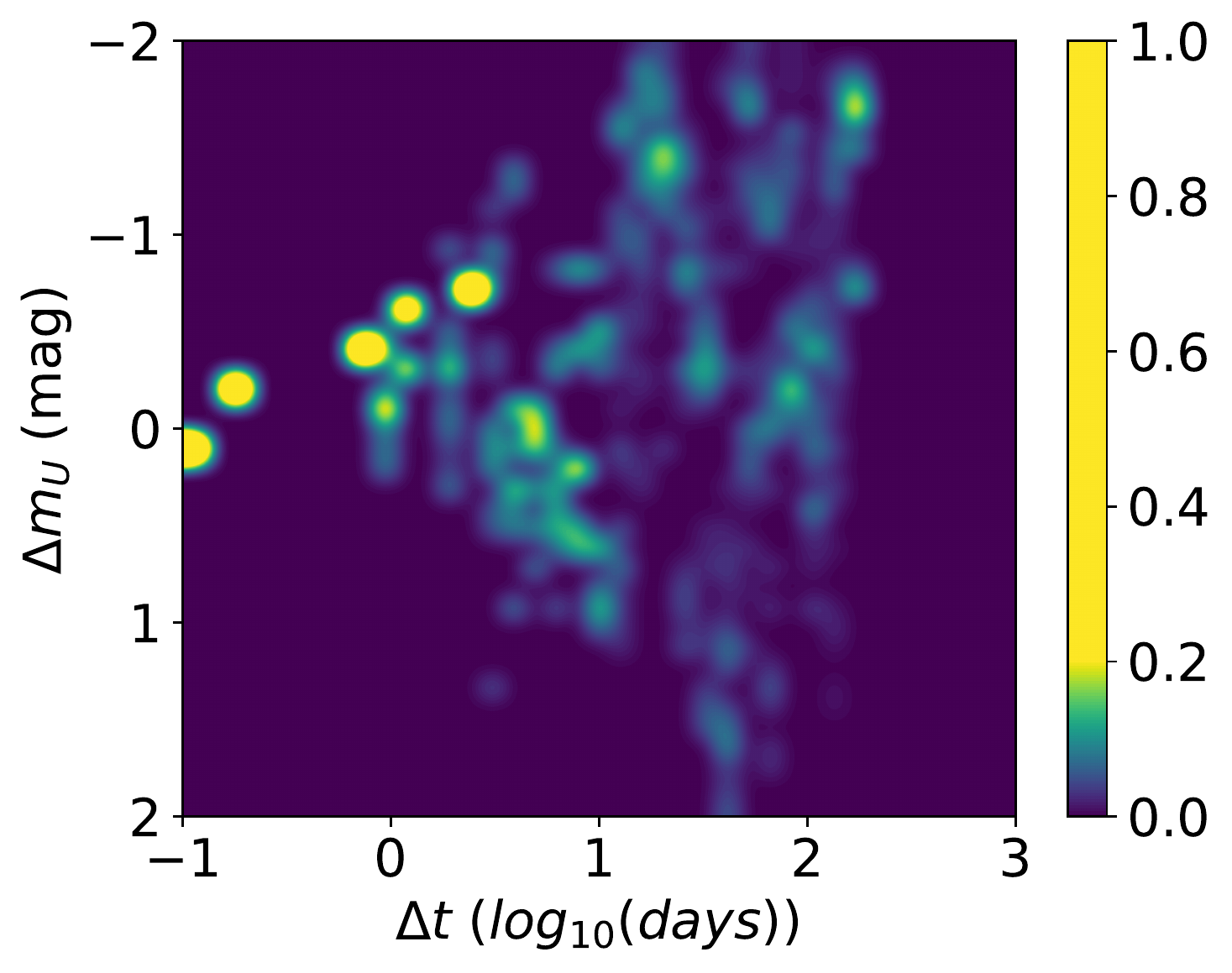} \hfill
\includegraphics[angle=0,width=0.5\columnwidth]{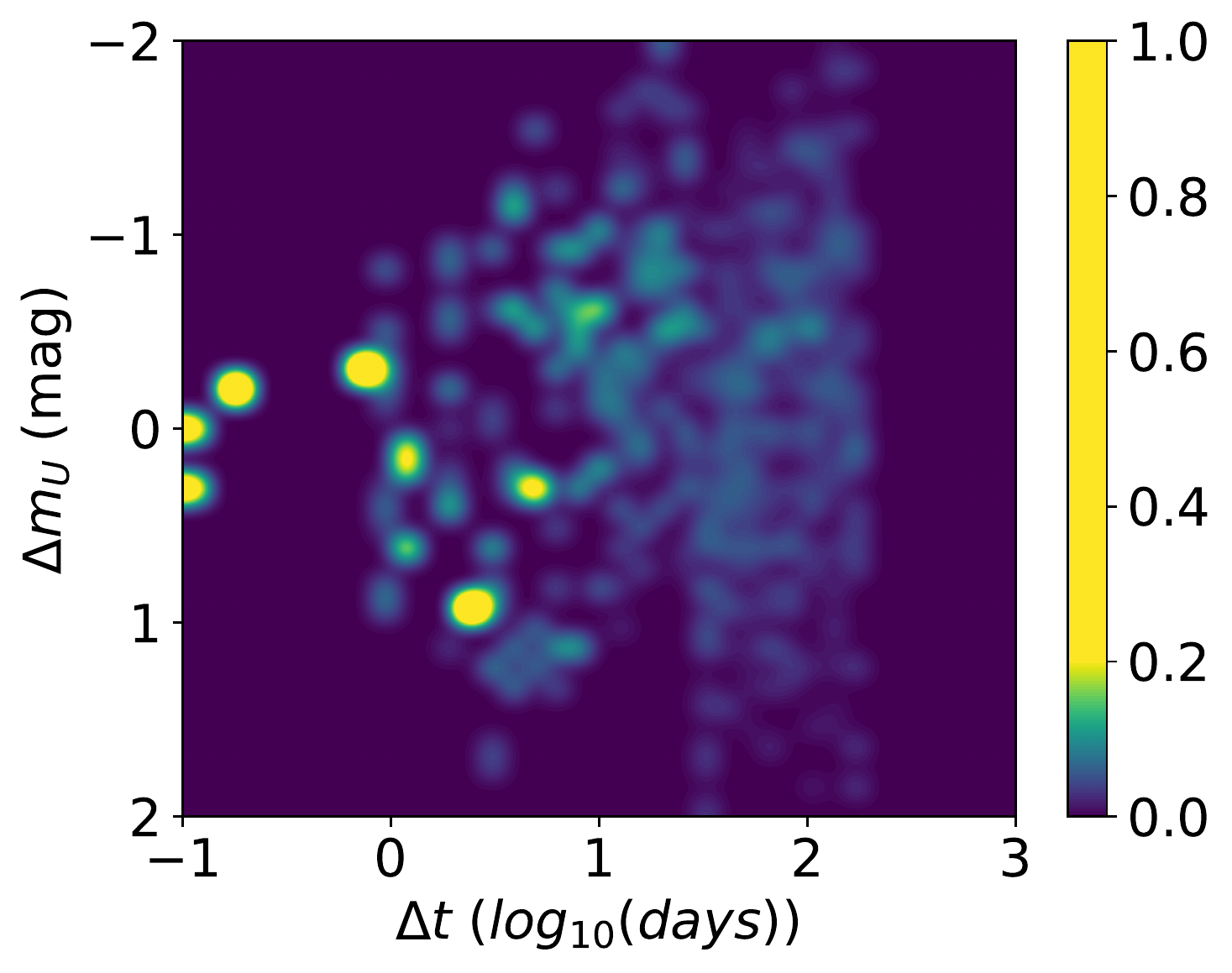} \hfill
\includegraphics[angle=0,width=0.5\columnwidth]{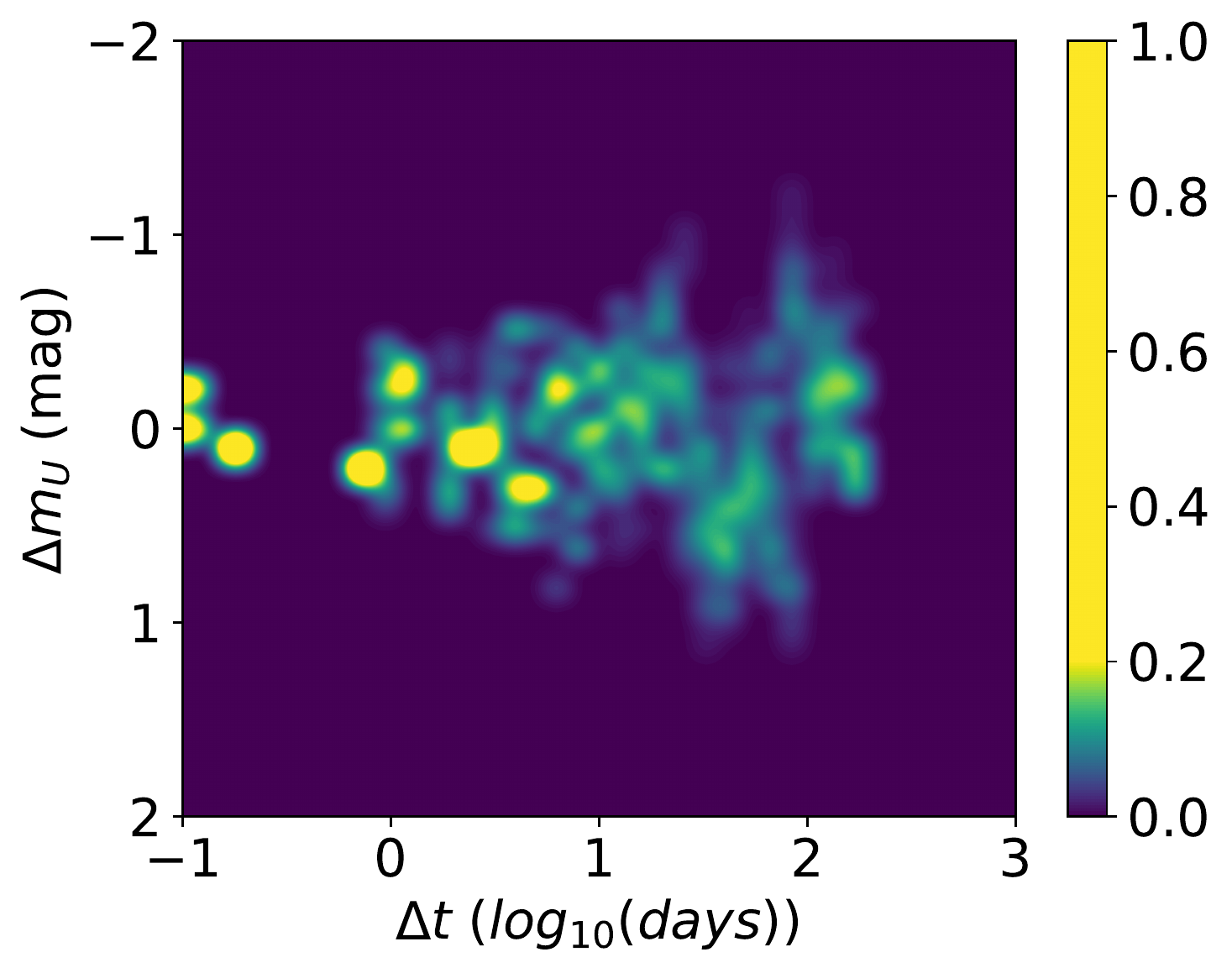} \hfill
\includegraphics[angle=0,width=0.5\columnwidth]{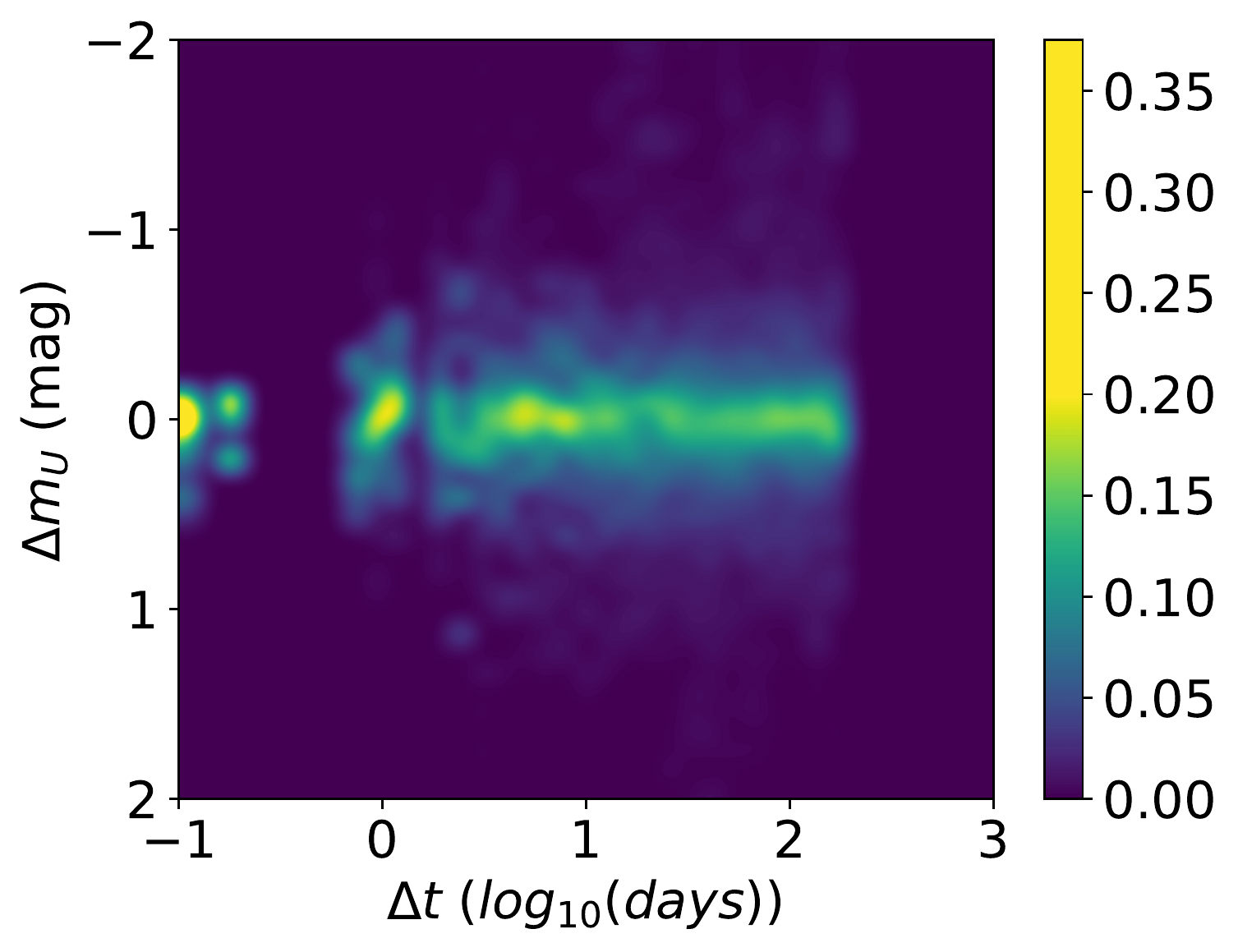} \\
\caption{Variability fingerprints for the three $\sigma$\,Ori ULLYSES targets based on all available HOYS data. {\bf 1st column:} TX\,Ori {\bf 2nd column:} V505\,Ori {\bf 3rd column:} V510\,Ori. In the {\bf 4th column}, we show the average variability fingerprints of the 140 other Gaia selected YSOs in the $\sigma$\,Ori field. From top to bottom the rows represent the plots for the I, R, V, B, and U data, respectively. Please see the text in Sect.\,\ref{fingerprints} for the details of the interpretation of the plots. \label{fig_fp}}
\end{figure*}

\subsection{Variability fingerprints}\label{fingerprints}

In this section, we investigate the variability of the ULLYSES targets over the entirety of the available HOYS data. Our data in principle allows us to characterise the variability from intra-night (hours) to multi-year time scales. In \citet{2020MNRAS.493..184E} we have developed variability fingerprints for this purpose. These are based on previous work by e.g. \citet{2004A&A...419..249S}, \citet{2015ApJ...798...89F} and \citet{2017MNRAS.465.3889R}. 

To generate those fingerprints we take all pairs of HOYS observations for a source in a particular filter and determine the time difference ($\Delta t$) and magnitude difference ($\Delta m$). We generate a two dimensional histogram in the $\Delta t$ vs. $\Delta m$ parameter space with a bin size of 0.1 for $\log_{10}(\Delta t[days])$ and 0.1\,mag for $\Delta m$. Each column, i.e. $\Delta m$ histogram, for a given value of $\Delta t$ is normalised to an integral of one. Thus, the fingerprint graphs show the probability that a given star varies by a given amount in magnitude when observed a time $\Delta t$ apart. When a star is observed over a significant duration, these diagrams hence represent the entire variability statistics on all time scales. 

We show the variability fingerprints of the ULLYSES targets in all broad band filters in the left three columns in Fig.\,\ref{fig_fp}. The colour scales in all panels are identical for ease of comparison. Probabilities above 0.2 are set to yellow. In the displayed plots we have applied a bi-cubic interpolation of the normalised two dimensional histograms. The investigated time differences range from 2.4\,hrs to 1000\,d and the magnitude changes cover $\pm$\,2\,mag. Note that there are a few rare occasions where the magnitude variations exceed this range. It is worth noting a few abnormalities in the plots. There are several columns, i.e. bins in $\Delta t$ which contain very few pairs of observations. These are in particular the bins with half integer day time differences, i.e. 0.5\,d, 1.5\,d, 2.5\,d - larger half integer time gaps are smoothed out by the bin size. This is due to the fact that most of our broad band data are from Chile, and if additional data from Europe is available in the same night, typically they are only taken 3\,hrs apart. To cover the half integer day time differences, a better longitudinal distribution of observatories is needed. Thus, one should not trust any features in the diagrams at these half integer day time gaps. 

In general, for all three ULLYSES targets the fingerprints show very similar structures in all broad band filters (except U). The typical amplitudes of the variations are increasing with decreasing wavelengths, in accordance with the expected behaviour for the physical causes of the general variations. Both, changing extinction and variable mass accretion rates cause higher amplitudes at shorter wavelengths. The U-Band panels do look very different from the others due to a much decreased number of data points in the filter - similar to the half integer day time gaps. While we typically have approximately 600 data points in the B, V, R, and I filters, there are only about 35 HOYS data points available in U. Thus, the interpretation of the U fingerprints needs to be done with caution, but they still allow to extract some basic variability information. We also note that we only have U data for the 2020/21 observing season, thus there is no data for time differences above 200\,d.

All fingerprints are dominated by the 2020/21 statistics. However, the very right hand side of the plots, for time differences above about 200\,d, represents the longer term, multi year behaviour of the sources. We note, that statistics of data for those larger time differences becomes worse, due to the low number of observations of the field in the first few years of our project. The very left side of the fingerprints represents the intra-night (less than about 6\,hrs) variability of the sources. Below we discuss some specific features in the variability fingerprints for each of the ULLYSES targets. The average variability fingerprints of all other 140 YSOs, which are shown in the right column of Fig.\,\ref{fig_fp} are discussed in Sect.\,\ref{context_var}.

\noindent {\bf TX\,Ori:} This object shows a steady increase in the range of variability with time scale. On intra-day and one day time scales the variations are low, in agreement with the photometric uncertainties. The variability range then increases to almost one magnitude in I and two magnitudes in B. Starting from timescales of about 100 days, there are some distinct features, which are most likely caused by the small amount of data in the early observing seasons. More data are needed to investigate these longer time scales. It is clear however, that we have not yet covered the full range of variability of the source in our data. The U data shows in principle a similar behaviour, but with much increased amplitudes and less statistical significance due to the small number of data points. 

\noindent {\bf V505\,Ori:} This source shows a rapid increase in the amount of variability with time scale. There is very little variability on intra-day  timescales, indeed it is consistent with the photometric uncertainties. The variability then rapidly increases at one day scales and reaches a maximum range at two to four days. The variability then clearly decreases towards the period of the source at seven days. Beyond that, the range of variability in the B, V, R, and I filters is consistent with the maximum at two to four days. This means, that the maximum peak to peak variations of the source can be explained by the periodic nature of the object. Hence, other causes of variability do not significantly contribute to the brightness changes during our observations. The U data, however, behaves very differently. The range of variability increases steadily with time scale. Thus, the U data seems dominated by stochastic accretion rate variations, and seems to be decoupled from the periodic variability.

\noindent {\bf V510\,Ori:} Of the three targets this is the least variable object in all filters. The variability increases steadily from the intra-day  scales to the duration of one observing season. There is an apparent decrease in variability at longer timescales, but this is caused by the paucity of data in the first few years of data. Thus, it seems evident that we have not yet covered the full range of typical variability of this source. There are indications in the plots of semi-periodic behaviour at time scales of about ten days and one month. These can be seen as increased probabilities of finding zero variation at those time scales. More data from future observing seasons is needed to verify these. 

\begin{figure*}
\centering
\includegraphics[angle=0,width=5.8cm]{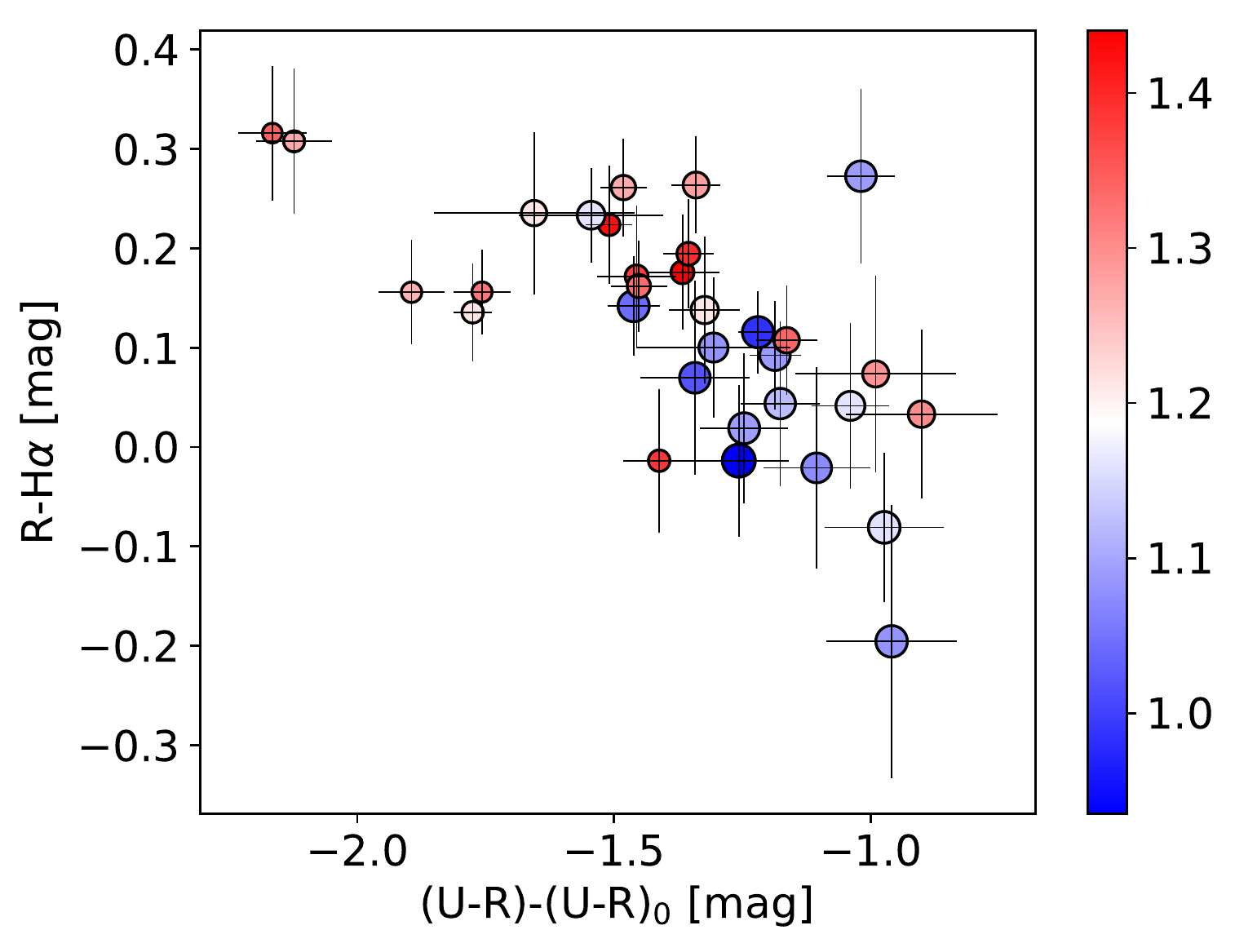} \hfill
\includegraphics[angle=0,width=5.8cm]{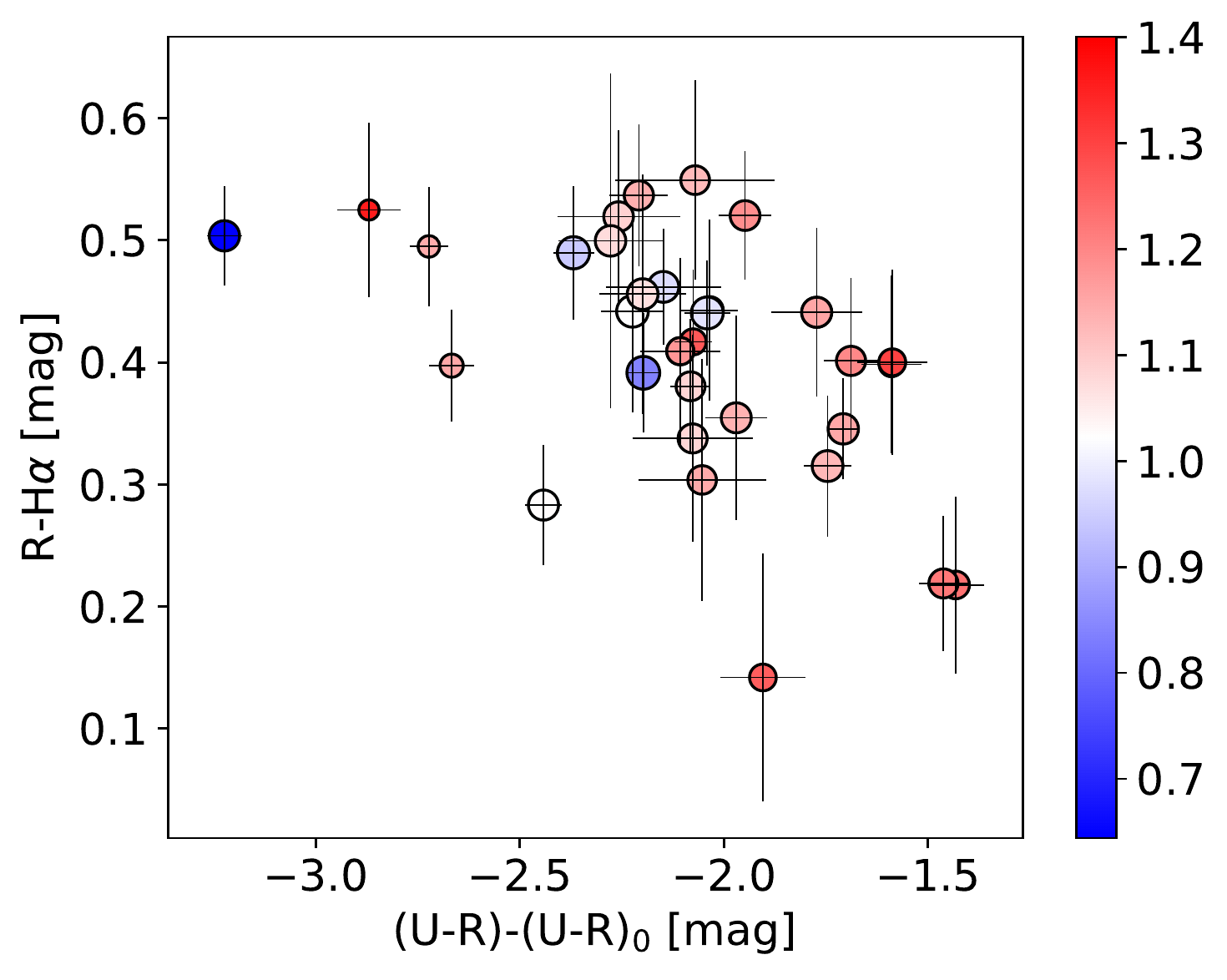} \hfill
\includegraphics[angle=0,width=5.8cm]{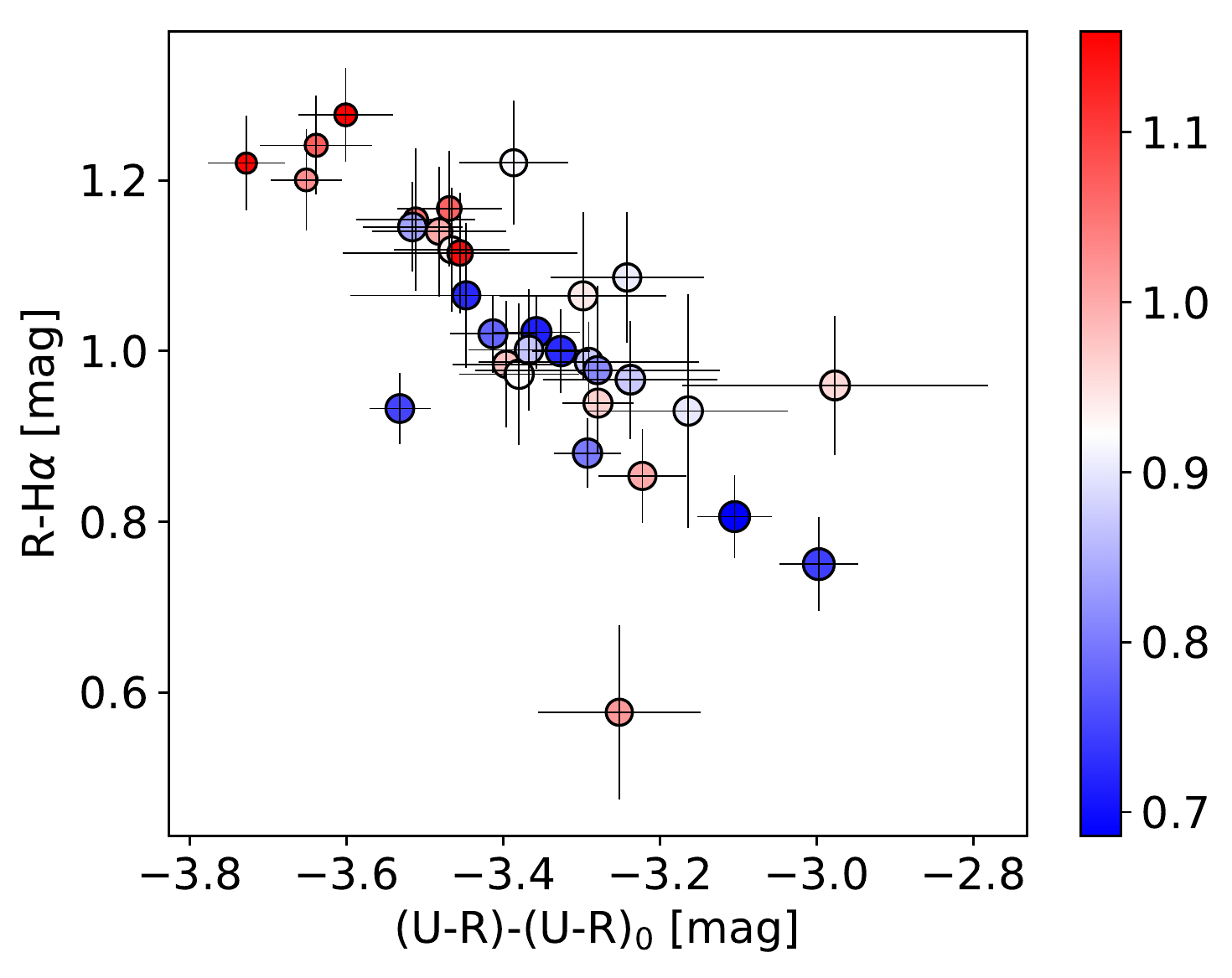} \\
\includegraphics[angle=0,width=5.8cm]{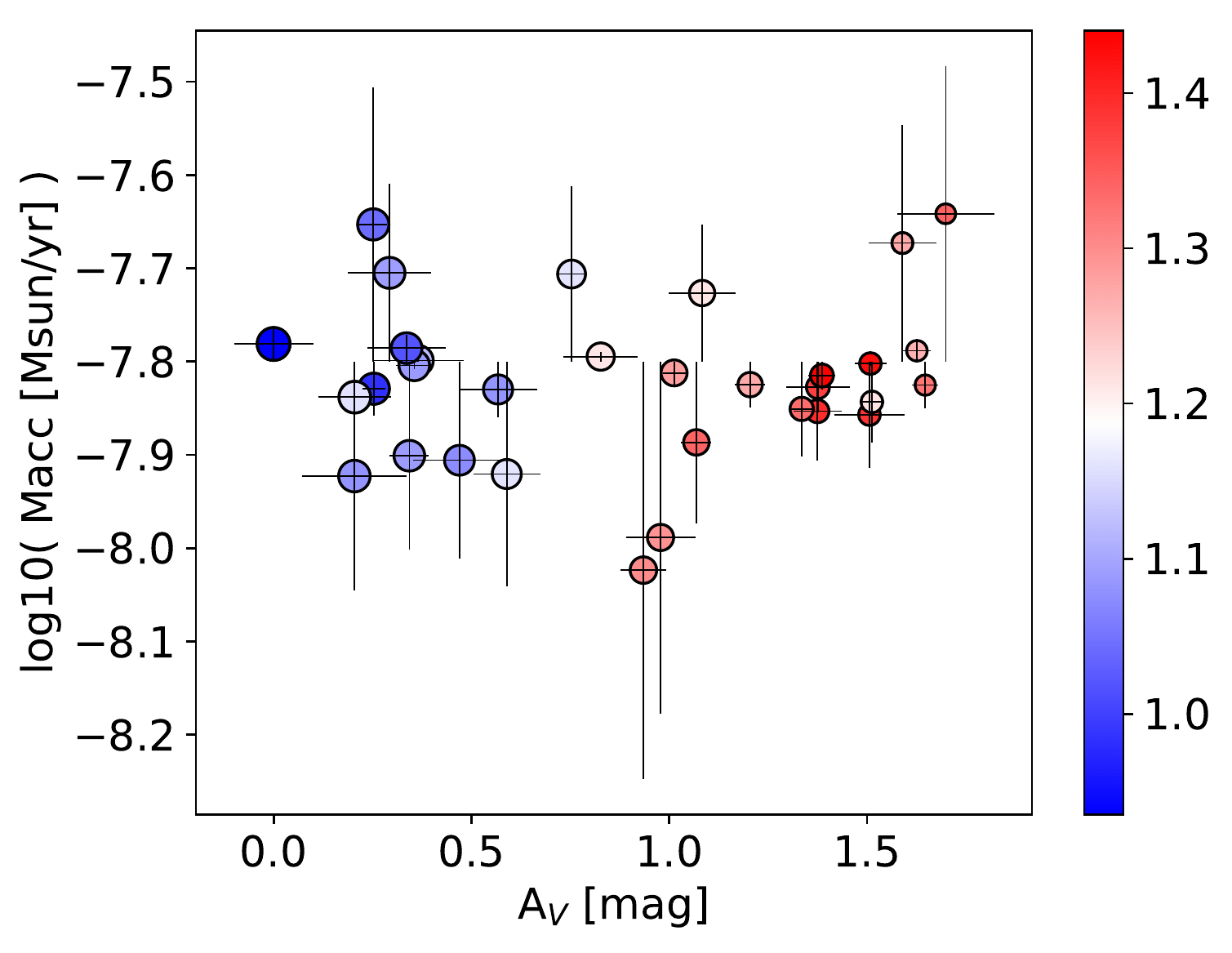} \hfill
\includegraphics[angle=0,width=5.8cm]{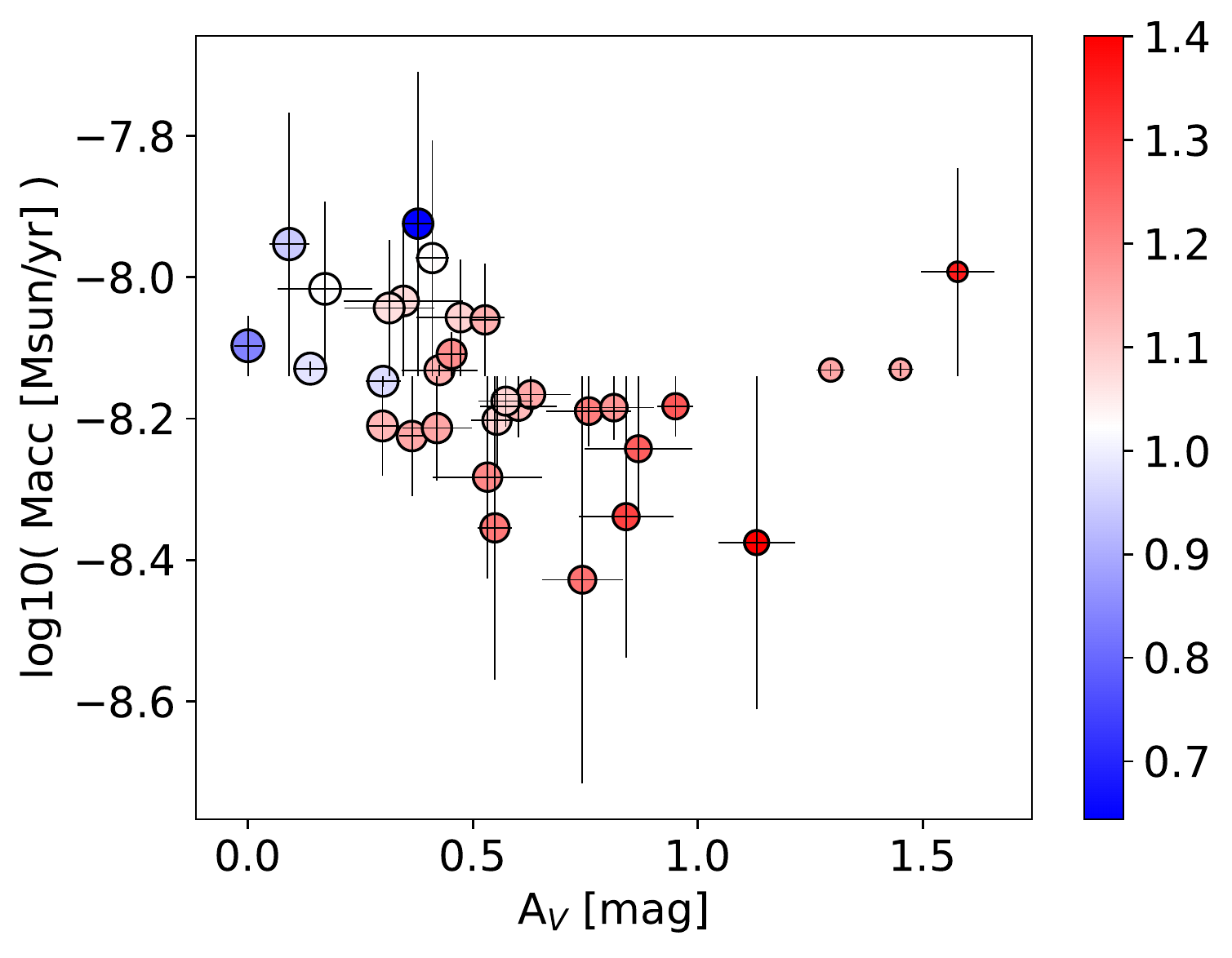} \hfill
\includegraphics[angle=0,width=5.8cm]{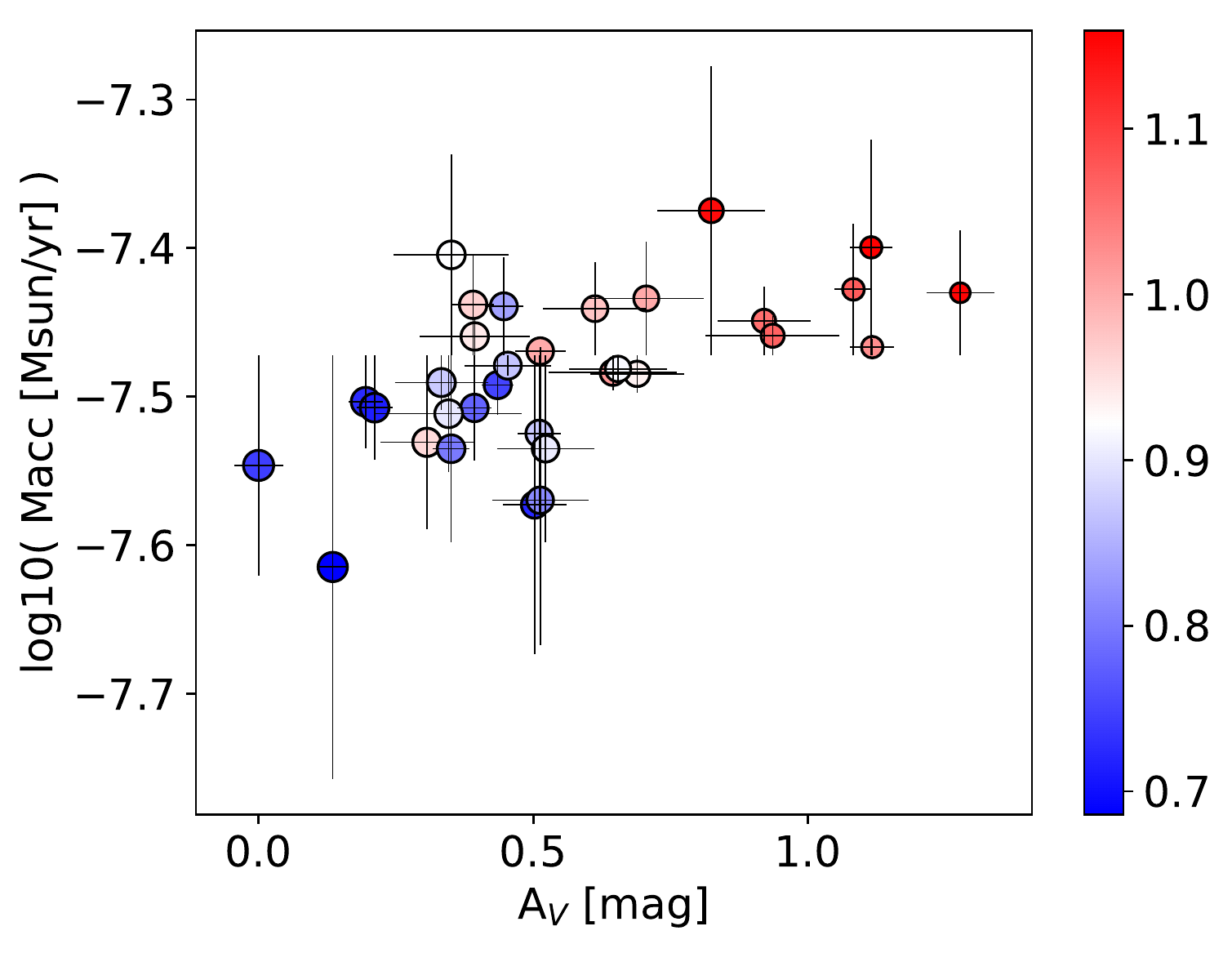} \\
\caption{ {\bf Top Row:} (U-R)-(U-R)$_0$ vs R-\ha\ colour-colour diagrams for the ULLYSES targets. The intrinsic colours are corrected for the extinction (see text in Sect.\,\ref{macc} for details). The colour scale indicates the V-I colours and the symbol size is proportional to the V brightness (small = faint). {\bf Bottom Row:} $A_V$ vs Mass accretion rates determined from U-band excess emission. The colour scale and symbol size are the same as in the top row. Each column represents one of the objects. {\bf Left:} TX\,Ori {\bf Middle:} V505\,Ori {\bf Right:} V510\,Ori. \label{colcol_hst}}
\end{figure*}

\subsection{Mass accretion rate indicators}\label{macc}

Accretion rate variations are common in YSOs \citep{2016ARA&A..54..135H}. Thus, we investigate the photometric accretion rate indicators for the $\sigma$\,Ori ULLYSES targets, such as the strength of the \ha\ emission line and the U-Band excess. The \ha\ line strength can be measured using the R-\ha\ colour. The value of this colour for non accreting stars is slightly dependent on the effective temperature (T$_{\rm eff}$) or the spectral type (SpT) of the star \citep[e.g.][]{2014MNRAS.440.2036D}. But generally larger values in R-\ha\ tend to indicate higher mass accretion rates. Another indicator for the mass accretion rate is the U-band excess and both typically correlate very well \citep{2015MNRAS.453.1026K}. We obtain this excess by investigating the U-R colour in the HOYS data sets for the three ULLYSES targets. This colour was chosen over the other possibilities, as this is the least variable (with T$_{\rm eff}$ or SpT) intrinsic colour for non accreting objects \citep[see e.g. Fig.\,4 in][]{2011A&A...525A..47R}. For the effective temperatures of the stars investigated here (see Table\,\ref{table:1}), the intrinsic colour (U-R)$_0$ should be about 3.5\,mag \citep{2011A&A...525A..47R}.

In the top row of Fig.\,\ref{colcol_hst} we show the U-band excess vs R-\ha\ colour diagrams for the three ULLYSES targets. The U-band excess is determined as the difference between the observed U-R colour and the intrinsic colour (U-R)$_0$. Thus, the values of (U-R)-(U-R)$_0$ directly indicate the U-Band excess. The colours are determined, as for the CMDs in Fig.\,\ref{cmd_hst}, for data points taken less than five hours apart. The intrinsic colour (U-R)$_0$ for each data point is corrected by the extinction for each data point. We estimate the extinction from the V magnitude of the source at the time of the U and R data and determine the (U-R)$_0$ correction using the \citet{1990ARA&A..28...37M} extinction law for R$_{\rm V}$\,=\,5.0. We use the brightest V magnitude of the source as the extinction free state in all cases and the difference to the V magnitude at the time of the U and R data are the extinction. Note that this assumes that all V-band variations are due to extinction changes and that we use the correct extinction law. We have already seen that the latter is not the case (see Sect.\,\ref{colmags}). But as we use the V-band to correct the (U-R)$_0$ colour, the systematic errors due to an inaccurate extinction law are minimised. The assumption that the V-band variations are caused by extinction changes is statistically correct \citep{2017MNRAS.472.2990L,2018MNRAS.478.5091F} and supported by the CMDs in Fig.\,\ref{cmd_hst}. In particular for V505\,Ori, we have shown that this is the most likely scenario. Furthermore, as evident in the CMDs the extinction values are generally less than one magnitude and thus the potential errors for the intrinsic U-R colour are much smaller than the U-band excess.

We show the photometry errors for the colours in Fig.\,\ref{colcol_hst}. They are only based on the uncertainties of the individual magnitudes and do not contain the potential systematic off-sets discussed above. The symbol size in the plots indicates the V-band magnitude and the colour scale shows the V-I colour of the objects. The plots, when viewed together, show a general trend that an increased U-band excess is correlated with an increase in R-\ha. This can also be seen more clearly in our discussion of the same plot for all $\sigma$\,Ori sources in Sect.\,\ref{results_context}. Furthermore, larger values of V-I colour, which indicate increased extinction, do not generally correlate with the U-band excess or R-\ha. 

We follow the methodology from \citet{2011A&A...525A..47R} to determine the mass accretion rates from the U-band excess emission. We use the individual GaiaEDR3 parallaxes for the excess flux-to-luminosity conversion, the flux zero point in U of 417.5\,$\times$\,10$^{-11}$\,erg\,cm$^{-2}$\,s$^{-1}$\,${\rm \mathring{A}}^{-1}$ \citep{1998A&A...333..231B}, $\Delta \lambda$\,=\,0.06\,$\mu$m for the U filter, and two solar radii for the stellar radius. Using the median U band and R band magnitudes, we find median mass accretion rates of 1.5, 0.7, and 3.3\,$\times$\,10$^{-8}$\,M$_\odot$/yr for TX\,Ori, V505\,Ori and V510\,Ori, respectively. The $\log_{10}$(M$_{\rm acc}$[M$_\odot$/yr]) values are $-7.8$, $-8.2$, and $-7.5$. Compared to Table\,\ref{table:1}, these are slightly higher than in the literature. These differences can be caused by a number of things, e.g. the assumptions for the stellar radii and also source variability. We note that e.g. the \citet{2021A&A...650A.196M} accretion rates for V505\,Ori are determined from spectroscopy during the faint state and their extinction values are quite small. Furthermore, our assumption that the brightest HOYS magnitude is the extinction free state of the source could be wrong, i.e. the extinction could be higher.

In the bottom row of Fig.\,\ref{colcol_hst} we show the determined mass accretion rates for all individual data points against the extinction estimated from the V data. For TX\,Ori there is no correlation of the mass accretion rate with the extinction. For the other two objects there is a weak trend, but in different directions. These trends can be caused by a number of our assumptions. The extinction used to de-redden the U and R magnitudes could be too low,  we might not have observed the extinction free state of the source in HOYS. We already discussed that the extinction law used is not the correct one, and that there could be contributions from scattered light. Finally, the brightness variations could be caused by accretion rate changes instead of variable extinction from disk material. In reality all these contributions will be at play for each source to a different extent. Hence, only a detailed modelling of each source (considering interstellar as well as grey extinction, scattered light and the structure of the inner disk and viewing geometry) are needed to explain all the observational data. Note that the mass accretion rate error bars are only based on the statistical photometric uncertainties and do not include any systematic uncertainties caused by our assumptions to convert the U-band excess into accretion rates. Below we discuss some additional details in the plots for each source.

\noindent {\bf TX\,Ori:} While the range of U-R is about one magnitude, this is the source with the least amount of U-Band excess. There are two states of the source, one with the extinction below 0.5\,mag, the other one with the extinction above 1.0\,mag. But these do not correlate with the U band excess or R-\ha\ colour, indicating that indeed the variability is caused by extinction changes. Contrary to the literature mass accretion rate values listed in Table\,\ref{table:1}, TX\,Ori is not the strongest accretor of the ULLYSES targets.

\noindent {\bf V505\,Ori:} The total range of U-band excess values for this object is about 1.5 to 3.0\,mag. The median U-band excess is just above 2\,mag. There is a group of four data points with stand-out high U-excess. They do not correlate with the source brightness, R-\ha\ or V-I colour. These could indicate photometric outliers or short duration accretion rate increases or flare like events. However, our data does not have the time resolution to reliably identify flares. 

\noindent {\bf V510\,Ori:} This is the object with the highest U-band excess (median about 3.3\,mag) and the largest R-\ha\ colour (1.0\,mag). It also shows the least amount of scatter/variability of the three objects in the U-band excess, R-\ha\ and mass accretion rates. It is worth noting that the accretion rates for TX\,Ori and V510\,Ori are about the same in our measurements but the \ha\ strength is very different, with almost going to absorption in TX\,Ori. 

\section{Comparison to other \texorpdfstring{$\sigma$\,Ori YSOs}{}}\label{results_context}

\begin{figure}
\centering
\includegraphics[angle=0,width=\columnwidth]{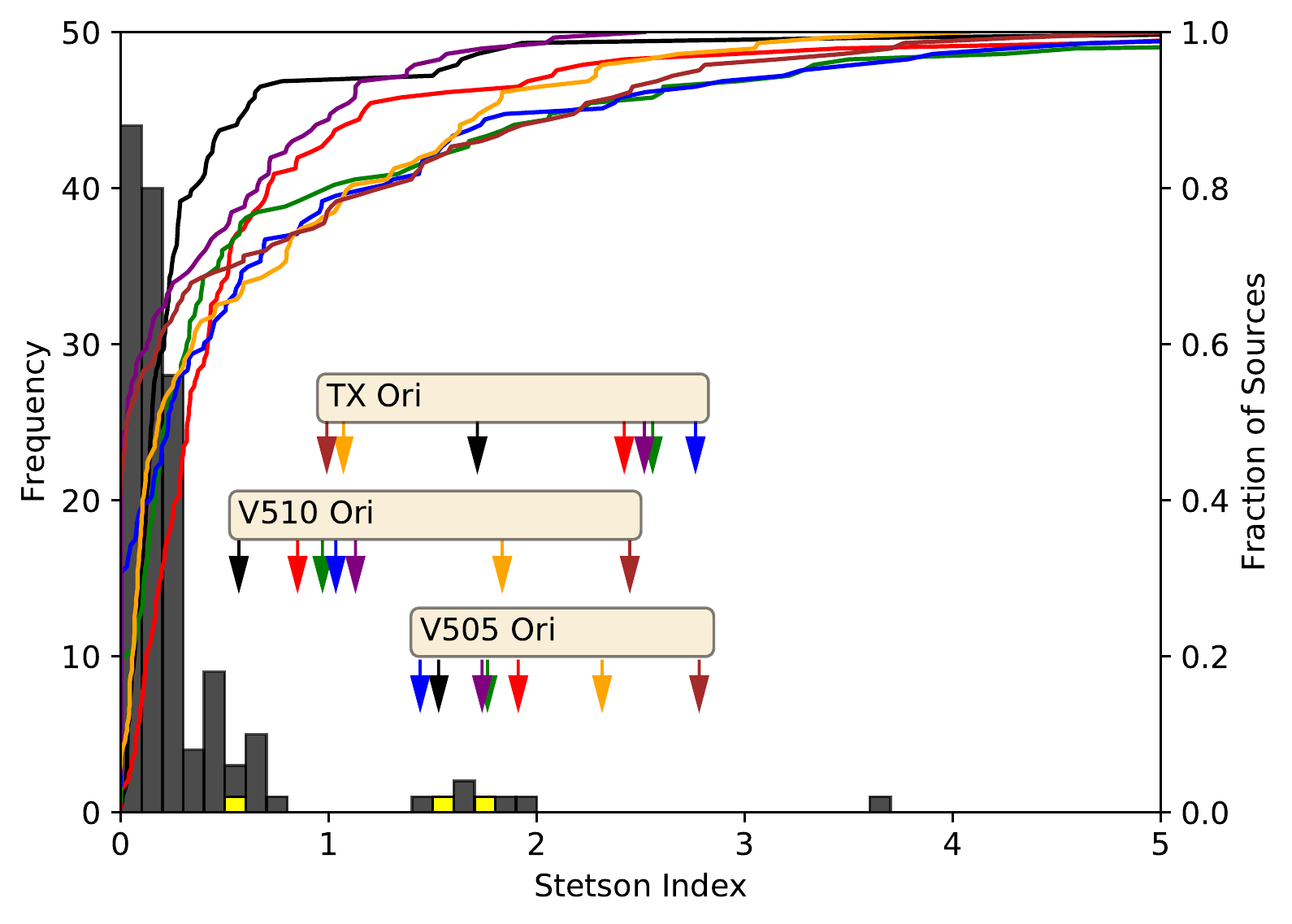} 
\caption{Histogram of the I-Band Stetson index distribution of the $\sigma$\,Ori YSOs (dark-grey) with the three $\sigma$\,Ori ULLYSES targets indicated (yellow). As solid coloured lines we also show the CDFs of all YSOs in W1 (orange), W2 (brown), I-band (black), R-band (red), V-band (green), B-band (blue), U-band (purple). For each of the ULLYSES targets the coloured arrows indicate the Stetson index in the respective filters. The W1/W2 indices are divided by three and the U-band indices by two in the plot for better visibility. \label{HST_context_var}}
\end{figure}

\begin{table*}
\caption{\label{table_varall} Stetson indices (I) for the three ULLYSES targets in all broad band filters, as well as the fraction (F$^{<}$, in percent) of $\sigma$\,Ori YSOs that are less variable than the ULLYSES target. Note that not all sources are detected in all filters. Thus, the fractions are based only the detected sources in each filter.}
\centering
\begin{tabular}{|l|ccccccc|ccccccc|}
\hline
Object & I$_{\rm U}$ & I$_{\rm B}$ & I$_{\rm V}$ & I$_{\rm R}$ & I$_{\rm I}$ & I$_{\rm W1}$ & I$_{\rm W2}$ & F$^{<}_{\rm U}$ & F$^{<}_{\rm B}$ & F$^{<}_{\rm V}$ & F$^{<}_{\rm R}$ & F$^{<}_{\rm I}$ & F$^{<}_{\rm W1}$ & F$^{<}_{\rm W2}$ \\ \hline
TX\,Ori   & 5.03 & 2.76 & 2.56 & 2.42 & 1.71 & 3.21 & 2.97 & 99 & 91 & 90 & 96 & 97 & 78 & 76 \\
V505\,Ori & 3.47 & 1.44 & 1.76 & 1.91 & 1.53 & 6.94 & 8.34 & 96 & 78 & 84 & 92 & 94 & 95 & 94 \\
V510\,Ori & 2.26 & 1.03 & 0.97 & 0.85 & 0.57 & 5.50 & 7.34 & 87 & 73 & 76 & 83 & 88 & 91 & 91 \\
\hline
\end{tabular}
\end{table*}

After characterising the three $\sigma$\,Ori ULLYSES targets in detail, we aim to place their properties into context. This means we compare their photometric properties with the sample of the 140 other YSOs selected in Sect.\,\ref{hoys_selection}. We have hence repeated the analysis done for the three ULLYSES targets for all other sources in the sample. As the purpose of this paper is to characterise the ULLYSES targets within their cluster, we will compare their properties to the other sources but do not discuss any other object in detail. This is beyond the scope of this paper.

\subsection{Object Variability}\label{context_var}

We first investigate how the ULLYSES targets compare to the other YSOs in the field in terms of their variability. One possible way to characterise generic variability is the Stetson index \citep{1996PASP..108..851S}. We determine the Stetson index for all sources in all filters where they have been detected at least five times. This includes the optical broad band filters (U, B, V, R, I) as well as the two (NEO)WISE bands (W1, W2). We show the results of this in Fig.\,\ref{HST_context_var}. This shows, as one example, the histogram of the I-band Stetson index of all objects in dark grey, and of the three ULLYSES targets in yellow. Additionally the plot also shows, as colour coded solid lines the cumulative distribution functions (CDFs) for the Stetson indices in all filters. Due to the smaller uncertainties in the WISE bands and the observing cadence, the Stetson indices in those bands are generally much larger. For better visibility we have hence divided the W1 and W2 indices by three and in U by two in the plot. For each of the three ULLYSES targets the colour coded arrows are placed at the value of the respective Stetson index for this particular filter. 

Interestingly, as can be seen from Fig.\,\ref{HST_context_var}, the optical and IR Stetson indices indicate a different behaviour for the ULLYSES targets. While in the optical the variability of TX\,Ori is most pronounced, it is the least variable object among the three targets in the IR. Remarkably, the behaviour is opposite for V510\,Ori and V505\,Ori. Both sources are more variable in the IR than in the optical. In case of TX\,Ori, a major part of the variability might be induced by temporal extinction variations which, because of the wavelength dependence of the dust extinction, is weaker at longer wavelengths. The converse finding for V510\,Ori and V505\,Ori seems to point at radial irregularities at the inner disk rim which cause a modulation of the IR flux but do not scatter efficiently.

In Table\,\ref{table_varall} we list the individual Stetson indices for the three ULLYSES targets in all the broad band filters. As the indices are not easily comparable between the different filters due to the change in photometry errors, we compare the three sources against all other sources in each filter. We determine the fraction of objects (in percent) that is less variable than the ULLYSES  target in all filters. These numbers are also listed in Table\,\ref{table_varall}. 

We have also determined the average variability fingerprint for all the $\sigma$\,Ori YSOs in all filters. These are shown in the right panel of Fig.\,\ref{fig_fp} with the same colour scaling as all other plots. We only include objects in these average plots that have at least 200 data points per filter, with the exception of the U-band where we require at least 30 data points per light curve. In those plots we see that the average YSO is less variable than the three HST targets. The variability increases with decreasing wavelength. One can also see that the variability increases with the time scale for each filter. This is less obvious than for the individual sources, but is evident from the decreasing maximum probability at zero variation for longer time scales.
 
Thus, the most obvious result from the plots in Figs.\,\ref{HST_context_var} and \ref{fig_fp} and the numbers in Table\,\ref{table_varall} is that the three ULLYSES targets are amongst the most variable young objects in the sample. With very few exceptions, in excess of 80\,\% of the YSOs in the field are less variable than the ULLYSES targets. Thus, the three sources are amongst the most variable YSOs in the $\sigma$\,Ori cluster. Below we discuss some specifics of the individual sources.

\noindent {\bf TX\,Ori:} This is the most variable of the three objects in all optical broad band filters. This is also reflected in the CMDs in Fig.\,\ref{cmd_hst}, which indicate scattered light dominating during the faintest states. Indeed there is only one single object in the field that is more variable in the U-band. However, in the (NEO)WISE filters the source is the least variable of the three ULLYSES targets, and about one quarter of all YSOs are more variable. This could be an indication that most of the variations of the source are caused by changing extinction by disk material.

\noindent {\bf V505\,Ori:} In the optical filters this is the second most variable of the three targets. Similar to TX\,Ori it is in particular variable in the U-band. The source is also the most variable of the three in the WISE bands, with only very few other sources more variable in those filters. Together with the periodicity of the source which cannot be explained by a surface spot, this indicates that changes at the inner disk rim, which is potentially warped, are causing the variations.

\noindent {\bf V510\,Ori:} As already evident from the CMDs in Fig.\,\ref{cmd_hst}, this is the least variable source in all the optical filters. While the object has the highest mass accretion rate of the three, the U-band variability is the weakest. In the (NEO)WISE filters the variability of the source is in-between the other two objects. With the exception of the periodicity, the object behaves similar to V505\,Ori in terms of its variability. Thus, changes in the inner disk are the most likely cause of the brightness changes.

We note that generally there is a low variability of the majority of the YSOs in the $\sigma$\,Ori field investigated. Thus, for most of the objects the variability is too small to accurately determine the slope in the CMDs. Hence, we refrain from analysing the statistics of the CMDs and reddening laws for the entire sample. 

\begin{figure*}
\centering
\includegraphics[angle=0,width=\columnwidth]{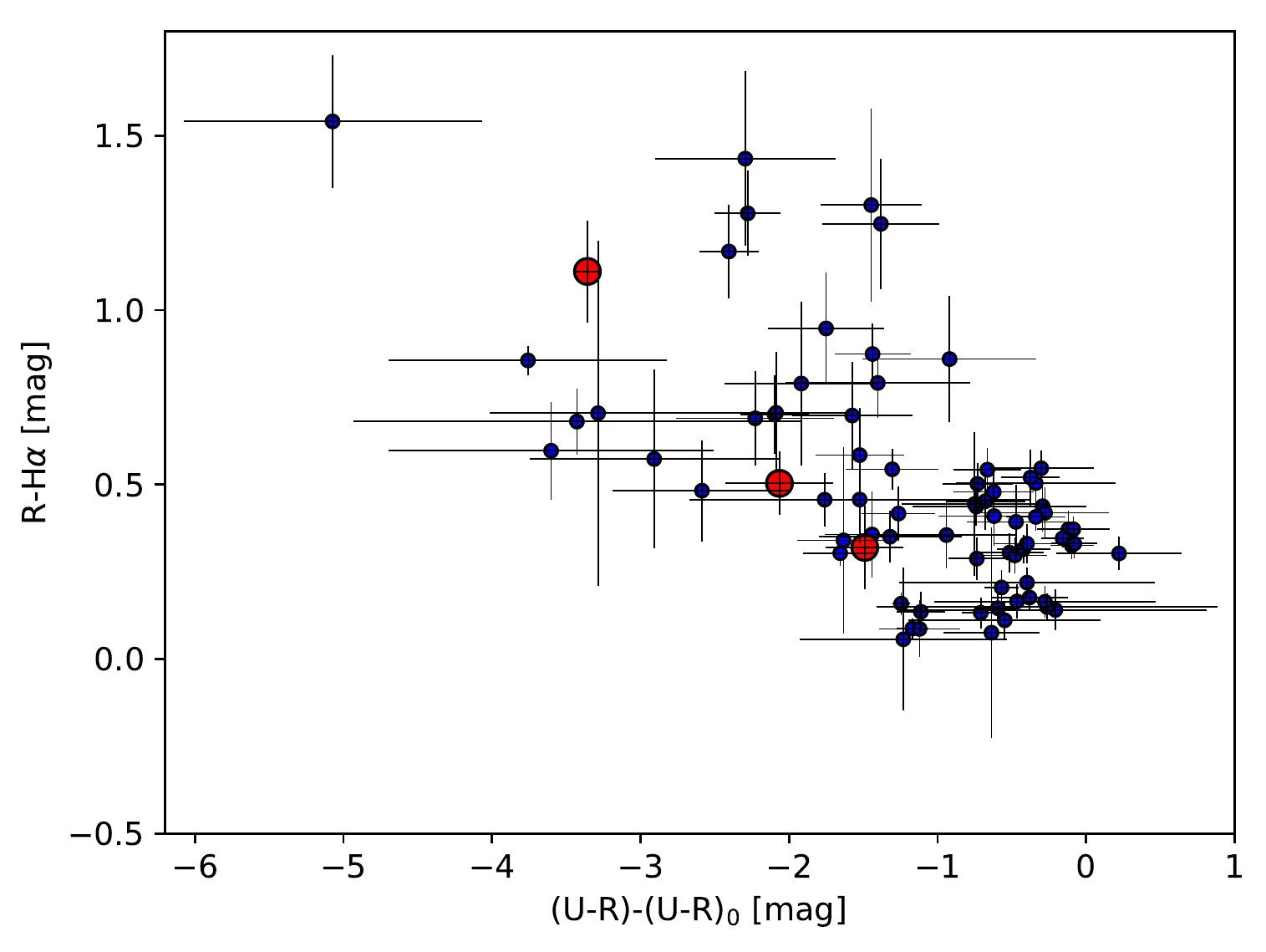} \hfill
\includegraphics[angle=0,width=\columnwidth]{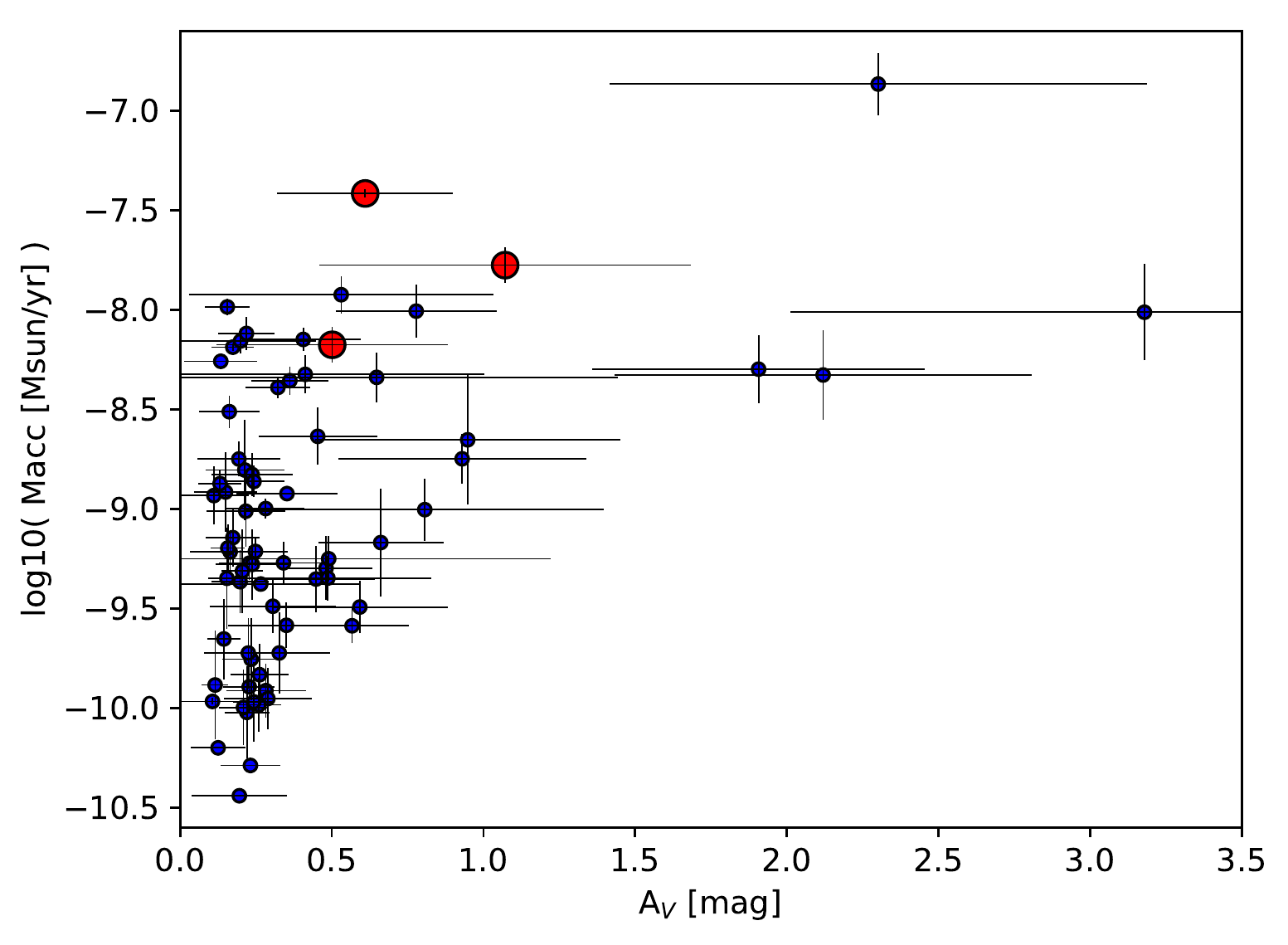} \\
\caption{Mass accretion rate indicators for all YSOs in $\sigma$\,Ori (blue) with the three ULLYSES  targets highlighted (red). {\bf Left:} Median U-Band excess vs R-\ha\ colour. {\bf Right:} Median extinction vs mass accretion rate. See text for details. \label{HST_context_macc}}
\end{figure*}

\subsection{Mass accretion rate indicators} 

As for the three ULLYSES targets, we analyse the R-\ha\ and U-band excess of all YSOs in the field. Contrary to the detailed analysis for every individual photometry data point conducted in Sect.\,\ref{macc}, we look at the median values for each source. All other assumptions made to determine the U-band excess, the extinction and mass accretion rates are unchanged. In other words, the stellar radius is assumed to be the same for all sources. This introduces a small bias in the accretion rates for stars less massive than the ULLYSES targets, making them slightly too low. All error bars are based on the scatter of the individual values from the median for each source.

In the left hand side panel of Fig.\,\ref{HST_context_macc} we show the median U-band excess against the median R-\ha\ values for all $\sigma$\,Ori YSOs as blue circles. The error bars represent the intrinsic standard deviation from the median from the individual measurements for each source. We indicate the three ULLYSES targets in the plot as larger red circles. 

As briefly described in the discussion for the individual targets in Sect.\,\ref{macc}, we observe the expected correlation of the U-band excess with the R-\ha\ colour. There is, however, a large scatter in the correlation, beyond the intrinsic variations of the objects. This is in part caused by the different intrinsic colours, effective temperatures and masses of the objects, but to a large extent also influenced by the different viewing geometries. The majority of objects has very low U-band excess values, less than one magnitude, and also low R-\ha\ values. The three ULLYSES targets are amongst the group with larger U-band excess values, indicating they are amongst the stronger accretors in the sample. 

The right hand side of Fig.\,\ref{HST_context_macc} shows the extinction of all objects plotted against the mass accretion rates. The colour coding is identical to the left hand panel. Given the way they are determined (see Sect.\,\ref{macc}), the values and scatter of the extinction values are in essence an indication of the variability of a source. In agreement with the discussion in the previous section, we find that most YSOs have very low extinction values. There is also the expected correlation of the extinction values and their scatter.

The three ULLYSES targets are clearly amongst the objects with the highest mass accretion rates. Indeed, V510\,Ori and TX\,Ori have the second and third highest mass accretion rate of all objects in the field. This is understandable, as the targets are amongst the brighter and hence more massive cluster members investigated (see bottom right panel in Fig.\,\ref{sigori_ysos}), and typically the mass accretion rates are higher for higher mass stars \cite[e.g.,][]{2020MNRAS.493..234W}. As discussed in the previous section, all ULLYSES targets are in the group of more variable sources. The right hand side of Fig.\,\ref{HST_context_macc} also shows that our HOYS data allow to trace mass accretion rates over a range of more than three orders of magnitude, down to as low as 10$^{-10}$\,M$_\odot$/yr.

In summary we find that when placing the three ULLYSES targets into context with the other YSOs that are detected in the same HOYS target field they are unusual. The targets are amongst the most variable sources and have amongst the highest mass accretion rates. Thus, when drawing conclusions based on the analysis of HST and VLT spectra about the general properties of CTTS, caution needs to be taken. The bias of the ULLYSES targets to have higher accretion rates is an obvious one, as they are amongst the brighter sources in the cluster. Similarly, one would expect a slight bias of fainter sources being less variable, because if they dim too much they will not be detected in the photometry any longer. However, the three ULLYSES targets stand out in terms of their variability. We encourage a similar analysis of other target clusters to see if the $\sigma$\,Ori ULLYSES targets are outliers or if indeed all ULLYSES targets do stand out in the same way.

\section{Conclusions}

Star and planet formation is an intrinsically complex process. Numerous processes, the stellar mass, multiplicity and line of sight orientation all significantly influence the observable spectral and photometric properties, all of which can show changes on a vast variety of time scales. In this paper we have analysed HOYS photometry data of three YSOs in the $\sigma$\,Ori cluster, which are targets of the HST ULLYSES program of low and medium resolution spectroscopy focused on the UV. Our analysis of the photometry time series places these spectra into context and investigates how the selected HST targets compare to the general YSO population in the cluster.

We used GaiaEDR3 data to establish a list of potential YSO members of the $\sigma$\,Ori cluster using only parallax and proper motion. We find that along the sight line there are two distinct populations of YSOs at distances of about 403\,pc and 368\,pc. The nearby group is less populated and forms a coherent group in proper motion space. The more distant main group of YSOs shows a much larger spread in proper motions. It also contains a smaller sub-group separated by approximately 7.5\,km\,s$^{-1}$ in transverse velocity. We selected all potential YSOs from the three sub-groups and applied photometry quality cuts in the associated HOYS data to select a sample of 140 YSOs in the field (in addition to the three ULLYSES targets) that have sufficient HOYS data for a detailed analysis. 

Our optical HOYS data show that the ULLYSES targets have been observed by HST and VLT during periods of changing brightness. Visual magnitudes have changed by 0.5\,mag (TX\,Ori), 1.7\,mag (V505\,Ori) and 0.7\,mag (V510\,Ori) during the three days in which the spectra are taken. The colour changes observed for all sources are not in agreement with any normal interstellar reddening law. For two objects, (TX\,Ori, V505\,Ori) the colours turn blue towards fainter magnitudes, thus indicating a significant contribution of scattered light to the brightness of the objects. The object V505\,Ori shows a clear periodic behaviour with a period of seven days. The amplitudes of about one magnitude in all optical filters cannot be explained by a surface spot model. Instead the object could be an AA\,Tau like source, i.e. it possesses a warped inner disk at the co-rotation radius. 

The variability of all sources shows a clear increase with decreasing wavelength. In particular all objects show much higher variability in the U-band compared to the other filters. As expected, the variability amplitudes also increase with increasing time scale, with the exception of the periodic source V505\,Ori, which plateaus after its seven day period. However, this source still shows U-band variability which indicates that variable accretion influences the brightness on top of the inner disk warp. Comparing the variability of the ULLYSES targets to the other YSOs in the field clearly shows that they are amongst the most variable YSOs in the $\sigma$\,Ori cluster.

The accretion rate indicators of U-band access and R-\ha\ colour show that all sources are active accretors with mass accretion rates of the order of a few 10$^{-8}$\,M$_\odot$/yr. Similar to the variability, the ULLYSES targets are amongst the strongest accretors in the cluster. In contrast, however, this is in part expected as the objects are amongst the brightest, hence more massive and thus stronger accretors. 

Future in-depth investigations of the exact nature of each of the ULLYSES targets will require more detailed modelling of the available HST and VLT spectra. These need to be placed in the context of the photometric state of the sources at the time the spectra are taken. Furthermore, they need to consider the long-term photometric properties, non-ISM extinction and contributions to the brightness from scattered light on the surface of the disks.

\section*{Acknowledgements}

%%%%%%%%%%%%%%%%%%%%%%%%%%%%%%%%%%%%%%%%%%%%%%%%%%%%%%%%%%%%%%%%%%%%%%%%%%%%%%%%%%%%%%%%%
We would like to thank all contributors of observational data for their efforts towards the success of the HOYS project.
%%%%%%%%%%%%%%%%%%%%%%%%%%%%%%%%%%%%%%%%%%%%%%%%%%
%Data Availability Statement

\section*{Data Availability Statement}

The data underlying this article are available in the HOYS database at http://astro.kent.ac.uk/HOYS-CAPS/.


\bibliographystyle{mnras}
\bibliography{bibliography} 

% Don't change these lines
\bsp	% typesetting comment
\label{lastpage}
\end{document}